\documentclass[fleqn,usenatbib]{mnras}

\usepackage[T1]{fontenc}

\DeclareRobustCommand{\VAN}[3]{#2}
\let\VANthebibliography\thebibliography
\def\thebibliography{\DeclareRobustCommand{\VAN}[3]{##3}\VANthebibliography}

\usepackage{graphicx}	
\usepackage{amsmath}	
\usepackage{amssymb}	

\usepackage{tabularx}
\usepackage{multirow}
\usepackage{textcomp}

\title[Thermal bomb supernova modelling and nickel production]{Parameterisations of thermal bomb explosions for core-collapse supernovae and $^{56}$Ni production}

\author[L. Imasheva et al.]{
Liliya Imasheva,$^{1,2}$\thanks{E-mail: imasheva@mpa-garching.mpg.de}
Hans-Thomas Janka,$^{1,3}$
and Achim Weiss$^{1,2}$
\\
$^{1}$Max-Planck Institut f\"ur Astrophysik, Karl-Schwarzschild-Str. 1, 85748 Garching, Germany\\
$^{2}$Ludwig-Maximilians-Universit\"at M\"unchen, Geschwister-Scholl-Platz 1, 80539 Munich, Germany\\
$^{3}$Physik-Department, Technische Universit\"at M\"unchen, James-Franck-Str. 1, 85748 Garching, Germany\\
}

\date{Accepted XXX. Received YYY; in original form ZZZ}

\pubyear{2022}

\begin{document}
\label{firstpage}
\pagerange{\pageref{firstpage}--\pageref{lastpage}}
\maketitle

\begin{abstract}
Thermal bombs are a widely used method to artificially trigger explosions of core-collapse 
supernovae (CCSNe) to determine their nucleosynthesis or ejecta and remnant properties.
Recently, their use in spherically symmetric (1D) hydrodynamic simulations led to the result
that $^{56,57}$Ni and $^{44}$Ti are massively underproduced compared to observational estimates
for Supernova~1987A, if the explosions are slow, i.e., if the explosion mechanism of CCSNe releases 
the explosion energy on long timescales. It was concluded that 
rapid explosions are required to match observed abundances, i.e., the explosion mechanism must 
provide the CCSN energy nearly instantaneously on timescales of some ten to order 100\,ms. 
This result, if valid, would disfavor the neutrino-heating mechanism, which releases the CCSN 
energy on timescales of seconds. Here, we demonstrate by 1D hydrodynamic simulations and 
nucleosynthetic post-processing 
that these conclusions are a consequence of disregarding the initial collapse of the stellar 
core in the thermal-bomb modelling before the bomb releases the explosion energy. We demonstrate 
that the anti-correlation of $^{56}$Ni yield and energy-injection timescale vanishes when the 
initial collapse is included and that it can even be reversed, i.e., more $^{56}$Ni 
is made by slower explosions, when the collapse proceeds to small radii similar to those 
where neutrino heating takes place in CCSNe. We also show that the $^{56}$Ni production
in thermal-bomb explosions is sensitive to the chosen mass cut and that a fixed mass layer 
or fixed volume for the energy deposition cause only secondary differences. Moreover, we propose
a most appropriate setup for thermal bombs. 
\end{abstract}



\section{Introduction}

Core-collapse supernovae (CCSNe) are one of the primary sources of heavy elements in the universe. They
modify and disseminate the products of the nucleosynthesis of their massive stellar progenitors and freshly 
produce radioactive and trans-iron species through various processes such as explosive burning in the shock-heated
ejecta, freeze-out from nuclear statistical equilibrium, neutrino-induced reactions, and neutron and proton 
capture chains \citep[e.g.,][]{2002RvMP...74.1015W,2016ApJ...821...38S,2019ApJ...870....2C,2020ApJ...888...91E,2021RvMP...93a5002C,2021PASA...38...62D}. Thus they play
a crucial role as one of the main drivers of galactic chemical evolution \citep[e.g.,][]{1995ApJS...98..617T,2003Ap&SS.284..539M,2015ApJ...808..132H,2020ApJ...900..179K,2021MNRAS.506.4131W}.

Large sets of progenitor models need to be surveyed with numerical simulations of CCSNe in order to 
account for a rich diversity of pre-collapse conditions, because the evolution of massive stars depends not 
only on the stellar mass and metallicity but also on the amount of rotation and the strength of internal magnetic 
fields, different prescriptions of mass loss rates through stellar winds as well as binary interactions and
mergers. Moreover, uncertainties connected to nuclear rates and the treatment of multi-dimensional effects 
such as angular momentum transport, convection, overshooting, and boundary mixing cause variations.
Systematic investigations of large model sets are therefore indispensable for characterising the electromagnetic 
signatures of CCSNe linked to different types of hydrogen-rich and stripped progenitors \citep[e.g.,][]{2016ApJ...821...38S,2021A&A...652A..64D,2019ApJ...880...59R,2021A&A...656A..61D,2021ApJ...921..143C,2022ApJ...934...67B}.
The same effort is also necessary for predicting the mass distributions of neutron stars and black holes as
the compact remnants of stellar core collapse events \citep[e.g.,][]{2012ApJ...757...69U,2015ApJ...801...90P,2016ApJ...821...38S,2016MNRAS.460..742M,2019ApJ...870....1E,2020ApJ...890...51E,2020ApJ...896...56W,2021A&A...645A...5S,2022ApJ...932L...3M}, 
which are responsible for the growing repository of measured gravitational-wave signals when they are components 
in close binary systems \citep{2021ApJ...913L...7A,2021arXiv211103606T}.

Although the mechanisms of CCSN explosions, either neutrino-driven or magneto-rotational, have been recognized to be 
generically multi-dimensional hydrodynamic phenomena \citep[see, e.g.,][for reviews of full-fledged state-of-the-art, multi-dimensional CCSN simulations]{2005NatPh...1..147W,2007PhR...442...38J,2005ARNPS..55..467M,2012ARNPS..62..407J,2014AIPA....4d1013H,2016ARNPS..66..341J,2016PASA...33...48M,2017RSPTA.37560271C,2020LRCA....6....3M,2021Natur.589...29B},  
three-dimensional (3D) simulations are still constrained by their prohibitive requirements of computational resources.
Therefore the enormous diversity of the progenitor conditions can currently be accounted for only by 
CCSN calculations in spherical symmetry (one dimension; 1D), which permit to follow the long-time evolution in
order to determine the explosion properties including nucleosynthesis and electromagnetic observables for 
large sets of stellar models.

Traditionally, this task has been undertaken by triggering the CCSN explosions artificially either by a so-called 
``thermal bomb'' mechanism \citep[e.g.,][]{1988A&A...196..141S,1989A&A...210L...5H,1990ApJ...349..222T,1996ApJ...460..408T,2001ApJ...555..880N,2006NuPhA.777..424N,2008ApJ...673.1014U,2010ApJ...719.1445M,2011ApJ...729...61B}, in which an outgoing shock wave is initiated by 
dumping thermal energy into a chosen volume around a chosen initial mass cut. This initial mass cut is picked by
nucleosynthesis constraints based on the electron fraction ($Y_e$) of the pre-collapse progenitor or by guessing
the mass of the compact remnant, and it is 
intended to define the borderline between this emerging compact object and the explosion ejecta before fallback 
happens later and possibly brings back matter that does not achieve to become gravitationally unbound. 
Or, alternatively, the outgoing shock was generated by a piston-driven mechanism \citep[e.g.,][]{1988ApJ...330..218W,1995ApJS..101..181W,2002RvMP...74.1015W,2007PhR...442..269W,2008ApJ...679..639Z}, where kinetic
energy is deposited by the outward motion of a piston, which is placed at a chosen Lagrangian mass shell corresponding
to the initial mass cut to push the overlying shells. Refinements of these methods concern, for example, a contraction of 
the location of the piston or initial mass cut to mimic the collapse that precedes the subsequent expansion, and
variations of the duration of the 
energy deposition by the thermal bomb instead of an instantaneous delivery of the energy. In yet another approach \citep[e.g.,][]{2003ApJ...592..404L,2006ApJ...647..483L,2012ApJS..199...38L,2013ApJ...764...21C,2018ApJS..237...13L}
a ``kinetic bomb'' approach was applied in 1D Lagrangian hydrodynamic simulations of CCSN explosions such that
the blast wave is started by imparting
an initial expansion velocity at a mass coordinate around 1\,$M_\odot$, which is still well inside the iron core,
and tuning the value of this velocity such that desired values of the ejected amount of $^{56}$Ni and/or of the final
kinetic energy of the ejecta are obtained. Also multi-dimensional
(2D, 3D) variants of the method of thermal (or kinetic) bombs exist to trigger highly asymmetric blast waves 
and jet-induced 
or jet-associated explosions \citep[see, e.g.,][for a few exemplary applications from a rich spectrum of publications]{1997ApJ...486.1026N,1999ApJ...524..262M,1999ApJ...524L.107K,2000ApJ...531L.119A,2003ApJ...596..401N,2003ApJ...598.1163M,2006ApJ...647.1255N,Ono+2020,Orlando+2020}. 

All of these methods of artificially exploding massive stars depend on numerous free parameters, for example 
the location of the initial mass cut, the width of the energy-deposition region and the timescale of energy
deposition for the thermal bomb,
the duration and depth of the collapse-like contraction, and the initial expansion velocity and coasting 
radius for the piston method, the initial velocity of the kinetic bomb, or the 2D/3D geometry of the energy input.
These parameters are chosen suitably to produce defined values for the explosion energy and the expelled $^{56}$Ni mass or to reproduce multi-dimensional properties of observed supernovae and supernova remnants.
Such degrees of freedom have an influence on the nucleosynthetic yields through the 
initial strength of the shock and the volume and extent of the heating achieved by the thermal energy injection, 
which determine the ejecta mass where sufficiently high peak temperatures for nuclear reactions are reached.  
Moreover, the traditional explosion recipes do not enable one to track the conditions in the innermost
ejecta, whose neutron-to-proton ratio gets reset by the exposure to the intense neutrino fluxes from the 
nascent neutron star or from an accretion torus around a new-born black hole \citep[see, e.g.,][]{2016ApJ...818..123B,2017MNRAS.472..491M,2019Natur.569..241S,2021ApJ...915...28B}.

For these reasons more modern CCSN explosion treatments by means of ``neutrino engines'' have been introduced that
attempt to capture essential effects of the neutrino-driven mechanism but replace the highly complex 
and computationally intense, energy and direction dependent neutrino transport used in full-fledged 
neutrino-hydrodynamical CCSN models by simpler treatments. This line of research has been pursued in
2D and 3D simulations either neglecting neutrino transport and 
replacing it by a so-called light-bulb approximation with chosen (time-dependent) neutrino luminosities 
and spectra \citep[e.g.,][]{1996A&A...306..167J,2000ApJ...531L.123K,2001ApJ...552..756S,2003A&A...408..621K,2006A&A...453..661K,2013ApJ...771...27Y} or by using an approximate, grey description of the neutrino transport with
a boundary condition for the neutrino emission leaving the optically thick, high-density core of the proto-neutron star \citep[e.g.,][]{2006A&A...457..963S,2010ApJ...725L.106W,2013A&A...552A.126W,2015A&A...577A..48W,2017ApJ...842...13W}.

Neutrino-engine treatments are also applied in 1D hydrodynamic CCSN simulations 
with neutrino transport schemes of different levels of refinement for determining the supernova and 
compact remnant properties as well as the associated nucleosynthetic outputs for large sets of 
stellar progenitor models. In these studies neutrino-driven explosions are obtained 
by parametrically increasing the neutrino-energy deposition behind the stalled bounce shock 
\citep{2011ApJ...730...70O}, by describing the neutrino emission of the newly formed neutron star via
a model with parameters that are calibrated to reproduce basic properties of the well-observed CCSNe of 
SN~1987A and SN~1054 (Crab) 
\citep[P-HOTB;][]{2012ApJ...757...69U,2016ApJ...818..124E,2016ApJ...821...38S,2020ApJ...890...51E},
by parametrizing additional energy transfer to the CCSN shock via muon and tau neutrinos (also using 
observational constraints) 
\citep[PUSH;][]{2015ApJ...806..275P,2019ApJ...870....1E,2019ApJ...870....2C,2020ApJ...888...91E}, 
or by also including the effects of convection and turbulence through a modified mixing-length 
theory approach with free parameters adjusted to fit the results of 3D simulations \citep[STIR;][]{2020ApJ...890..127C}.
Alternatively to these novel simulation approaches, semi-analytic descriptions have been applied, either by
using spherical, quasi-static evolutionary sequences to determine the explosion threshold and energy input to
the explosion via a neutrino-driven wind \citep{2015ApJ...801...90P} or by parametrically phrasing the elements 
of multi-dimensional processes that play a role in initiating and powering CCSNe via the neutrino-heating
mechanism \citep{2016MNRAS.460..742M,2021A&A...645A...5S,2022arXiv220400025A}.

Despite these more advanced modelling efforts, which generally reflect more of the physics of the CCSN
explosion mechanism than thermal-bomb or piston models, the latter are still widely used. In fact, thermal
bombs have experienced an increase in popularity in 1D applications recently, because they are applied in 
the open-source codes MESA \citep{2011ApJS..192....3P,2015ApJS..220...15P} and SNEC \citep{2015ApJ...814...63M}. They have 
the advantage of simplicity and great flexibility in their usage, allowing one to control the dynamics 
of the explosion by choosing the value, timescale, mass layer or volume of the energy deposition, and
the evolution of the inner boundary, i.e., if and how the collapse of the stellar core is taken into account.

The sensitivities of the traditional thermal or kinetic bombs and piston mechanisms and of the associated
nucleosynthesis to the involved parameterisations and the corresponding limitations of these methods
have been investigated in previous works, though never comprehensively \citep{1991ApJ...370..630A,2007ApJ...664.1033Y}. 
In a seminal study \citet{1991ApJ...370..630A} discussed the 
parameters employed in the numerical recipes to artificially launch the explosion of a 20\,$M_\odot$ 
progenitor in 1D. They initiated explosions at different locations of enclosed mass, and compared the
ejecta conditions (especially the peak temperatures reached behind the outgoing shocks) as well as the
explosively created nuclear yields. In particular, they considered thermal bomb and piston calculations for two
variations, namely when the inner core was allowed to collapse prior to shock initiation or not. We will call 
such cases ``collapsed'' (C) versus ``uncollapsed'' (U) models. They concluded that the former are a better
representation of the CCSN physics, which is governed by the iron-core collapse to a neutron star. However,
in their study the C-cases also showed more differences between piston and bomb results. Their main concerns
were the uncertainties in the choice of the mass-cut location and in the assumed duration of the initial collapse 
phase, and the differences in the peak temperature because of too much kinetic energy being connected to the piston 
and too much thermal energy to the bomb mechanism. Moreover, they expressed concerns that the instantaneous
energy deposition assumed in their simulations might not be appropriate if the CCSN mechanism is delayed and
the shock receives energy input by neutrino heating for several seconds \citep[as indeed seen in state-of-the-art self-consistent CCSN simulations, e.g.,][]{2021ApJ...915...28B}. 

In a subsequent study, \citet{2007ApJ...664.1033Y} arrived at similar conclusions and found not only a strong
sensitivity of the elemental and isotopic yields of silicon and heavier elements to the assumed explosion
energy, but also considerable differences of the abundances of these nuclei between piston-driven and 
thermal-bomb type explosions even for the same explosion energy. In particular, they considered a 23\,$M_\odot$ 
star, whose collapse, bounce-shock formation, and shock stagnation were followed by a 1D neutrino-hydrodynamics
simulation. Their work was focused on triggering explosions of different energies by thermal energy injection
over time intervals of 20\,ms, 200\,ms, and 700\,ms, starting at 130\,ms after bounce (corresponding to 380\,ms 
after the start of the collapse simulation) and leading to explosions at 150\,ms, 330\,ms, and 830\,ms after bounce,
respectively. The authors reported a considerable increase of intermediate-mass and Fe-group yields with the 
longer delay times of the explosion (i.e., longer duration of the energy deposition) and, in particular 
significantly more (orders of magnitude!) $^{56}$Ni and several times more $^{44}$Ti production 
for models with $1.5\times 10^{51}$\,erg explosion energy and 200\,ms and 700\,ms delay time compared
to a case with the same explosion energy but a short energy injection time of only 20\,ms.

Recently, \citet{2019ApJ...886...47S} (in the following SM19) published a study where they came to exactly 
the opposite conclusion 
based on 1D hydrodynamic CCSN models with a thermal-bomb prescription to trigger the explosions of 15, 
20, and 25\,$M_\odot$ progenitors. They found that the produced amount of $^{56}$Ni {\em decreases} 
with longer timescales of the energy deposition; observational constraints for nucleosynthesis products of
CCSNe could be fulfilled only by 
rapid explosions when the final blast-wave energy was reached within $\lesssim$\,250\,ms, and best compatibility
was obtained for nearly instantaneous explosions where the energy was transferred within $\lesssim$\,50\,ms.  
They interpreted their results as a serious challenge for the neutrino-heating mechanism, which delivers 
the explosion energy in progenitors as massive as those considered by SM19 only on 
timescales that are significantly longer than 1\,s \citep[see][]{2016ApJ...818..123B,2017MNRAS.472..491M,2021ApJ...915...28B,2021Natur.589...29B}.

However, the opposite trends reported by \citet{2007ApJ...664.1033Y} and SM19 for the dependence of the 
$^{56}$Ni yields on the energy-deposition timescale do not need to contradict each other.
In this context it is important to remember that the former study considered collapsed (C) models, whereas 
SM19 did not collapse their stars (using U models) before switching on the thermal energy deposition. This 
is likely to have important consequences for the hydrodynamic response of the stellar gas when the energy 
input happens on different timescales. With the expansion of the heated gas setting in, which is easier 
in an uncollapsed star, expansion cooling takes place. Therefore slow energy injection in a star that has
not collapsed will not be able to achieve sufficiently high temperatures in sufficiently large amounts of
ejecta to enable any abundant production of $^{56}$Ni. 

In our work we aim at investigating this question quantitatively by means of 1D hydrodynamical simulations 
within the framework of the thermal-bomb method. Two different aspects serve us as motivation. 
First, SM19 and also \citet{2019MNRAS.483.3607S} claimed that long 
energy transfer timescales or slow growth rates of the blast-wave energy (``slow explosions'') 
suppress the $^{56}$Ni production. The authors interpreted this proposition as a problem for 
current self-consistent neutrino-driven explosion models and the neutrino-driven mechanism itself. 
Second, our study is supposed to assist the design of suitable thermal-bomb treatments that can serve 
as easy-to-implement methods to conduct systematic CCSN simulations in 1D for large progenitor sets 
without the need of a detailed treatment of neutrinos. Naturally, such approaches can never capture
all aspects of ``realistic'' multi-dimensional CCSN models, in particular not with regard to the 
innermost, neutrino-processed ejecta. Nevertheless, such simplified explosion treatments can still
be useful to answer many observationally relevant questions, in particular since the explosive
nucleosynthesis past the outer edge of the silicon shell is mostly determined by the explosion 
energy and the progenitor structure, but little sensitive to the initiation method of the explosion
\citep{1991ApJ...370..630A}.\footnote{According to present-day understanding, this statement better 
holds good for the outer edge of the oxygen layer instead of the silicon shell.}
Similarly, the explosive nucleosynthesis in these layers is also 
unlikely to depend strongly on the neutrino 
physics and the multi-dimensional hydrodynamic processes that play a crucial role in the CCSN 
mechanism and that determine the observable asymmetries of the explosions.

In this paper we thus investigate the influence of the energy-deposition timescale for thermal 
bombs in collapsed as well as uncollapsed models.
But instead of conducting a complete survey of all free parameters needed to steer the thermal bombs, 
we will stick to simple and well-tested prescriptions already applied in previous publications.
For a diagnostic property we will focus on the produced mass of $^{56}$Ni before any effects of
fallback could modify the ejecta, because fallback will also depend on the radially outward mixing 
of metals and thus on multi-dimensional effects that can be accounted for in 1D models only with 
additional assumptions for parametric treatments. The amount of $^{56}$Ni produced by the CCSN
``engine'' is not only a crucial characteristic of the early dynamics of the explosion but also
a primary observable that governs the light curve and the electromagnetic display of CCSNe
from weeks to many years \citep[e.g.][]{1989ARA&A..27..629A,1994ApJ...437L.115I}.
In a follow-up paper we plan to explore a wider range of thermal-bomb parameterisations and check
them against piston-triggered and neutrino-driven CCSN explosion models. Moreover, in this
subsequent work we will compare the results for a greater selection of products of explosive
nucleosynthesis. 

Our paper is organised as follows. In Section~\ref{section:methods} we briefly describe the stellar
evolution models considered in our study, the methodology of the hydrodynamic explosion modelling, 
the small nuclear reaction network used in the hydrodynamic simulations and the 
large network applied in a more detailed post-processing of the nucleosynthesis. 
In Section~\ref{section:setups} we describe our setup for reference models, 
guided by the calculations reported by SM19, i.e., uncollapsed models, as well
as the variations investigated by us, i.e., collapsed models and different mass 
layers vs. radial volumes for the energy deposition. 
In Section~\ref{section:results} we present our results,
followed by a summary and discussion in Section~\ref{section:conclusions}.


\section{Methods and inputs}
\label{section:methods}

In this section we describe the three aspects of our calculations: the progenitors used as
input models, the corresponding explosion simulations including the definition of the thermal bomb method, and the nucleosynthetic post-processing with an extended nuclear-reaction network. Our progenitors were taken from the work of \citet{2014ApJ...783...10S}, the explosion modelling was performed using the hydrodynamic code \textsc{Prometheus-HOTB} \citep{1996A&A...306..167J,2003A&A...408..621K,2006A&A...457..963S,2007A&A...467.1227A,2012ApJ...757...69U,2016ApJ...818..124E}, but without making use of the neutrino-transport module associated with this code, and the detailed explosive nucleosynthesis was calculated with the SkyNet open-source nuclear network code \citep{2017ApJS..233...18L}.

\begin{table}
	\centering
	\caption{Properties of the progenitors used in this work. $M_{\rm pre}$ is the total pre-collapse mass, $M_{\rm He}$ is the mass of the helium core, $M_{\rm CO}$ the mass of CO core, $M_{s=4}$ is the enclosed mass where the dimensionless entropy $s/k_\mathrm{B} = 4$, and $M_{Y_e=0.48}$ is the enclosed mass where the electron fraction is equal to 0.48. All the masses are in $M_\odot$.}
	\label{tab:psn}
\begin{tabularx}{\columnwidth}{lccccc}
\hline
\hline
$M_{\rm ZAMS}$ & $M_{\rm pre}$ & $M_{\rm He}$ & $M_{\rm CO}$& $M_{s=4}$ & $M_{Y_e=0.48}$ \\ 
\hline
$12.3$  & $11.0599$   &  $3.29162$  &   $2.22902$  & $1.59102$  & $1.23017$ \\ 
$19.7$  & $15.7490$   &  $6.09592$  &   $4.85410$  & $1.53298$  & $1.25635$ \\ 
$21.0$  & $16.1109$   &  $6.62284$  &   $5.37384$  & $1.48435$  & $1.27209$ \\ 
$26.6$  & $15.3093$   &  $8.96794$  &   $7.69495$  & $1.73833$  & $1.38264$ \\ 
\hline
\end{tabularx}
\end{table}

\begin{figure}
	\includegraphics[width=\columnwidth]{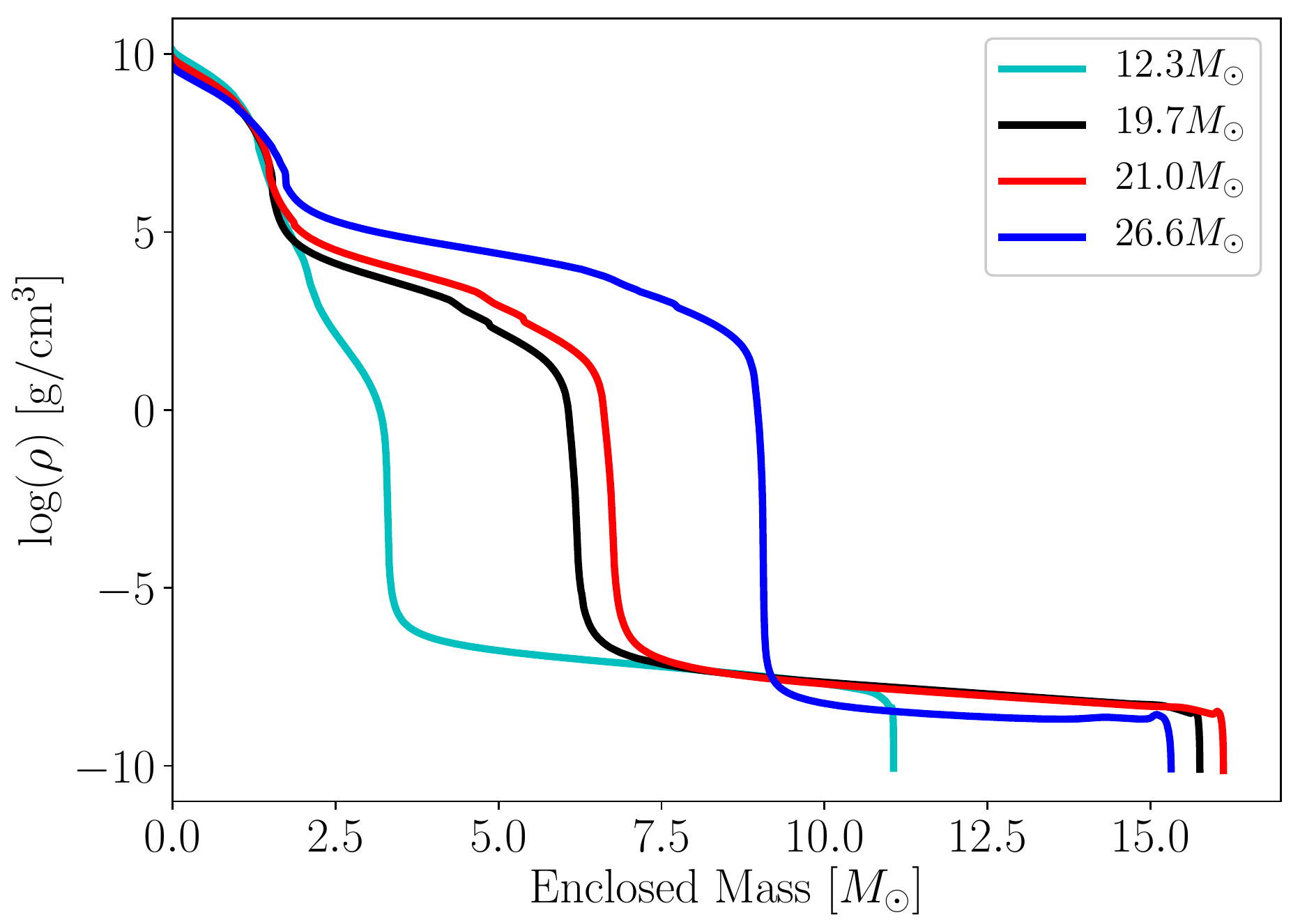}
    \caption{Density structure as a function of enclosed mass for the considered progenitors with $M_{\rm ZAMS}=12.3\,M_\odot$ (cyan line), $19.7\,M_\odot$ (black line), $21.0\,M_\odot$ (red line), and $26.6\,M_\odot$ (blue line). The color convention for the progenitors is kept the same throughout our paper.}
    \label{fig:psn}
\end{figure}

\begin{figure*}

\includegraphics[width=\columnwidth]{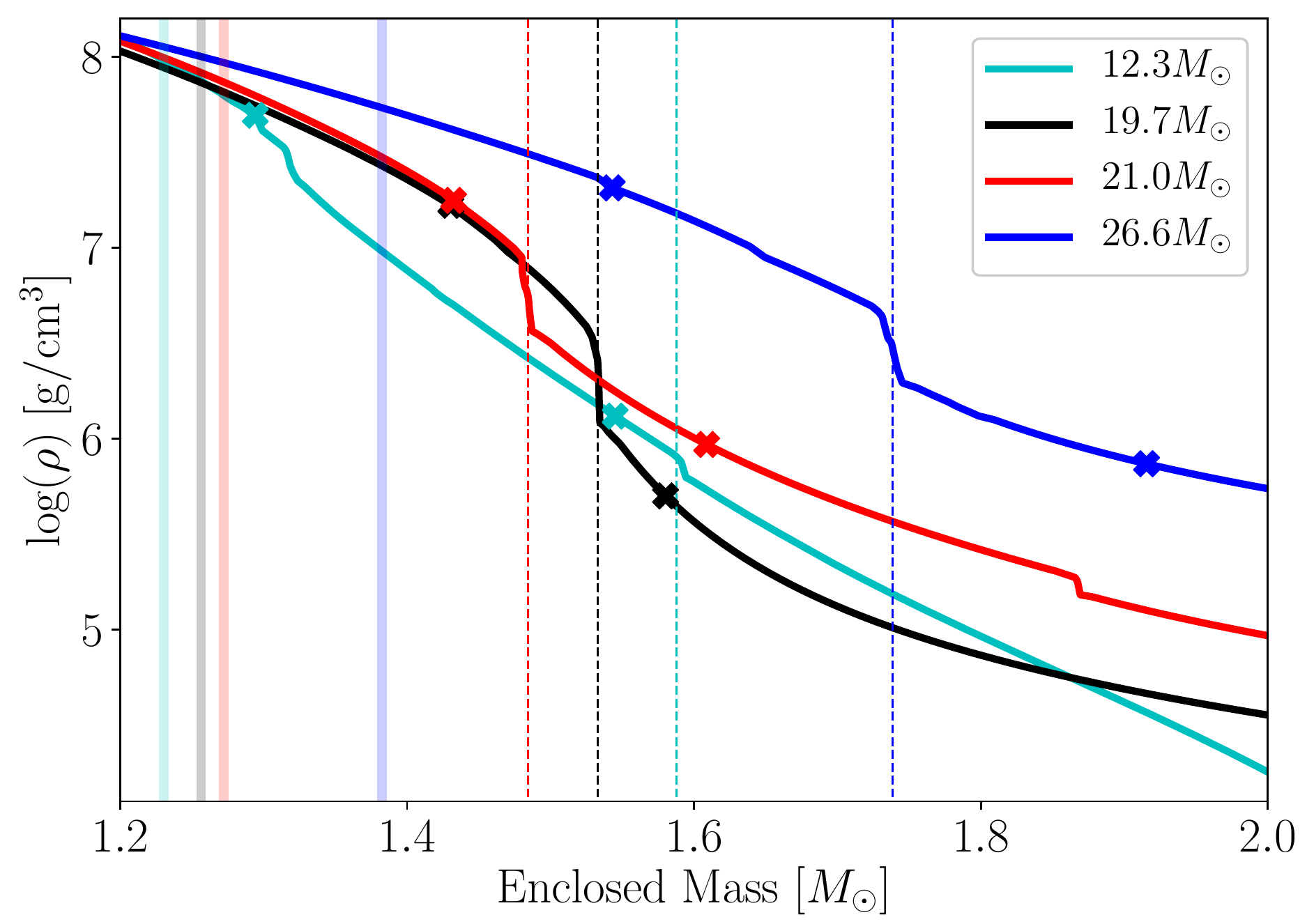}
\includegraphics[width=\columnwidth]{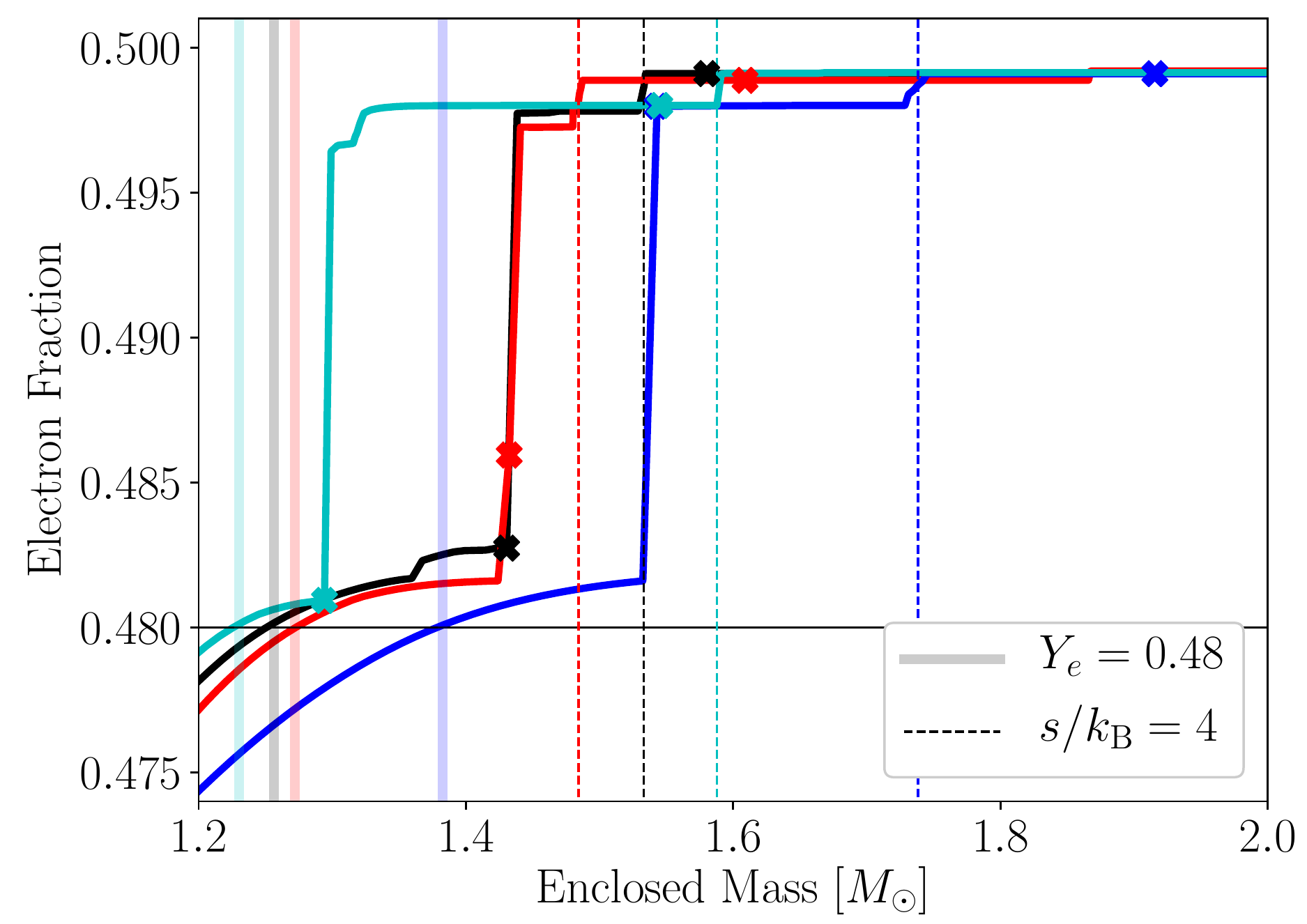}

\includegraphics[width=\columnwidth]{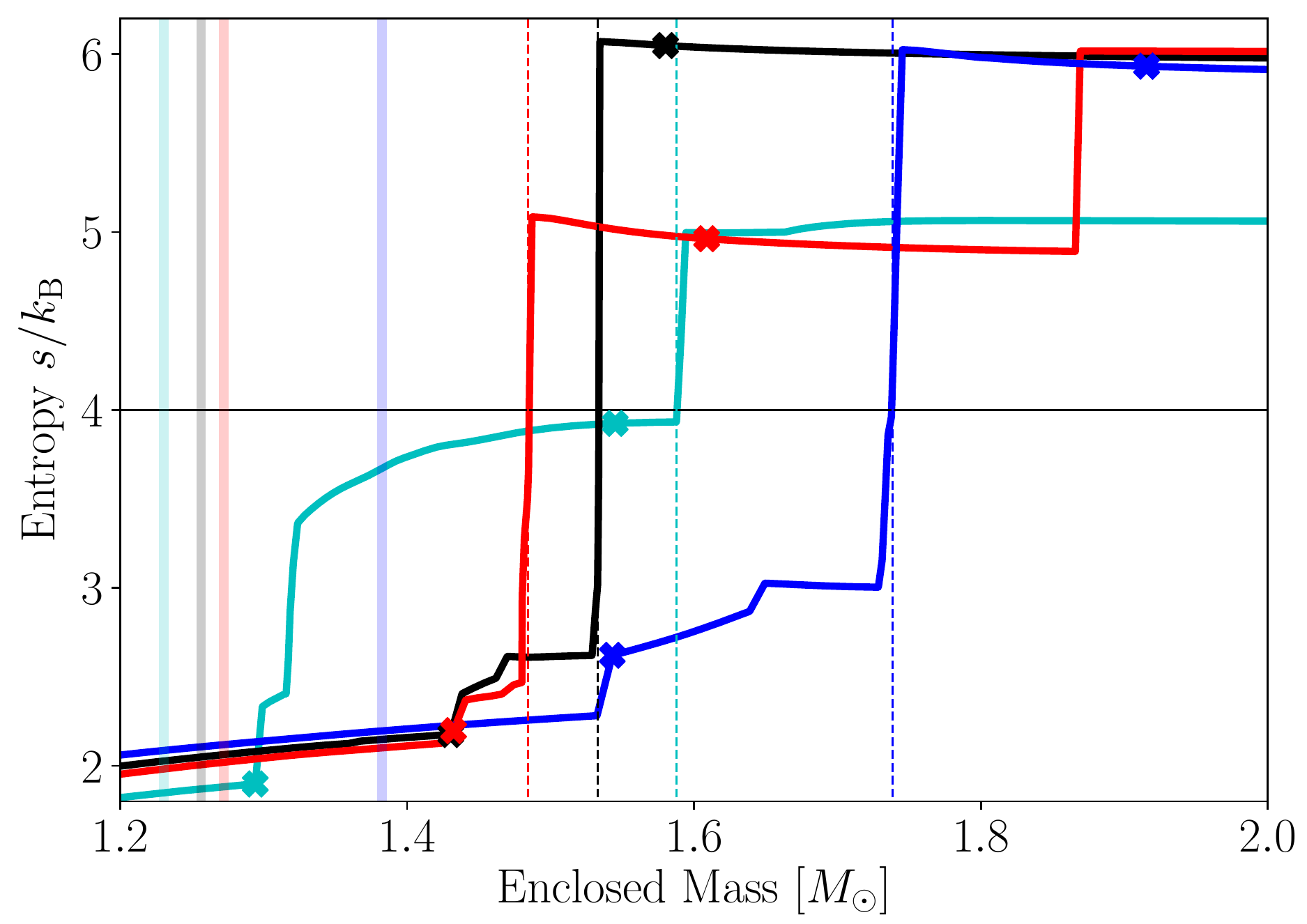}
\includegraphics[width=\columnwidth]{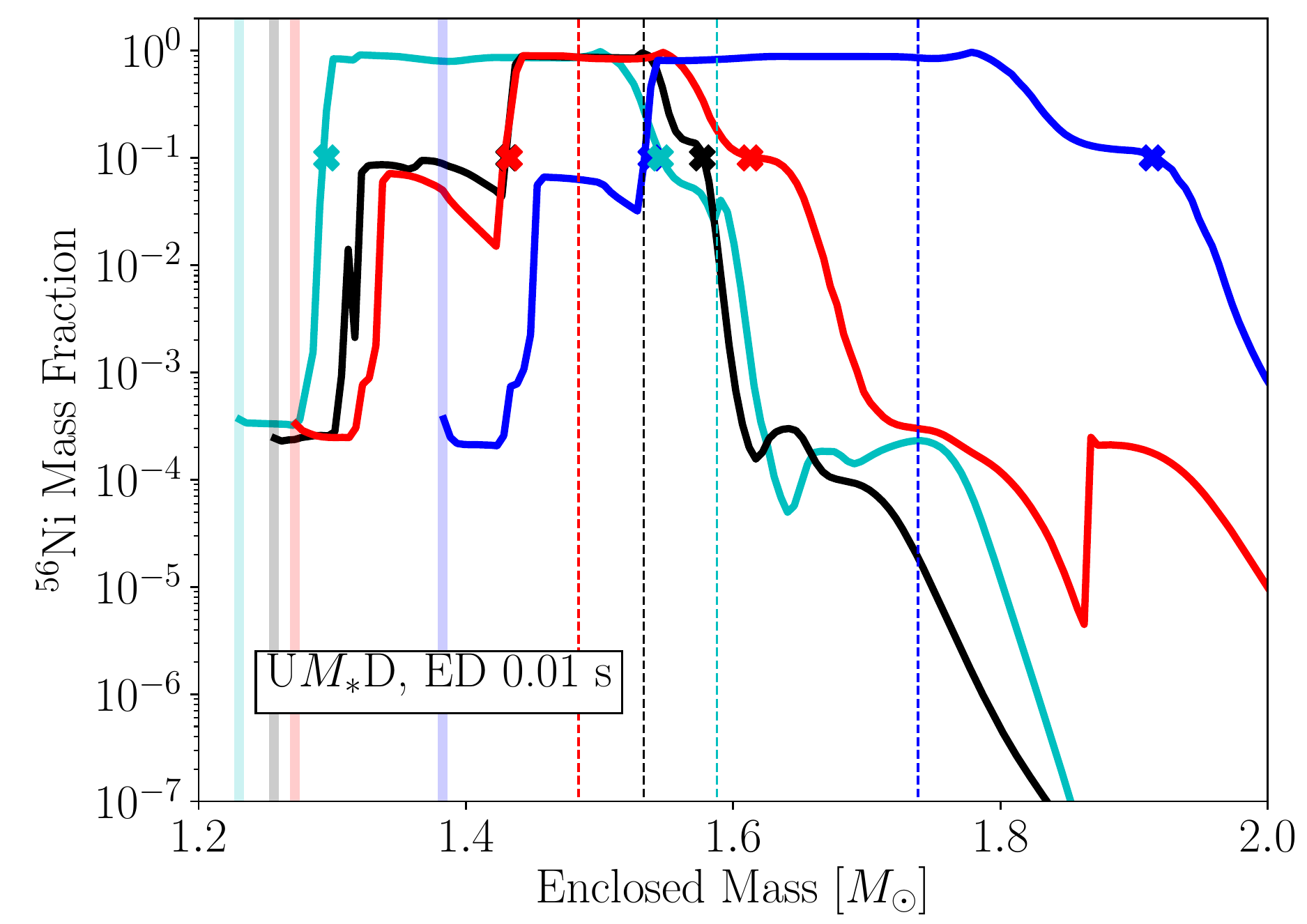}
\caption{Pre-collapse structure of the progenitors used in this work, namely the density (top left), the dimensionless entropy per nucleon $s/k_\mathrm{B}$ (bottom left), and the electron fraction $Y_e$ (top right) versus enclosed mass. Vertical lines indicate the inner grid boundaries chosen in our explosion simulations, with the line colors corresponding to the colors chosen for the four stellar models: the pale solid lines mark the deeper locations where $Y_e=0.48$, which is also indicated by the horizontal black line in the $Y_e$ plot, and the short-dashed lines define the points where the dimensionless entropy per nucleon $s/k_\mathrm{B}$ equals 4, which can also be seen by the horizontal black line in the $s/k_\mathrm{B}$ plot. The lower right panel displays the mass fraction of $^{56}$Ni obtained as function of enclosed mass for our default setup of uncollapsed models with deep inner boundary; the energy-deposition (ED) timescale assumed for the displayed case is $t_{\rm inj}=0.01$\,s. The crosses on the stellar profiles in all panels mark the locations of the inner and outer edges of the main production region of $^{56}$Ni (see Section~\ref{section:psn} for the definition of this region). Note that due to the similarity of the profiles the red and black crosses in the two left panels and the lower right panel partly overlap.}
\label{fig:psn_closer}
\end{figure*}

\subsection{Presupernova models}
\label{section:psn}

The progenitor models for this work were computed with the 1D hydrodynamics code KEPLER \citep{1978ApJ...225.1021W} and are a subset of the large model set published by \citet{2014ApJ...783...10S}. They represent non-rotating stars with solar metallicity, which were evolved from the main sequence until the onset of the iron-core collapse. The physics of this set of progenitors was discussed in detail in the literature \citep[e.g.][]{2002RvMP...74.1015W, 2007PhR...442..269W}.

In order to investigate basic features of the nickel production using different setups for the thermal bomb triggering the CCSN explosion, we selected four progenitors with zero-age-main-sequence (ZAMS) masses of $M_{\rm ZAMS} = 12.3$, $19.7$, $21.0$, and $26.6\,M_\odot$. Their characteristic properties are listed in Table~\ref{tab:psn}, where $M_{\rm pre}$ is the total pre-collapse mass, $M_{\rm He}$ is the helium-core mass defined by the mass coordinate where $X({\rm H})\le0.2$, $M_{\rm CO}$ is the mass of the carbon-oxygen core associated with the location where $X({\rm He})\le0.2$, $M_{s=4}$ is the mass enclosed by the radius where the value of the dimensionless entropy per nucleon is $s/k_\mathrm{B}=4$ (where $k_\mathrm{B}$ is the Boltzmann constant), and $M_{Y_e=0.48}$ is the enclosed mass
where the electron fraction is $Y_e=0.48$. 

This selection of the progenitors is motivated by the aim to cover approximately the same range of progenitor masses as considered by SM19. For the lighter progenitors, we investigated two models with $M_{\rm ZAMS}=12.3\,M_\odot$ and $19.7\,M_\odot$, representing two extreme cases with respect to their density declines at mass coordinates $m \gtrsim 1.5\,M_\odot$ and differing from each other by the shape of their corresponding density profiles (see Figures~\ref{fig:psn} and \ref{fig:psn_closer}). 
Our simulations are intended to explore the uncertainties in the thermal-bomb modelling, and these progenitor models exhibit a different behavior in the explosive nickel production based 
on their structure and our calculations, as will be discussed in Section~\ref{section:results}.

The upper two panels and the lower left one in Figure~\ref{fig:psn_closer} visualize the progenitor structures in more details by showing density, electron fraction $Y_e$, and dimensionless entropy per nucleon as functions of enclosed mass. The crosses indicate the inner and outer edges of the regions where most of the $^{56}$Ni is produced, based on the results given in the lower right panel of Figure~\ref{fig:psn_closer}. This last panel displays, as an exemplary case, the nickel mass fractions for one of our setups (namely the uncollapsed models with deep inner boundary and an energy deposition timescale of 0.01\,s, see below).
The main region of $^{56}$Ni production is defined by the requirement that the mass fraction of this isotope is greater than 0.1 and consequently at least 90\% of its total yield are produced between the limits marked by two crosses.

Nickel and other heavy elements are mainly produced in the close vicinity of the inner grid boundaries of the simulations (for the relevant models these are marked by vertical pale solid lines in Figure~\ref{fig:psn_closer}), i.e., close to the mass region that is assumed to end up in the newly formed neutron star. Therefore differences in the $^{56}$Ni production will be connected to differences in the progenitor structures between the inner grid boundary and below roughly $2\,M_\odot$. 

\subsection{Hydrodynamic explosion modelling} 
\label{section:boom}
 
The progenitor models were exploded by making use of the 1D hydrodynamics code \textsc{Prometheus-HOTB}, or in short P-HOTB, which
solves the hydrodynamics of a stellar plasma including evolution equations for the electron fraction and the nuclear species in a conservative manner on an Eulerian grid, employing a higher-order Godunov scheme with an exact Riemann solver. The code employs a micro-physical model of the equation of state that includes a combination of non-relativistic Boltzmann gases for nucleons and nuclei, arbitrarily degenerate and arbitrarily relativistic electrons and positrons, and energy and pressure contributions from trapped photons. Although the hydrodynamics is treated in the Newtonian limit, the self-gravity of the stellar matter takes into account general relativistic corrections. Relevant details of the code and its upgrades over time can be found
in the papers of \citet{1996A&A...306..167J,2003A&A...408..621K,2006A&A...457..963S,2007A&A...467.1227A,2012ApJ...757...69U,2016ApJ...818..124E,2020ApJ...890...51E}.
The CCSN models discussed in this paper were computed with a radial mesh of 2000 zones, geometrically distributed
from the inner grid boundary at radius $R_\mathrm{ib}$ to the stellar surface with a resolution of $\Delta r/R_\mathrm{ib} = 10^{-3}$ in the innermost grid cell and $\Delta r/r < 0.013$ everywhere on the grid.

The central volume ($r < R_\mathrm{ib}$) was excluded from the computational mesh and replaced by an inner grid boundary at $R_\mathrm{ib}$ plus a gravitating point mass at the grid center. This introduces a first 
parameter into the artificial explosion modelling, namely the enclosed mass at the location of this inner boundary 
(sometimes called the (initial) mass cut), which is identified with the initial mass of the compact remnant.
In our calculations we considered two cases for the choice of the position of the inner boundary. In a first case, following SM19, it was placed where $Y_e=0.48$ in the outer regions of the progenitor's iron core. This deep location, indicated by the letter ``D'' in the names of the corresponding explosion models, is extreme because the ejection of matter with $Y_e$ as low as 0.48 is severely constrained by observational bounds on the $^{58}$Ni production in CCSNe \citep[see, e.g., SM19 and][]{2015ApJ...807..110J}. In a second case we placed the inner grid boundary at the location where the dimensionless entropy per nucleon rises to $s/k_\mathrm{B}=4$, which corresponds to the base of the oxygen shell. This position is thus farther out in mass (see Table~\ref{tab:psn}) and is indicated by the letter ``O'' in the names of the corresponding explosion simulations.
This location was also used in 1D piston-driven CCSN models by \citet{2007PhR...442..269W} and \citet{2008ApJ...679..639Z} 
and is better compatible with the initial mass cut developing in neutrino-driven explosions \citep[see, e.g.,][]{2016ApJ...818..124E}.
In Figure~\ref{fig:psn_closer} these two choices of the inner boundary position are indicated by vertical lines 
for each progenitor. 
Realistically, the surface of the proto-neutron star is likely to be located somewhere between these two positions and will also be determined only after possible fallback has taken place. 
The mass of the proto-neutron star cannot be significantly larger than the base of the oxygen shell (``O'' location), because otherwise the typical neutron star masses will be too big to be compatible with observations \citep[][]{2007PhR...442..269W}. 

The temporal behavior of the inner boundary is likely to affect the dynamics of the explosion, because the effect of the 
deposition of energy by the thermal-bomb method will depend on the state of the matter the energy is transferred to. If the boundary radius was kept constant at its initial value, i.e., if the stellar core was not collapsed and the explosion was initiated right away, this corresponds to uncollapsed models and is denoted by the initial letter ``U'' in the model names. Alternatively, if the boundary was first contracted to mimic the collapse of the progenitor's degenerate core, this allowed the matter just exterior to the inner boundary to move to the higher densities and deeper into the gravitational potential of the central mass before the bomb was started. This approach defines our collapsed models and is indicated by the initial letter ``C'' in the names of the corresponding explosion models.

In the thermal bomb method, the CCSN explosion is triggered by thermal energy input into a chosen layer around the inner boundary, either instantaneously \citep[e.g.][]{1991ApJ...370..630A} or over a chosen interval in time \citep[e.g., SM19 and][]{2007ApJ...664.1033Y}.
The injected energy $E_\mathrm{inj}$, the mass layer $\Delta M$ or volume $\Delta V$ where the energy is deposited, and the timescale of the energy injection $t_\mathrm{inj}$ are free parameters of such a procedure. 
These parameters define energy transfer rates per unit of mass or volume, respectively:
\begin{eqnarray}
\dot e_\mathrm{inj, M} &=& \frac{E_\mathrm{inj}}{\Delta M\,t_\mathrm{inj}} \,, \label{eq:edotm} \\
\dot e_\mathrm{inj, V} &=& \frac{E_\mathrm{inj}}{\Delta V\,t_\mathrm{inj}} \,. \label{eq:edotv} 
\end{eqnarray}
The expressions of Equations~(\ref{eq:edotm}) and (\ref{eq:edotv}) assume that, for simplicity, the energy input rate is constant in time and thus the deposited energy grows linearly with time.

The total injected energy $E_\mathrm{inj}$ was varied in order to obtain a chosen value for the terminal explosion energy $E_\mathrm{exp}$ at infinity. In our study we considered CCSN models with an explosion energy close to $E_\mathrm{exp} = 10^{51}$\,erg and determined this value at $t \ge 80$\,s, at which time it had saturated in each model.
The layer of the energy deposition is characterized by two fixed Lagrangian mass coordinates in the case of $\Delta M$ and two fixed radii in the case of $\Delta V$. In our simulations the inner boundary of the energy-deposition layer (IBED) was set to be the inner boundary of the computational grid, and the outer boundary of the energy-deposition layer (OBED) depends on the choice of $\Delta M$ or $\Delta V$. 
The last parameter here is the timescale of the energy deposition $t_\mathrm{inj}$, which defines how fast the shock will be developing and which we varied in our study, following SM19.

During the CCSN simulations carried out for our investigation, we employed a reflecting inner boundary condition in order to maintain the pressure support while the explosion was still developing. This setting is motivated by the continued push of the CCSN ``engine'' (either neutrino-driven or magneto-rotational) over the period of time when the blast-wave energy builds up. We note in passing that we do not intend to discuss any effects of fallback, which typically play a role only on timescales longer than those considered for nucleosynthesis in the present work.

\begin{figure*}
    \centering
	\includegraphics[width=0.92\textwidth]{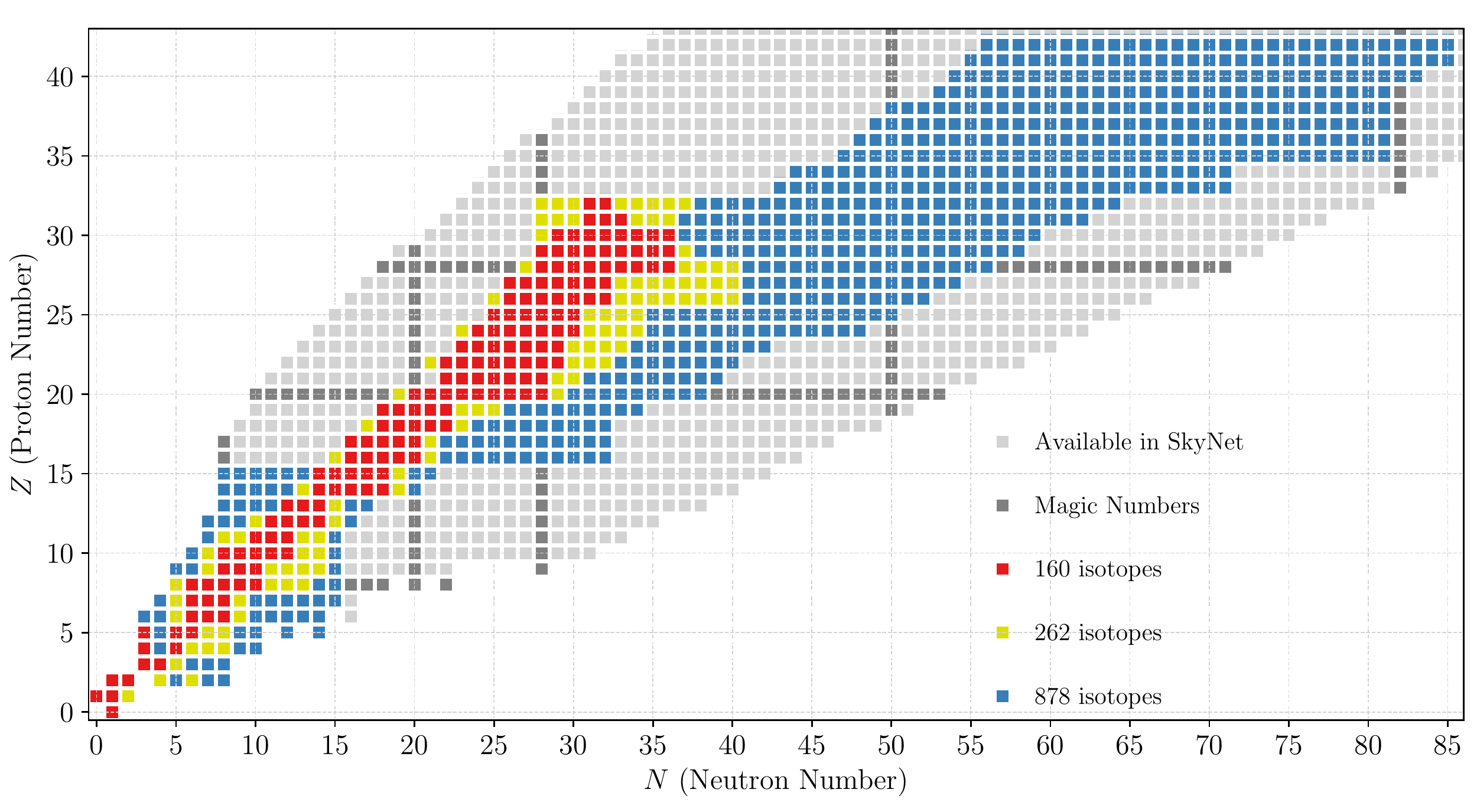}
    \caption{Nuclear chart visualizing the three sets of isotopes used in this work for testing the final nucleosynthetic outputs. The test calculations were done under extreme conditions of density, $Y_e$, and entropy, and were carried out until $t=10$\,s. Their results showed convergence in the final yields of the 50 most abundantly produced isotopes between the sets with $262$~isotopes and $878$~isotopes.}
        \label{fig:network}

\end{figure*}

\begin{table}
	\centering
	\caption{Nuclear species used for the nucleosynthetic post-processing of our thermal-bomb CCSN models with SkyNet.}
	\label{tab:list}
\begin{tabularx}{\columnwidth}{c}
\hline
\hline	
Nuclei used in the 262-species network\\
\end{tabularx}
\begin{tabularx}{\columnwidth}{ccccc}
\hline
n & $^{1-3}$H & $^{3-4,6,8}$He& $^{6-8}$Li& $^{7,9-12}$Be \\
 $^{8,10-13}$B & $^{11-15}$C & $^{12-16}$N& $^{13-21}$O & $^{16-23}$F \\
 $^{17-24}$Ne & $^{19-25}$Na & $^{22-27}$Mg & $^{25-28}$Al & $^{27-33}$Si \\
 $^{29-34}$P & $^{31-37}$S& $^{33-38}$Cl & $^{35-41}$Ar & $^{37-44}$K \\
 $^{39-49}$Ca& $^{43-51}$Sc& $^{43-54}$Ti& $^{46-56}$V & $^{47-58}$Cr \\
 $^{50-59}$Mn & $^{51-66}$Fe& $^{53-67}$Co& $^{55-68}$Ni& $^{57-66}$Cu\\
 $^{58-66}$Zn & $^{59-67}$Ga& $^{60-69}$Ge\\
\hline
\end{tabularx}
\end{table}

\subsection{Reaction Networks}
\label{section:reactions}

A small $\alpha$-network is consistently coupled to the hydrodynamic modelling with \textsc{P-HOTB}. It is described in the relevant details by \citet{1986A&A...162..103M} and is capable of tracking the bulk nucleosynthesis and thus to account for the contribution to the explosion energy provided by explosive nuclear burning.
The network includes the $13$ isotopes of the alpha-chain, $^{4}$He, $^{12}$C, $^{16}$O, $^{20}$Ne, $^{24}$Mg, $^{28}$Si, $^{32}$S, $^{36}$Ar, $^{40}$Ca, $^{44}$Ti, $^{48}$Cr, $^{52}$Fe, and $^{56}$Ni, plus a ``tracer nucleus'' $^{56}$Tr, which is connected to the network with the reaction rates of $^{56}$Ni and is supposed to keep track of the formation of neutron-rich species in matter with considerable neutron excess, i.e., when $Y_e < 0.49$ \citep{2000ApJ...531L.123K,2001AIPC..561...21K,2003A&A...408..621K}. The network calculations made use of the reaction rates of \citet{1996ApJ...460..408T} and they were applied for temperatures between $0.1$\,GK and $9$\,GK, whereas for higher temperatures nuclear statistical equilibrium (NSE) was assumed.

In order to perform more detailed nucleosynthesis calculations of our models in a post-processing step, we made use of the modular nuclear reaction network library SkyNet \citep{2017ApJS..233...18L}. 
For this purpose we extracted the temperature and density evolution of selected mass-shell trajectories from our CCSN explosion simulations with \textsc{P-HOTB} and applied the SkyNet network to each of these shells, starting out with shells closest to the mass cut between ejecta and proto-neutron star and constraining the network calculations to the same regime in temperature as used for the small network in \textsc{P-HOTB}, namely to the interval between $0.1$\,GK and $9$\,GK. Adding up the nuclear abundances obtained for all mass shells that ended up to be ejected (i.e. that expanded outward continuously until the end of the hydrodynamic simulation) provided the integrated yields of chemical elements and isotopes. If mass shells reached a peak temperature above $T_{\rm NSE}=9$\,GK during their infall or explosive expansion, the network calculations were started only at the time when the temperature finally dropped below 9\,GK, using the local NSE composition as initial condition.\footnote{Note that any preceding nuclear composition is erased when NSE is established.} Otherwise, if mass shells did not reach temperatures as high as 9\,GK, the composition evolution of these mass shells was followed with SkyNet from the beginning of their infall through their shock heating and ejection, and the initial composition was taken from the progenitor data. The mass resolution for post-processing the nucleosynthesis was chosen to
be $10^{-4}\,M_\odot$ for the innermost part of the ejecta below a stellar mass coordinate of
$2\,M_\odot$, and $0.005\,M_\odot$ farther out.

SkyNet allows to define any selection of isotopes of interest and to define their relevant reactions. We took great care to employ a sufficiently big set of isotopes and to include all of their important reactions. 
To arrive there we started with three different sets of isotopes, inspired by their use in the literature: a small network with 160 isotopes \citep{2021ApJ...921..113S}, a medium-sized network with 204 isotopes \citep{2015ApJS..220...15P}, and a large network with 822 isotopes \citep{1992ApJ...395..202W}. We modified the medium and the large ones in a way that every next-bigger list included the previous one. On top of that we added more light isotopes; for the largest network, for example, we included all nuclear species available in SkyNet with $Z\le15$ and $N\le15$. After these modifications, we ended up with selections of 160, 262, and 878 isotopes (see Figure~\ref{fig:network}). With all of these three versions of the network we performed nucleosynthesis calculations for about 20
trajectories with the most extreme conditions (in density, $Y_e$, and temperature) picked from the set of our CCSN models. 
We found that the yields were well determined with an accuracy of better than 1\% for the 25 most abundantly produced isotopes when including 262 species compared to the case with 878 isotopes. Therefore we continued all further analyses with this medium-sized network, whose selection of nuclei is listed in Table~\ref{tab:list}.

In our present work, we will only discuss the production of $^{56}$Ni based on our network calculations with the 262-isotope setup of SkyNet. We focus on this nickel isotope and aim at exploring the dependence of its production on the parameterisation of the thermal-bomb treatment, because the mass of $^{56}$Ni ejected in the explosion is an important diagnostic quantity for CCSN observations \citep[e.g.,][]{1989ARA&A..27..629A,2017ApJ...841..127M,2021A&A...655A..90Y,Valerin+2022}. Any implementation of a method to artificially trigger explosions in CCSN models should therefore be checked for its ability to provide reasonable predictions of the $^{56}$Ni yield and for the robustness of these predictions concerning changes of the (mostly rather arbitrarily) chosen values of the parameters steering the trigger mechanism. The produced amount of $^{56}$Ni is particularly useful to assess these questions, because the isotope is made in the innermost CCSN ejecta. Therefore it is potentially most immediately and most strongly affected by the artificial method (or by the physical mechanism) that is responsible for initiating the explosion.


\begin{table*}
	\centering
	\caption{Properties of the thermal-bomb models computed in this work. $M_\mathrm{ZAMS}$ is the ZAMS mass of the progenitor star, ``Model'' is our name for the specific CCSN simulation (see text for our naming convention), ``Inner Grid Boundary'' specifies the criterion for placing the inner grid boundary, $M_\mathrm{ib}$ is the corresponding enclosed mass, $t_\mathrm{coll}$ is the collapse time, $r_\mathrm{min}$ is the minimum radius for the collapse phase, $\Delta M$ is the mass of the energy-injection layer or, respectively, the initial mass in the volume where the energy is injected, $t_\mathrm{inj}$ is the range of energy-deposition timescales considered, and $E_\mathrm{exp}$ is the range of final explosion energies to which the CCSN models for different energy-injection timescales were calibrated (see Section~\ref{section:setups} for details). Note that per construction all 26.6 $M_\odot$ models have identical values for $\Delta M$ in this listing (unless
	$\Delta M = 0.005\,M_\odot$). }
	\label{tab:explosions}
\begin{tabular}{lcccccccc}
\hline
\hline
$M_\mathrm{ZAMS}$ & Model & Inner Grid   & $M_\mathrm{ib}$ & $t_\mathrm{coll}$ & $r_\mathrm{min}$ & $\Delta M$  & $t_\mathrm{inj}$ & $E_\mathrm{exp}$ \\
$[M_\odot]$       &       & Boundary     & [$M_\odot$]     & [s]               & [cm]             & [$M_\odot$] & [s]              & [$10^{51}$\,erg] \\
\hline
\hline
$12.3$ & U$12.3$D  & $Y_e = 0.48$         & $1.230$ & no collapse & $-$     & $0.05$  & $0.01-2.0$    & $1.0099-1.0170$ \\
$12.3$ & U$12.3$DM\textquotesingle  & $Y_e = 0.48$         & $1.230$ & no collapse & $-$     & $0.005$  & $0.01-2.0$    & $0.9834-1.0241$ \\
\hline 
$19.7$ & U$19.7$D  & $Y_e = 0.48$         & $1.256$ & no collapse & $-$     & $0.05$  & $0.01-2.0$    & $1.0003-1.0178$ \\
$19.7$ & C$19.7$D  & $Y_e = 0.48$         & $1.256$ & $0.45$ & $5\cdot10^7$ & $0.05$  & $0.01-2.0$    & $1.0067-1.0125$ \\
$19.7$ & C$19.7$O  & $s/k_\mathrm{B} = 4$ & $1.533$ & $0.45$ & $5\cdot10^7$ & $0.05$  & $0.01-2.0$    & $1.0048-1.0160$ \\
$19.7$ & xC$19.7$O & $s/k_\mathrm{B} = 4$ & $1.533$ & $0.45$ & $1.5\cdot10^7$ & $0.05$ & $0.01-2.0$   & $0.9977-1.0260$ \\
\hline 
$19.7$ & U$19.7$DM & $Y_e = 0.48$         & $1.256$ & no collapse & $-$     & $0.043$ & $0.01-2.0$    & $1.0018-1.0177$ \\
$19.7$ & C$19.7$DM & $Y_e = 0.48$         & $1.256$ & $0.45$ & $5\cdot10^7$ & $0.044$ & $0.01-2.0$    & $1.0016-1.0169$ \\
$19.7$ & C$19.7$OM & $s/k_\mathrm{B} = 4$ & $1.533$ & $0.45$ & $5\cdot10^7$ & $0.027$ & $0.01-2.0$    & $1.0000-1.0151$ \\
$19.7$ & U$19.7$DM\textquotesingle & $Y_e = 0.48$         & $1.256$ & no collapse & $-$     & $0.005$ & $0.01-2.0$    & $0.9889-1.0188$ \\
$19.7$ & C$19.7$OM\textquotesingle & $s/k_\mathrm{B} = 4$ & $1.533$ & $0.45$ & $5\cdot10^7$ & $0.005$ & $0.01-2.0$    & $1.0061-1.0394$ \\
\hline
$19.7$ & C$19.7$OV & $s/k_\mathrm{B} = 4$ & $1.533$ & $0.45$ & $5\cdot10^7$ & $0.027$ & $0.01-0.5$ & $0.9982-1.0302$ \\
$19.7$ & xC$19.7$OV & $s/k_\mathrm{B} = 4$ & $1.533$ & $0.45$ & $1.5\cdot10^7$ & $0.027$ & $0.01-2.0$ & $1.0009-1.0400$ \\
\hline
$21.0$ & U$21.0$D  & $Y_e = 0.48$         & $1.272$ & no collapse & $-$     & $0.05$  & $0.01-2.0$    & $1.0185-1.0334$ \\
$21.0$ & C$21.0$D  & $Y_e = 0.48$         & $1.272$ & $0.45$ & $5\cdot10^7$ & $0.05$  & $0.01-2.0$    & $1.0161-1.0302$ \\
$21.0$ & C$21.0$O  & $s/k_\mathrm{B} = 4$ & $1.484$ & $0.45$ & $5\cdot10^7$ & $0.05$  & $0.01-2.0$    & $1.0160-1.0266$ \\
$21.0$ & xC$21.0$O & $s/k_\mathrm{B} = 4$ & $1.484$ & $0.45$ & $1.5\cdot10^7$ & $0.05$ & $0.01-2.0$ & $1.0210-1.0363$ \\
\hline
$21.0$ & U$21.0$DM & $Y_e = 0.48$         & $1.272$ & no collapse & $-$     & $0.042$ & $0.01-2.0$    & $1.0207-1.0334$ \\
$21.0$ & C$21.0$DM & $Y_e = 0.48$         & $1.272$ & $0.45$ & $5\cdot10^7$ & $0.041$ & $0.01-2.0$    & $1.0205-1.0319$ \\
$21.0$ & C$21.0$OM & $s/k_\mathrm{B} = 4$ & $1.484$ & $0.45$ & $5\cdot10^7$ & $0.068$ & $0.01-2.0$    & $1.0196-1.0247$ \\
$21.0$ & U$21.0$DM\textquotesingle & $Y_e = 0.48$         & $1.272$ & no collapse & $-$     & $0.005$ & $0.01-2.0$    & $1.0251-1.0545$ \\
$21.0$ & C$21.0$OM\textquotesingle & $s/k_\mathrm{B} = 4$ & $1.484$ & $0.45$ & $5\cdot10^7$ & $0.005$ & $0.01-2.0$    & $1.0067-1.0417$ \\
\hline
$21.0$ & C$21.0$OV & $s/k_\mathrm{B} = 4$ & $1.484$ & $0.45$ & $5\cdot10^7$ & $0.068$ & $0.01-1.0$ & $1.0321-1.0503$ \\
$21.0$ & xC$21.0$OV & $s/k_\mathrm{B} = 4$ & $1.484$ & $0.45$ & $1.5\cdot10^7$ & $0.068$ & $0.01-2.0$ & $1.0101-1.0346$ \\
\hline
$26.6$ & U$26.6$D  & $Y_e = 0.48$         & $1.383$ & no collapse & $-$     & $0.05$  & $0.01-2.0$    & $1.0677-1.0811$ \\
$26.6$ & C$26.6$D  & $Y_e = 0.48$         & $1.383$ & $0.45$ & $5\cdot10^7$ & $0.05$  & $0.01-2.0$    & $1.0652-1.0784$ \\
$26.6$ & C$26.6$O  & $s/k_\mathrm{B} = 4$ & $1.738$ & $0.45$ & $5\cdot10^7$ & $0.05$  & $0.01-2.0$    & $1.0652-1.0775$ \\
$26.6$ & xC$26.6$O & $s/k_\mathrm{B} = 4$ & $1.738$ & $0.45$ & $1.5\cdot10^7$ & $0.05$  & $0.01-2.0$ & $1.0595-1.0904$ \\
\hline
$26.6$ & U$26.6$DM & $Y_e = 0.48$         & $1.383$ & no collapse & $-$     & $0.05$  & $0.01-2.0$    & $1.0677-1.0811$ \\
$26.6$ & C$26.6$DM & $Y_e = 0.48$         & $1.383$ & $0.45$ & $5\cdot10^7$ & $0.05$  & $0.01-2.0$    & $1.0652-1.0784$ \\
$26.6$ & C$26.6$OM & $s/k_\mathrm{B} = 4$ & $1.738$ & $0.45$ & $5\cdot10^7$ & $0.05$  & $0.01-2.0$    & $1.0652-1.0775$ \\
$26.6$ & U$26.6$DM\textquotesingle & $Y_e = 0.48$         & $1.383$ & no collapse & $-$     & $0.005$  & $0.01-2.0$    & $1.0492-1.0992$ \\
$26.6$ & C$26.6$OM\textquotesingle & $s/k_\mathrm{B} = 4$ & $1.738$ & $0.45$ & $5\cdot10^7$ & $0.005$  & $0.01-2.0$    & $1.0562-1.1010$ \\
\hline
$26.6$ & C$26.6$OV & $s/k_\mathrm{B} = 4$ & $1.738$ & $0.45$ & $5\cdot10^7$ & $0.05$  & $0.01-1.0$ & $1.0666-1.0855$ \\
$26.6$ & xC$26.6$OV & $s/k_\mathrm{B} = 4$ & $1.738$ & $0.45$ & $1.5\cdot10^7$ & $0.05$  & $0.01-2.0$ & $1.0738-1.0985$ \\
\hline
\hline
\end{tabular}
\end{table*}

\section{Thermal-bomb setups}
\label{section:setups}

In order to investigate the effects of the thermal-bomb parameterisation, we simulated models without a collapsing
central core as well as models including the core collapse, varied the timescale $t_\mathrm{inj}$ of the energy deposition, changed the location of the inner grid boundary, and tested models with the volume $\Delta V$ for the energy deposition fixed in time instead of the mass layer $\Delta M$ being kept unchanged with time. Our naming convention for the CCSN models is the following:
\begin{enumerate}
\item U and C are used as first letters to discriminate between the uncollapsed and collapsed models.
\item Numerical values refer to the ZAMS masses (in units of $M_\odot$) of the progenitor models. They
are replaced by $M_*$ as a placeholder in generic model names.
\item Letters D or O are appended to distinguish the CCSN models with deep inner grid boundary at 
the progenitor's location where $Y_e = 0.48$ from the models with the inner grid boundary farther out 
where $s/k_\mathrm{B} = 4$.
\item Letters M or M\textquotesingle\ at the end of the model names denote two different types 
of test simulations where the fixed mass value $\Delta M$ of the energy-injection layer is changed 
compared to the standard case with $\Delta M = 0.05\,M_\odot$ (see Section~\ref{sec:bombvariations}).
\item Letters V instead of M at the end of the model names denote those simulations where the energy is
injected into a fixed volume $\Delta V$ instead of a fixed mass shell $\Delta M$.
\item Letters xC at the beginning of the model names indicate that the collapse of these models was
prescribed to reach an ``extreme'' radius, smaller than in the C-models.

\end{enumerate}
A summary of all CCSN simulations studied for the four considered progenitor stars is given in Table~\ref{tab:explosions}.
The explosion energy $E_\mathrm{exp}$ listed in this table is defined as the integral of the sum of the kinetic, internal,
and gravitational energies for all unbound mass, i.e., for all mass shells that possess positive values of the binding 
energy at the end of our simulation runs.
We exploded our progenitors with an explosion energy of approximately $E_{\rm exp}\approx 1\,\mathrm{B} = 10^{51}$\,erg, guided by the values of 1.01\,B for 
the $12.3\,M_\odot$ and $19.7\,M_\odot$ progenitors, 1.03\,B for the $21.0\,M_\odot$ star, and 1.07\,B for the $26.6\,M_\odot$
model.\footnote{These energies are slightly different in order to compare the thermal bomb models discussed here to existing neutrino-driven 1D explosion models from the study by \citet{2016ApJ...821...38S} in a follow-up project.} In all cases and setups, the energy was 
calibrated to the mentioned values with an accuracy of 3\%, which is a good compromise 
between accuracy needed and effort required by the iterative process for the calibration to such 
a precision. The corresponding ranges of the explosion energies for each set of models with different
energy-injection timescales are provided in the last column of Table~\ref{tab:explosions}.
The slight differences in the explosion energies between the models of each set as well as
between the different progenitors are of no relevance for the study reported here.
 
In detail, the different setups and corresponding simulations are as follows.

\begin{table*}
	\centering
	\caption{Parameters for our thermal-bomb models with fixed energy-deposition volume $\Delta V$ and models with variations of $\Delta M$ (except those with an extremely small value of $\Delta M = 0.005\,M_\odot$). $R_\mathrm{IBED}$ and $R_\mathrm{OBED}$ are the inner and outer boundary radii of $\Delta V$, $\Delta M$ is the initial mass in this volume, and the ratio gives the value of $R_\mathrm{OBED}/R_\mathrm{IBED}$. Since for each setup the $26.6\,M_\odot$ model, uncollapsed or collapsed, was taken to calculate the radius ratio, $\Delta M=0.05\,M_\odot$ in all of the cases for this progenitor.  
	}
	\label{tab:dR}
\begin{tabular}{l|ccc|ccc|ccc|ccc}
\hline
\hline
\multirow{2}{*}{\hspace{-0.2cm}$M_{\rm ZAMS}$} & \multicolumn{3}{c|}{U$M_*$DM}  & \multicolumn{3}{c|}{C$M_*$DM} & \multicolumn{3}{c|}{C$M_*$OM, C$M_*$OV} & \multicolumn{3}{c}{xC$M_*$OV}\\
 & $\Delta M$  & $R_\mathrm{IBED}$ & $R_\mathrm{OBED}$ & $\Delta M$  & $R_\mathrm{IBED}$ & $R_\mathrm{OBED}$ & $\Delta M$  & $R_\mathrm{IBED}$ & $R_\mathrm{OBED}$ & $\Delta M$  & $R_\mathrm{IBED}$ & $R_\mathrm{OBED}$\\
$[M_\odot]$                                          & $[M_\odot]$ & {[}cm{]}          & {[}cm{]}          & $[M_\odot]$ & {[}cm{]}          & {[}cm{]}          & $[M_\odot]$ & {[}cm{]}          & {[}cm{]}       & $[M_\odot]$ & {[}cm{]}          & {[}cm{]}      \\ \hline 
$19.7$                                    & $0.043$             & $1.066\cdot 10^8$      & $1.15\cdot 10^8$       & $0.044$             & $5\cdot 10^7$          & $5.4\cdot 10^7$        & $0.027$             & $5\cdot 10^7$          & $17.6\cdot 10^7$   & $0.027$             & $1.5\cdot 10^7$          & $15.88\cdot 10^7$      \\
$21.0$                                    & $0.042$             & $1.058\cdot 10^8$      & $1.14\cdot 10^8$       & $0.041$             & $5\cdot 10^7$          & $5.4\cdot 10^7$        & $0.068$             & $5\cdot 10^7$          & $17.6\cdot 10^7$      & $0.068$             & $1.5\cdot 10^7$          & $15.88\cdot 10^7$       \\
$26.6$                                    & $0.050$             & $1.278 \cdot 10^8$     & $1.38\cdot 10^8$       & $0.050$             & $5\cdot 10^7$          & $5.4\cdot 10^7$        & $0.050$             & $5\cdot 10^7$          & $17.6\cdot 10^7$  & $0.050$             & $1.5\cdot 10^7$          & $15.88\cdot 10^7$       \\ \hline
ratio                                     & \multicolumn{3}{c|}{$1.080$}                                            & \multicolumn{3}{c|}{$1.081$}                                            & \multicolumn{3}{c|}{$3.519$}                                           & \multicolumn{3}{c}{$10.587$}  \\ \hline
\end{tabular}
\end{table*}

\subsection{Models for comparison with SM19}
\label{sec:compSM19models}

We started our investigation with a setup that was guided by models discussed in SM19, i.e.,
the CCSN simulations did not include any collapse of the central core of the progenitors.
These U-models were supposed to permit a comparison with the results presented by SM19. 

In all of the discussed U-models the inner boundary was placed at the location where $Y_e=0.48$, 
and in our default setup the explosion energy was injected into a fixed mass layer 
with $\Delta M =0.05\,M_\odot$, which was the same in all CCSN models for the set of progenitors.
The inner boundary of this energy-deposition layer (IBED) was therefore chosen to be identical to 
the inner grid boundary. The entire mass exterior to the IBED, i.e., including the matter 
in the energy-deposition layer between the IBED and the outer boundary of the energy-deposition 
layer (OBED), was considered to be ejected, provided it became gravitationally
unbound by the energy injection. Note that in models with fixed
energy-deposition layer $\Delta M$, the outer radius of this shell, $R_\mathrm{OBED}$, moves 
outward as the heated mass $\Delta M$ expands, whereas the inner radius, $R_\mathrm{IBED}$, is 
set to coincide with the inner grid boundary $R_\mathrm{ib}$ and does not change with time.

Our thus chosen setup differs in two technical aspects from the choices made in SM19.
First, SM19 reported that they injected the thermal-bomb energy into
a fixed mass of 0.005\,$M_\odot$ (corresponding to the innermost 20 zones of their 1D Lagrangian
hydrodynamics simulations). In contrast, we adopted $\Delta M = 0.05\,M_\odot$ as our default
value. This larger mass appears more appropriate to us, at least in the case of the more realistic
collapsed models and in view of the neutrino-driven mechanism, where neutrinos transfer energy
to typically several 0.01\,$M_\odot$ to more than 0.1\,$M_\odot$ of circum-neutron star matter.
Second, SM19 did not count the mass in the heated layer as ejecta, which means that they considered
only the entire mass above the energy-deposition layer, i.e., exterior to the OBED, as ejecta.
We did not join this convention, because we chose a 10 times larger mass for $\Delta M$ than SM19.
In addition, again in view of the neutrino-driven mechanism, we do not see any reason why heated
matter that can also be expelled should not be added to the nucleosynthesis-relevant CCSN ejecta.
Moreover, we performed test calculations with $\Delta M = 0.005\,M_\odot$ and found no
significant differences in the $^{56}$Ni yields, at least not in the case of uncollapsed models
that served for a direct comparison with SM19. (This will be discussed in 
Section~\ref{sec:massvariations}.)

The timescale of the energy deposition used in Equation~(\ref{eq:edotm}) was varied from 
0.01\,s to 2\,s, using the following values:
\begin{equation}
\quad\quad\quad t_\mathrm{inj} = 0.01,\ 0.05,\ 0.2,\ 0.5,\ 1.0,\ 2.0\ \mathrm{s}\,.
\label{eq:timesinj}
\end{equation}
We thus tested the influence of different durations of the energy injection on the explosion 
dynamics and $^{56}$Ni production. 
Although our progenitors are different from those used by SM19 and also our setup for the 
CCSN simulations differs in details from the one employed by SM19, the modelling approaches
are sufficiently similar to permit us to reproduce the basic findings reported by SM19.

In Table~\ref{tab:explosions} the corresponding models are denoted by U$M_*$D, where $M_*$ 
stands here as a placeholder for the mass value of the model. While our standard setup uses
$\Delta M =0.05\,M_\odot$, we also performed test runs with 
$\Delta M \approx 0.04\,M_\odot$ for the U-setup. These models are denoted by U$M_*$DM in
Table~\ref{tab:explosions}. We also ran test cases with the SM19 value of 
$\Delta M =0.005\,M_\odot$; the corresponding models are named U$M_*$DM\textquotesingle\ in
Table~\ref{tab:explosions}, but they are not prominently discussed in the following, because 
such a small mass in the energy-deposition layer does not appear to be realistic for common 
CCSNe. It is most important, however, to note that all of these changes of $\Delta M$ led to secondary and never dominant differences in the produced amount of $^{56}$Ni compared to the changes connected to introducing a collapse phase or shifting the inner grid boundary (see Section~\ref{sec:bombvariations}). We did not consider any cases 
U$M_*$O, because moving the inner grid boundary farther out will lead to lower densities in 
the ejecta (Figure~\ref{fig:psn_closer}). This will significantly reduce the nucleosynthesized 
amount of $^{56}$Ni in this setup, and in particular for long $t_\mathrm{inj}$ it will lead to
even more severe underproduction of $^{56}$Ni compared to the yields inferred from observations 
of CCSNe with energies around $10^{51}$\,erg (see Section~\ref{sec:SM19results}).

\subsection{Variations of thermal-bomb setups}
\label{sec:bombvariations}

Instead of releasing thermal energy in the uncollapsed progenitor as assumed by SM19, we extended 
our setup in a next step by forcing the progenitor's core to contract before depositing the energy.
Adding such a collapse phase will change the dynamics of the explosion, even with the same explosion
energy and the same location of the inner boundary. 

To this end the inner grid boundary was moved inward for a time interval $t_\mathrm{coll}$, thus
mimicking the collapse phase that precedes the development of the explosion. The time-dependent
velocity for contracting the inner boundary was prescribed as in 
\citet{1995ApJS..101..181W,2002RvMP...74.1015W,2007PhR...442..269W} (who applied this prescription 
within the framework of the classical piston method):
\begin{equation}
\frac{\mathrm{d}r}{\mathrm{d}t}(t) =  v_0 -a_0t \quad \mathrm{for} \quad  t<t_\mathrm{coll} \,,
\label{eq:collapse}
\end{equation}
where $v_0 < 0$ is the initial velocity of the inner boundary (following the infall of
the progenitor model at the onset of its core collapse), and 
$a_0=2(r_0-r_\mathrm{min}+v_0t_\mathrm{coll})/t_\mathrm{coll}^2$ is a constant acceleration calculated in 
order to reach the minimum radius $r_\mathrm{min}$ after the collapse time $t_\mathrm{coll}$, with $r_0$ 
being the initial radius of the inner boundary. After this phase, the boundary contraction is stopped, 
matter begins to pile up around the grid boundary, and a shock wave forms at the interface to the still 
supersonically infalling overlying shells. Concomitantly, the deposition of internal energy by our 
thermal bomb was started.

Equation~(\ref{eq:collapse}) defines the inward movement of the constant Lagrangian mass shell 
corresponding to the closed inner grid boundary. The collapse is basically controlled by the parameters 
$t_\mathrm{coll}$ and $r_\mathrm{min}$, whereas the explosion phase is controlled by the thermal-bomb 
parameters $E_\mathrm{inj}$, $\Delta M$ (or $\Delta V$), and $ t_\mathrm{inj}$ (Equations~\ref{eq:edotm} 
and \ref{eq:edotv}). Again following the literature mentioned above, we adopt for our default collapse 
simulations $t_\mathrm{coll}=0.45$\,s and the minimum radius $r_\mathrm{min} = 5\cdot 10^7$\,cm. 
In Table~\ref{tab:explosions} the models with this collapse
setup and the deep inner boundary are denoted by C$M_*$D. In these models the central (and maximum) densities lie between $7\cdot10^8$\,g\,cm$^{-3}$ and $2\cdot10^9$\,g\,cm$^{-3}$.

In a variation of the setup for the C-models, we relocated the inner grid boundary outward to the base of the 
oxygen shell in the progenitor, i.e., to the radial position where $s/k_\mathrm{B} = 4$, with the goal of
studying the influence on the $^{56}$Ni production. These models are denoted by C$M_*$O in
Table~\ref{tab:explosions}. The central (and maximum) densities of these models are between $3\cdot10^7$\,g\,cm$^{-3}$ and $2\cdot10^8$\,g\,cm$^{-3}$. A variant of these models, named xC$M_*$O, considered the collapse to proceed
to a smaller radius of $r_\mathrm{min} = 1.5\cdot 10^7$\,cm, using the same value of $t_\mathrm{coll}=0.45$\,s
for the collapse time. In this case the central (and maximum) densities reach the values between $3\cdot10^9$\,g\,cm$^{-3}$ and $9\cdot10^9$\,g\,cm$^{-3}$.

As in the U-models, the inner boundary of the grid and the inner boundary of the energy-deposition layer 
(IBED) were chosen to coincide in all simulations. 
In both model variants, U-models as well as C-models, our standard runs were done with energy
being dumped into a fixed mass layer of mass $\Delta M = 0.05\,M_\odot$. For the C-models we also simulated
some test cases with different values of $\Delta M$ between about 0.03\,$M_\odot$ and roughly 0.07\,$M_\odot$.
The corresponding models are denoted by C$M_*$DM or C$M_*$OM in Table~\ref{tab:explosions}. We also tested
$\Delta M = 0.005\,M_\odot$ in simulations with collapse and the IBED at $s/k_\mathrm{B} = 4$, listed as 
models C$M_*$OM\textquotesingle\ in Table~\ref{tab:explosions}.
These variations turned out to have no relevant influence on the $^{56}$Ni yields in the D-boundary cases,
in agreement with what we found for the U-models. However, the change of $\Delta M$ caused some interesting, 
though secondary, differences in those cases that employed the O-boundary. We will briefly discuss these 
results in Section~\ref{sec:massvariations}.

In yet another variation we investigated cases for our more realistic setup of C-models with O-boundary,
where the volume of the energy deposition, $\Delta V$, was fixed 
instead of the mass layer $\Delta M$. Such a change might potentially affect the $^{56}$Ni production in
CCSN models with steep density profile near the inner grid boundary. This time-independent volume of the energy
deposition was determined for the different progenitors by a simple condition, connecting it to the initial
values of the outer boundary radius $R_\mathrm{OBED}$ and of the inner boundary radius 
$R_\mathrm{IBED} = R_\mathrm{ib}$ of our standard setup with $\Delta M = 0.05\,M_\odot$ in the 
26.6\,$M_\odot$ CCSN models. Specifically, the volume $\Delta V$, which is bounded
by $R_\mathrm{IBED}$ and $R_\mathrm{OBED}$, was defined by the requirement that the ratio of these two 
radii should have the same value as in the $26.6\,M_\odot$ model in all of the CCSN runs (i.e., for all
progenitors) of each considered setup:
\begin{equation}
\frac{R_\mathrm{OBED}}{R_\mathrm{IBED}}(26.6M_\odot)= \frac{R_\mathrm{OBED}}{R_\mathrm{IBED}}(21.0M_\odot)=\frac{R_\mathrm{OBED}}{R_\mathrm{IBED}}(19.7M_\odot)\,. 
\label{eq:rratio}
\end{equation}
This condition means that the inner radius of the deposition region, $R_\mathrm{IBED}$, was pre-defined  
by $R_\mathrm{ib}$ in the O-cases, and the outer radii $R_\mathrm{OBED}(21.0M_\odot)$ 
and $R_\mathrm{OBED}(19.7M_\odot)$ were calculated from the equation above. 
The chosen condition of Equation~(\ref{eq:rratio}) was also applied
more generally for defining variations of $\Delta M$ (or $\Delta V$) in collapsed or uncollapsed models
with deep or outer location of $R_\mathrm{ib}$ (Table~\ref{tab:dR}).
Such a procedure should ensure that the distance between $R_\mathrm{IBED}$ and $R_\mathrm{OBED}$ adjusts 
to the size of $R_\mathrm{ib}$ and thus accounts for the higher density in its vicinity instead of being 
rigid without any reaction to the progenitors' radial structures.

The models with fixed energy-deposition volume $\Delta V$ thus determined are denoted by C$M_*$OV 
or xC$M_*$OV in Table~\ref{tab:explosions} for standard and extreme collapse cases, respectively, 
and the values of $R_\mathrm{IBED}$ and $R_\mathrm{OBED}$ in our different 
model variations are listed in Table~\ref{tab:dR}. 
The latter table also provides numbers for the initial masses $\Delta M$ that correspond to
the volumes bounded by $R_\mathrm{IBED}$ and $R_\mathrm{OBED}$. Note that Equation~(\ref{eq:rratio}) implies 
that $\Delta M$ is still 0.05\,$M_\odot$ for the 26.6\,$M_\odot$ models, but the initial masses in the 
heating layers are not the same in the runs with fixed $\Delta V$ for the other progenitors. 
Of course, for fixed volume $\Delta V$, the radii $R_\mathrm{IBED}$ and 
$R_\mathrm{OBED}$ do not evolve with time, but the mass $\Delta M$ in this heated radial shell decreases 
with time as the heated gas expands outward.

Table~\ref{tab:dR} also provides the $\Delta M$ values that were obtained via Equation~(\ref{eq:rratio}) 
and apply for our tests performed with variations of the fixed heated mass-layer $\Delta M$ in models 
U$M_*$DM (see Section~\ref{sec:compSM19models}) as well as models C$M_*$DM and C$M_*$OM mentioned above. 
These subsets of models are interesting despite their small differences in $\Delta M$ compared to our 
default choice of $\Delta M = 0.05\,M_\odot$, because in the C-cases the initial volumes of the heated 
masses are the same for all progenitors instead of being different from case to case. Thus, these model
variations check another aspect of potential influence on the nucleosynthesis conditions in the innermost
ejecta.


\begin{figure}
\centering
\includegraphics[width=\columnwidth]{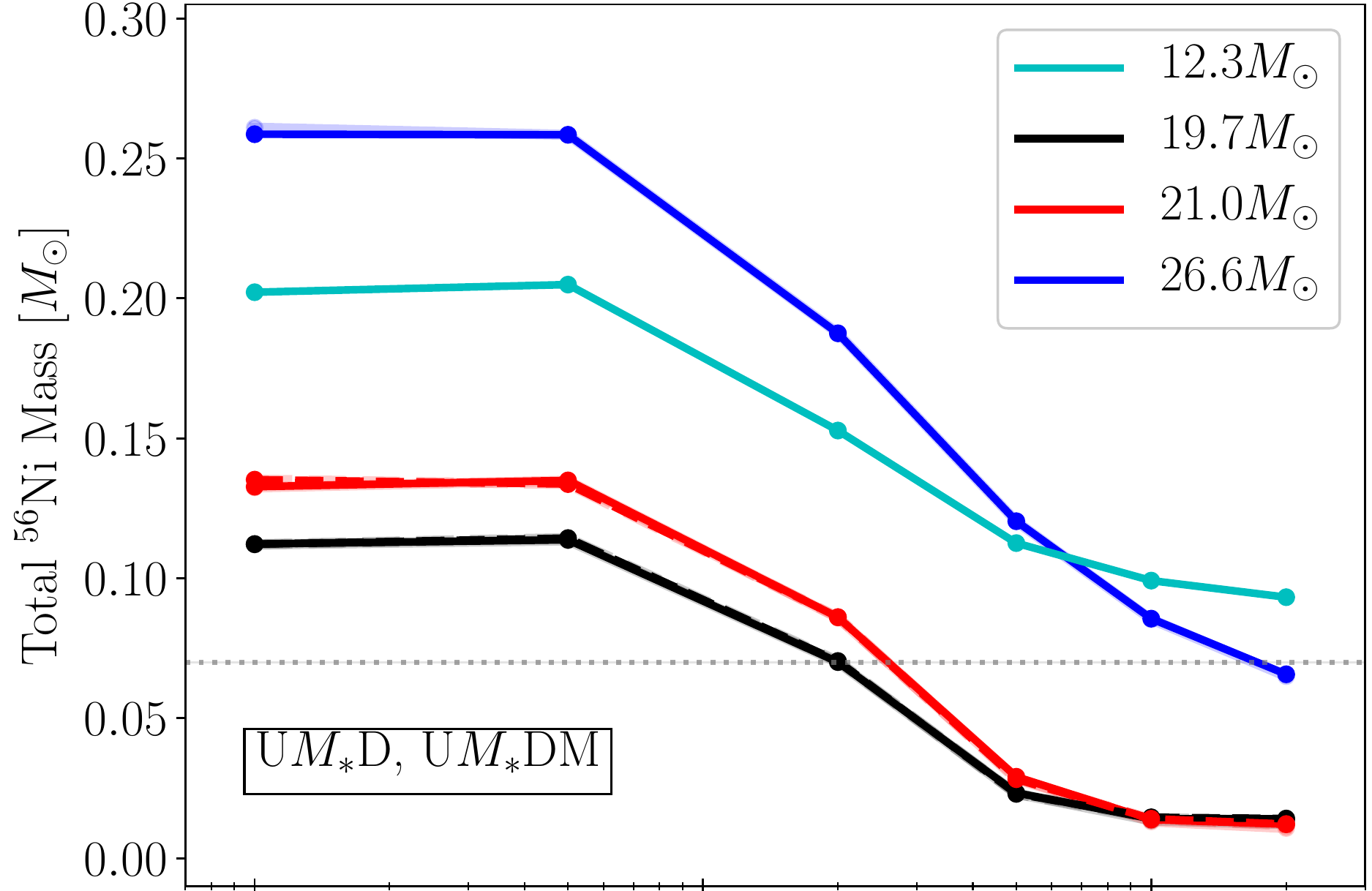}

\includegraphics[width=\columnwidth]{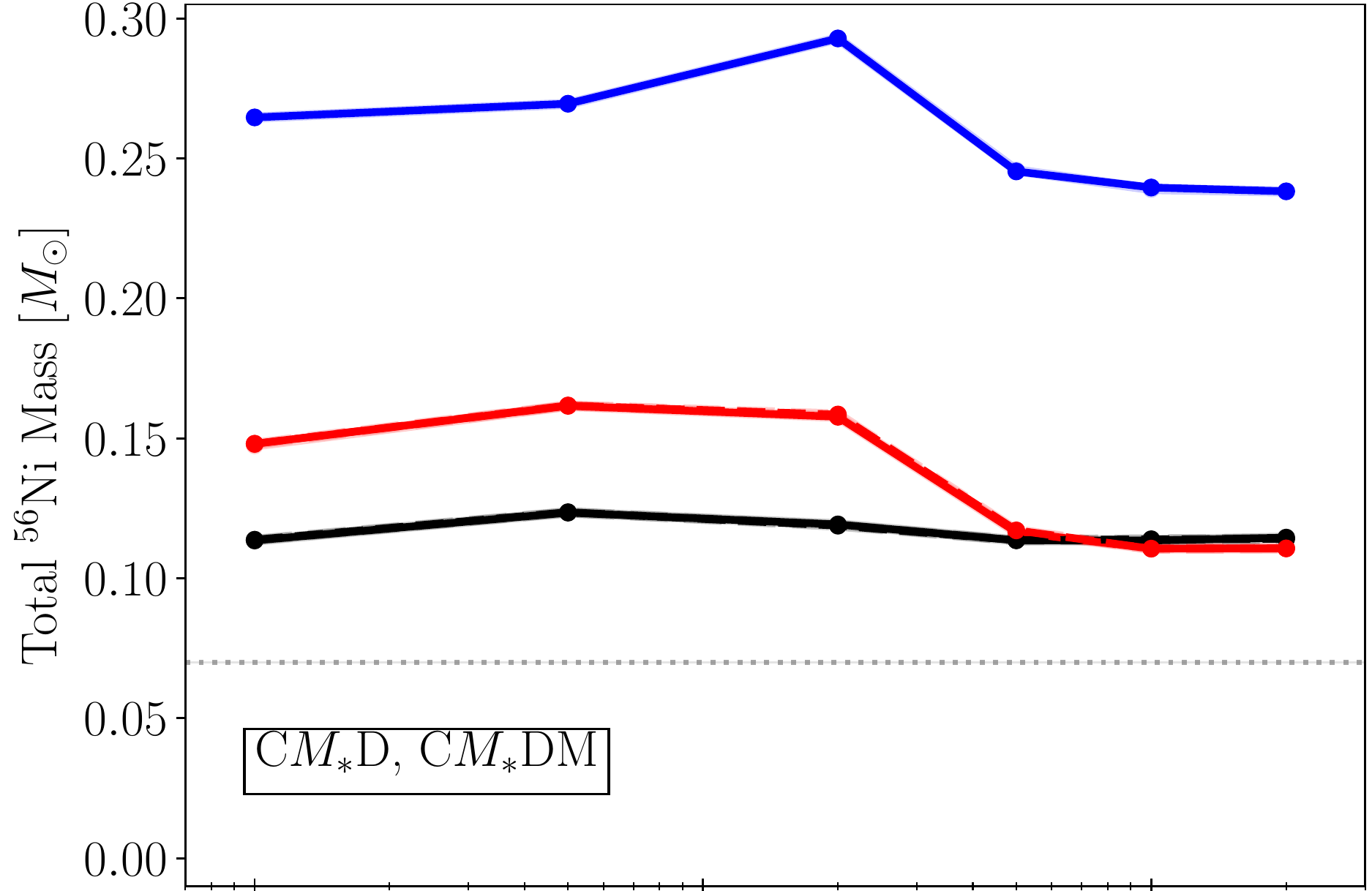}

\includegraphics[width=\columnwidth]{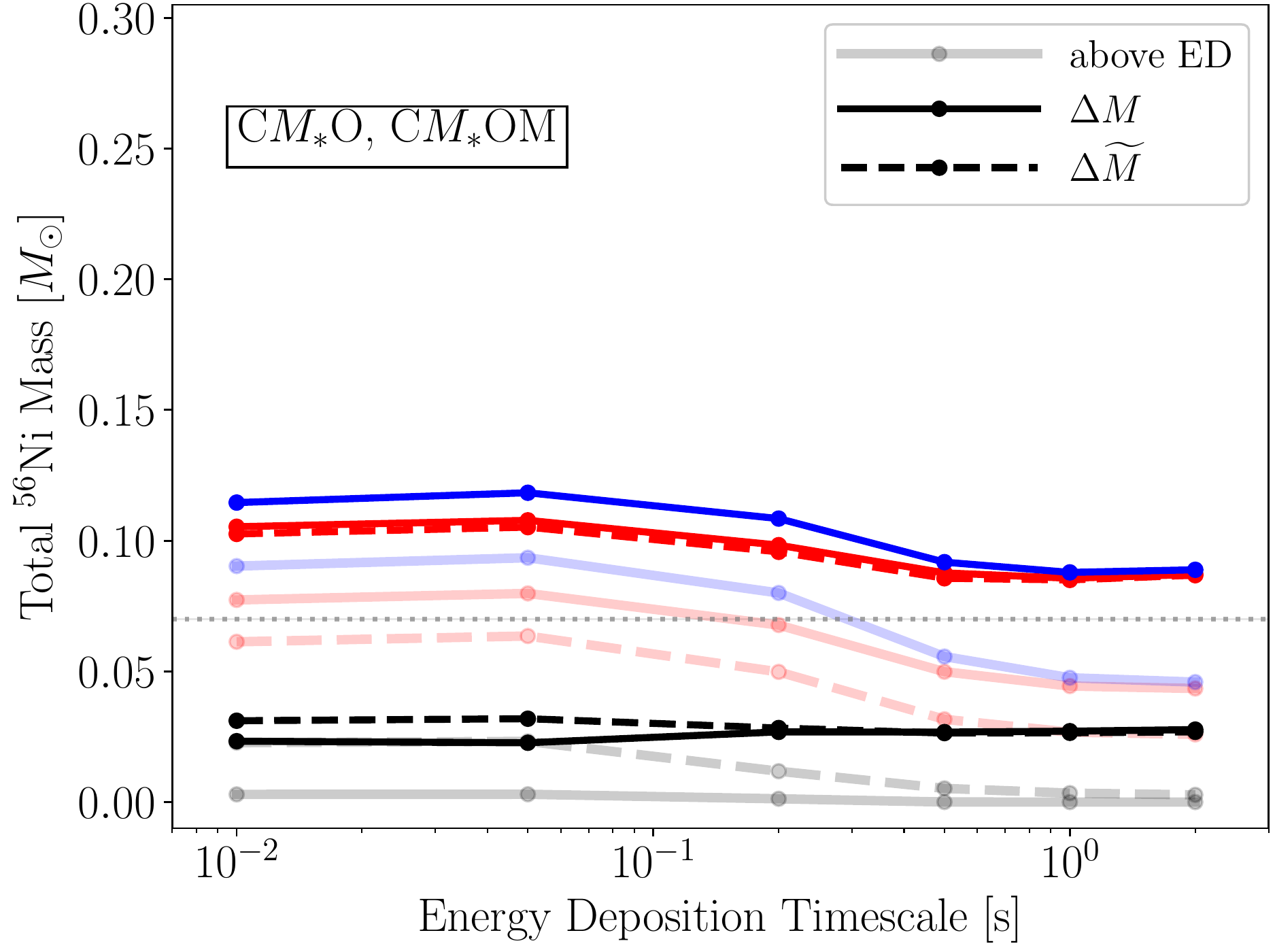} 

      \caption{$^{56}$Ni yields as functions of energy-injection timescale for uncollapsed CCSN models (top panel) and collapsed models (middle panel) with deep inner grid boundary, and collapsed CCSN models with the inner grid boundary shifted farther out (bottom panel). The different colors correspond to the different progenitors as labelled in the top panel. Solid lines belong to our standard choice of $\Delta M = 0.05\,M_\odot$ for the fixed mass in the energy-deposition layer and dashed lines refer to varied mass values $\Delta \widetilde{M}$ (models with unprimed M in their names; see Table~\ref{tab:explosions}). Note that in the top and middle panels the solid and dashed lines overlap and are almost completely indistinguishable. In all panels the blue solid and dashed lines fall on top of each other by definition. The light-colored lines (solid and dashed) in the bottom panel show the $^{56}$Ni yields when the mass in the energy-injection layer is excluded from the ejecta instead of adding unbound matter of this layer to the ejecta. The horizontal grey dotted line indicates the $^{56}$Ni yield of 0.07\,$M_\odot$ for a $\sim$\,$10^{51}$\,erg explosion, e.g., SN~1987A \citep{1989ARA&A..27..629A}.}
     \label{fig:mni0}   
\end{figure}

\begin{figure}
\centering

\includegraphics[width=\columnwidth]{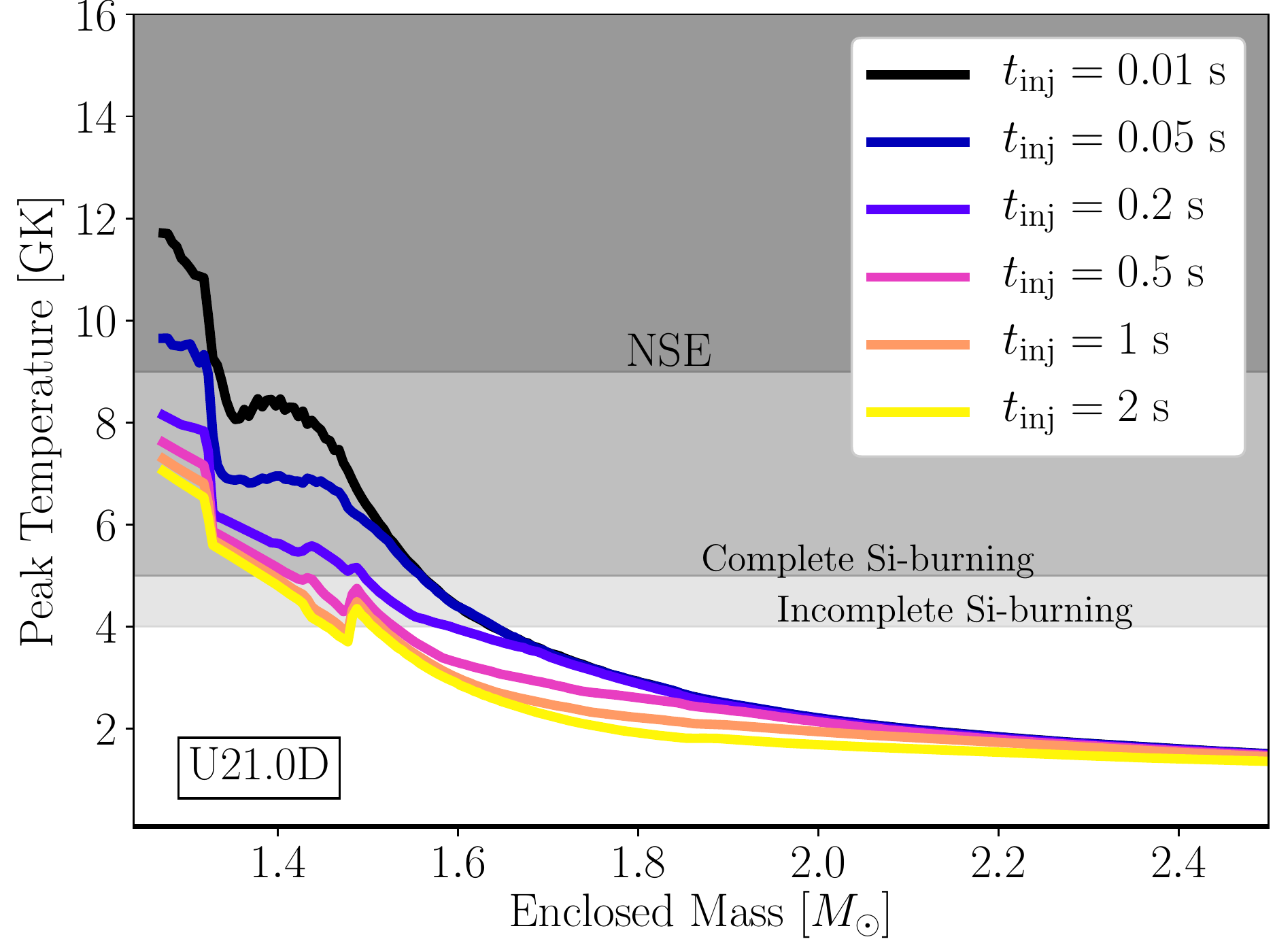}
\includegraphics[width=\columnwidth]{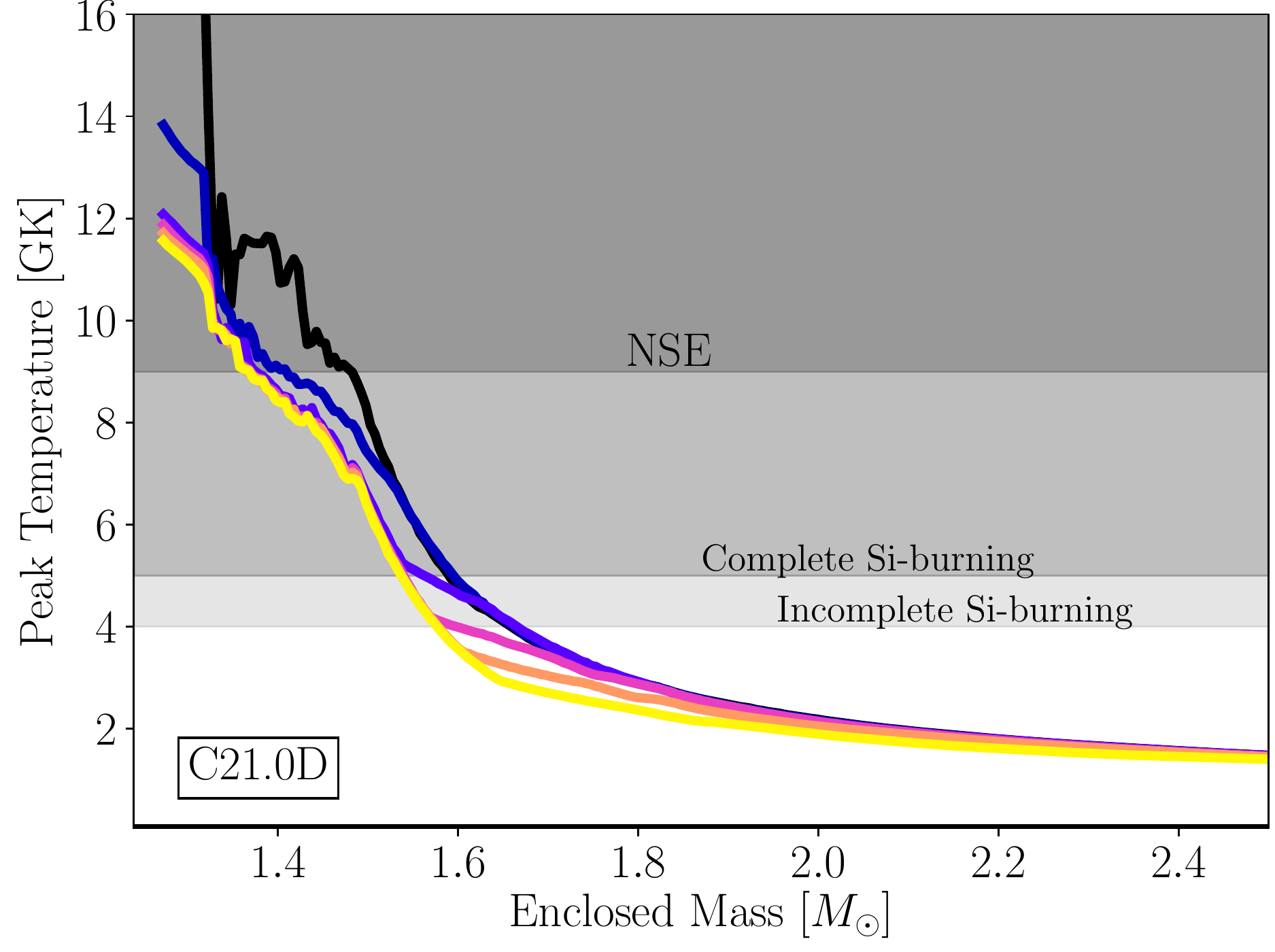}
\includegraphics[width=\columnwidth]{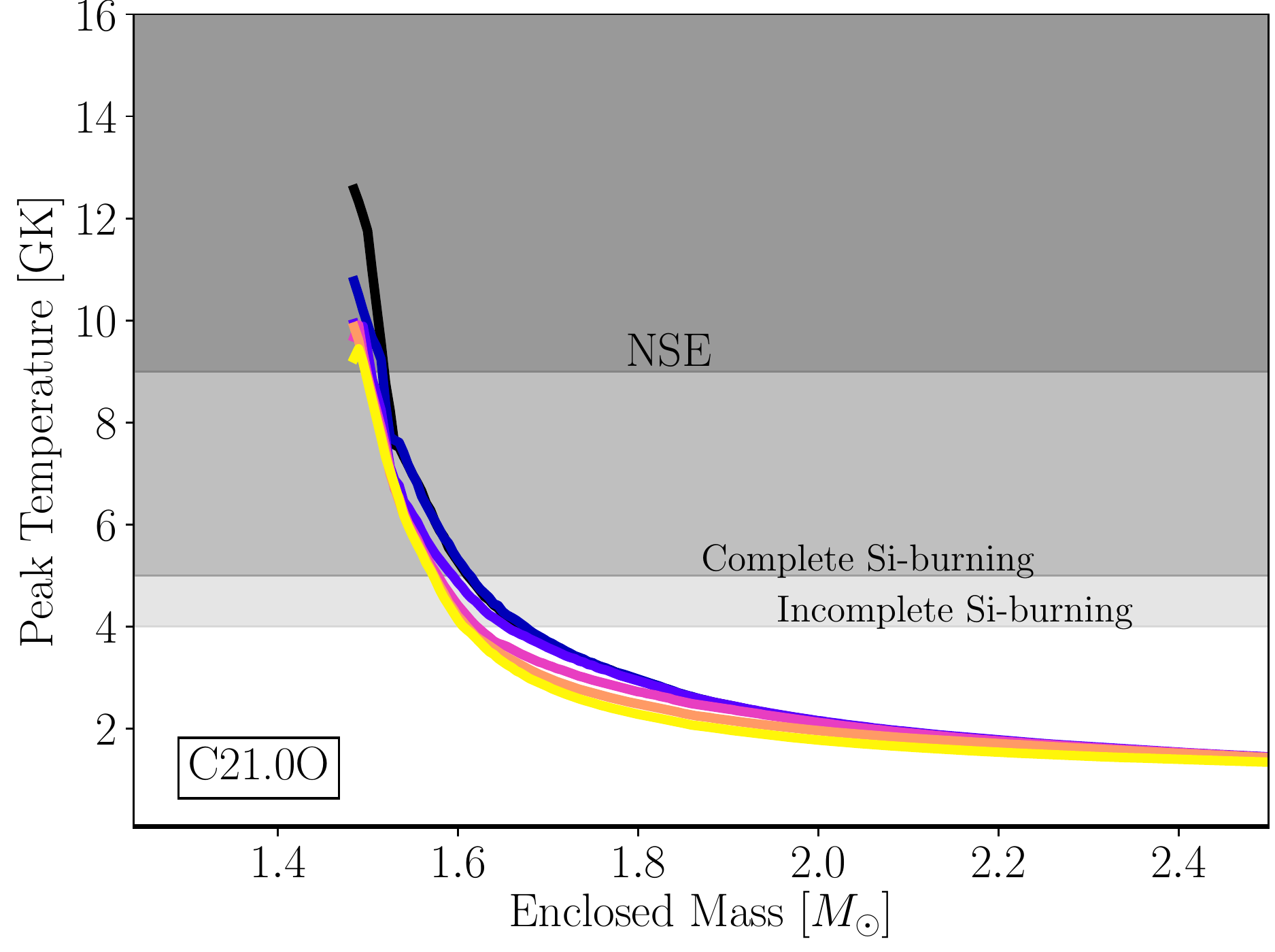}

       \caption{Peak temperatures as functions of enclosed mass for the CCSN runs with the 21\,$M_\odot$ progenitor and different energy-injection timescales for the same modelling setups shown in Figure~\ref{fig:mni0}: uncollapsed (top), collapsed (middle), and collapsed with inner grid boundary shifted farther out (bottom). Different intensities of grey shading indicate different regimes of explosive nucleosynthesis as labelled. Note that the peak temperatures are displayed only for the runs with our standard value of $\Delta M = 0.05\,M_\odot$ for the fixed mass in the energy-injection layer, because the differences compared 
       to the other choices of $\Delta M$ are effectively indistinguishable.}
\label{fig:peakTtinj}
\end{figure}

\begin{figure}

\includegraphics[width=\columnwidth]{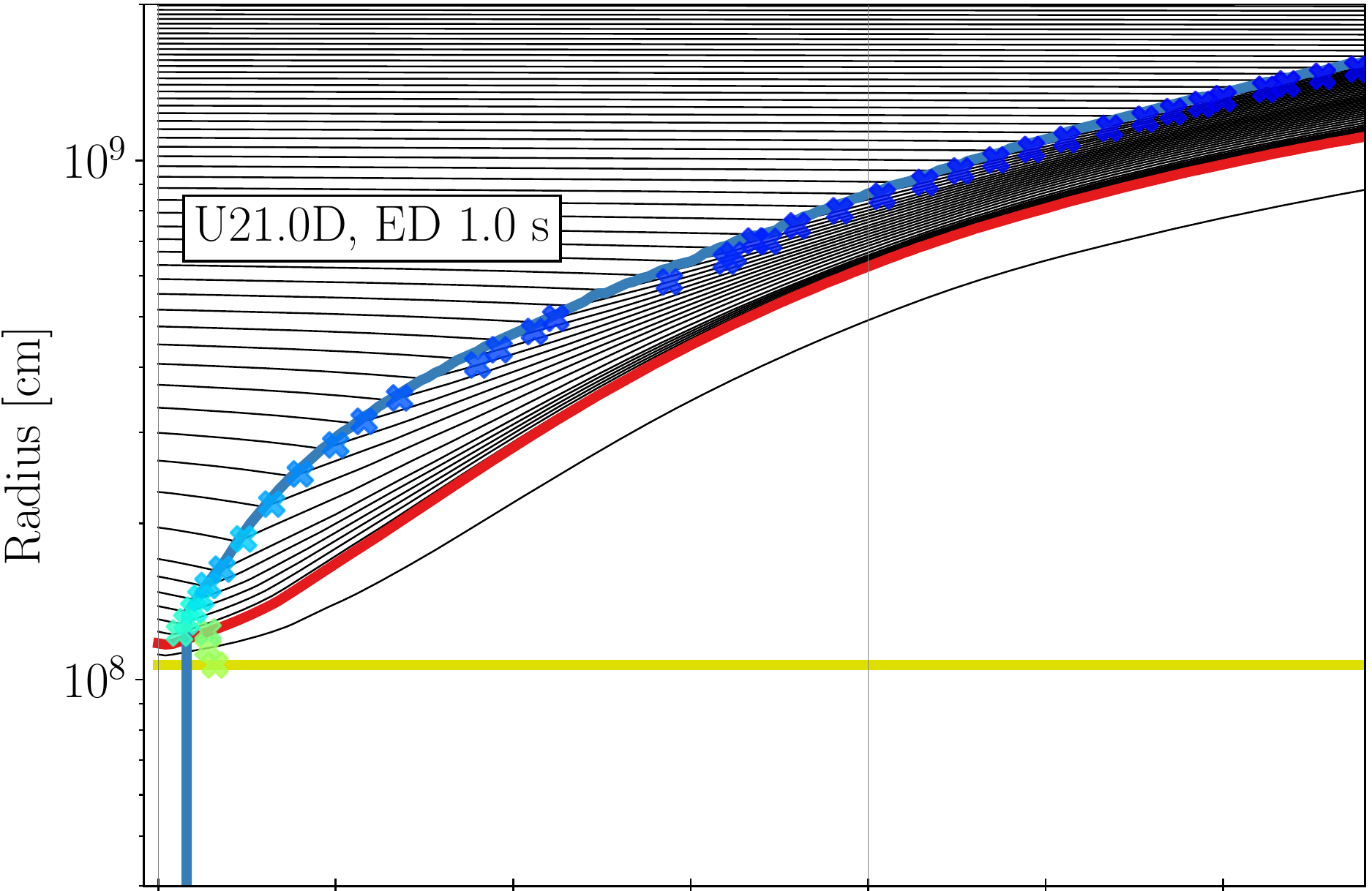}

\includegraphics[width=\columnwidth]{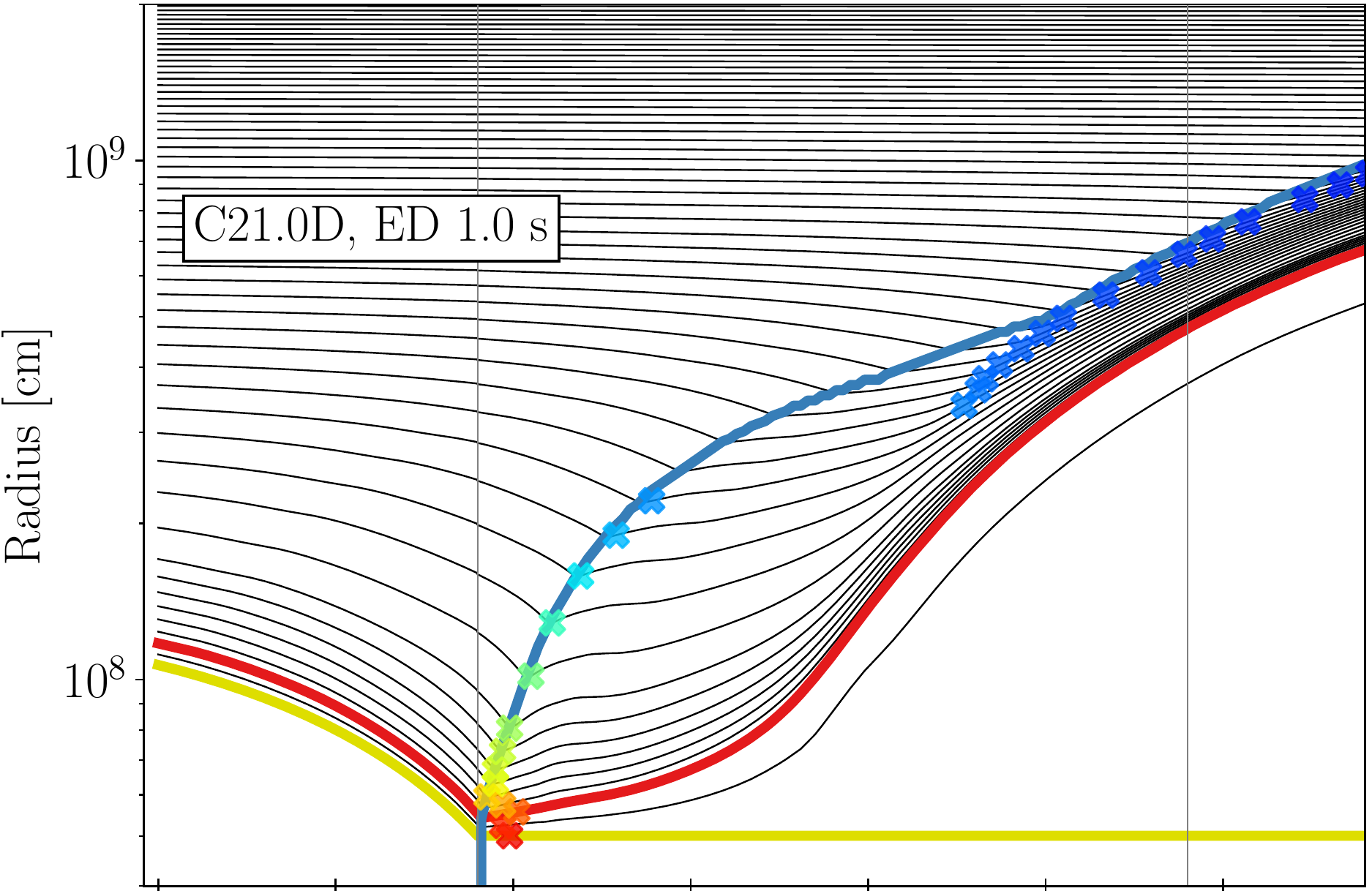}

\includegraphics[width=\columnwidth]{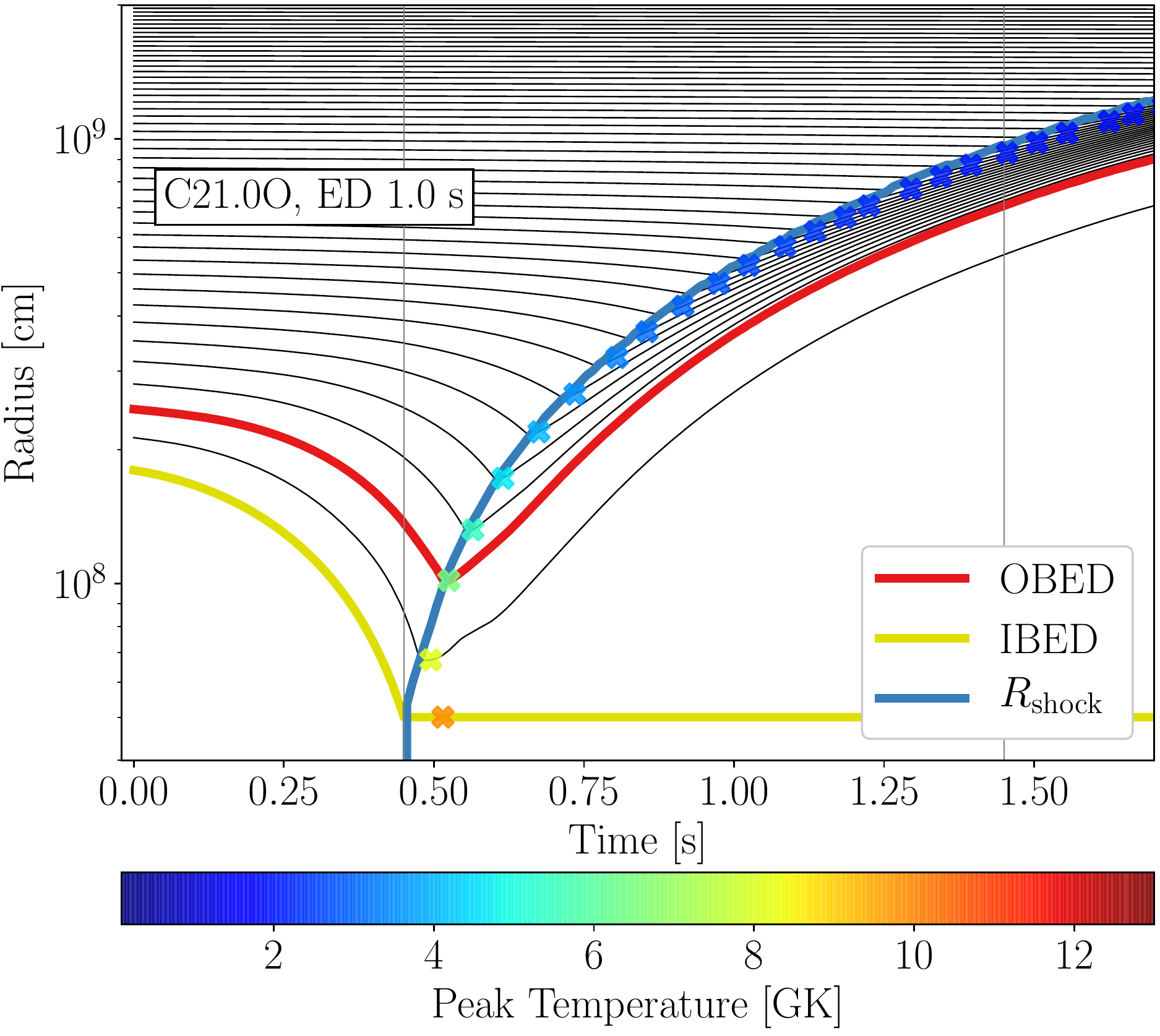}
\vspace{-0.5cm}

    \caption{Radius evolution of Lagrangian mass shells with time for the CCSN runs of the 21\,$M_\odot$ progenitor with standard value of $\Delta M = 0.05\,M_\odot$ for the fixed mass in the energy-injection layer and a representative energy-deposition timescale of 1.0\,s: uncollapsed (top) and collapsed (middle) with deep inner grid boundary, and collapsed with inner grid boundary shifted farther out (bottom). The thin black solid lines are the mass shells, spaced in steps of 0.025\,$M_\odot$, the blue line marks the shock radius, the red line indicates the radius of the outer edge of the energy-injection layer ($R_\mathrm{OBED}$), and the yellow line the radius of the inner grid boundary, $R_\mathrm{ib}$, which is chosen as the inner edge of the energy-injection layer ($R_\mathrm{IBED}$) when the thermal bomb is switched on. Crosses indicate the moments when the peak temperature of each mass shell is reached; their colors correspond to temperature values as given by the color bar. Vertical lines mark the beginning and the end of the energy deposition.}
    \label{fig:second_wave}
\end{figure}

\section{Results of thermal-bomb simulations}
\label{section:results}

In this section we present the results of our study, focusing on the mass of $^{56}$Ni 
produced in the ejecta as computed in a post-processing step with the 262-isotope version
of SkyNet (see Section~\ref{section:reactions}). 
These yields were determined after 10\,s of simulated evolution and,
different from SM19, we usually (unless explicitly stated differently) considered as ejecta 
also unbound matter contained in the energy-deposition layer. We stress, however, that 
for models with the deep inner boundary $R_\mathrm{ib} = R_\mathrm{IBED}$ at $Y_e = 0.48$, 
there is no relevant difference in the $^{56}$Ni yields when including or excluding the 
mass in the heating layer. The reason is seen in Figure~\ref{fig:psn_closer}, upper and
lower right panels: Since $Y_e < 0.485$ in the innermost $0.05\,M_\odot$ just outside of
$R_\mathrm{ib}$, i.e., in the mass between $R_\mathrm{IBED}$ and $R_\mathrm{OBED}$, the
$^{56}$Ni production is negligibly small in the energy-deposition layer.

In Section~\ref{sec:SM19results} we will first report on our models of the U-setup in comparison
to SM19. Then, in Section~\ref{sec:collapsedmodels}, we will discuss the differences when
our models included an initial collapse before the thermal bomb was switched on. In 
Section~\ref{sec:shiftedIBED} we will describe the influence of shifting the inner grid
boundary, $R_\mathrm{ib} = R_\mathrm{IBED}$, from the deep default location at $Y_e = 0.48$ 
to the outer location at the base of the oxygen shell where $s/k_\mathrm{B} = 4$. In 
Section~\ref{sec:massvariations} we will briefly summarize the consequences of changing 
the fixed mass $\Delta M$ of the energy-deposition layer, in 
Section~\ref{sec:fixedvolume} we will discuss the influence of changing from a 
fixed mass $\Delta M$ to a fixed volume $\Delta V$ of the energy-injection layer, and in Section~\ref{sec:minradius} we will finally
present results for different minimum radii prescribed for the collapse phase.


\subsection{Uncollapsed models compared to SM19}
\label{sec:SM19results}

When we consider uncollapsed models with deep inner grid boundary and the thermal-bomb energy 
injection into a fixed mass $\Delta M$ (the U$M_*$D simulations), following SM19, our results 
confirm the findings of this previous study (Figure~\ref{fig:mni0}, top panel): 
One can witness a clear anti-correlation between the amount of $^{56}$Ni produced and the 
timescale of the energy deposition for the explosion runs of all of the four considered
progenitors; slower energy injection leads to a clear trend of reduced $^{56}$Ni production.

Our set of CCSN models exhibits the same qualitative behavior as visible in Figure~7 (left panel)
of SM19, although there are significant quantitative differences. These are most likely
connected to the different core structures of the progenitor models, because the mentioned technical
differences in the explosion modelling (i.e., the choice of the value of $\Delta M$ for
the energy-injection layer and the inclusion of the heated mass in the ejecta) turned out
to have no significant impact on the $^{56}$Ni yields in the uncollapsed models with deep
inner boundary, see Sections~\ref{sec:shiftedIBED} and \ref{sec:massvariations}. 
For example, we investigated the effects of changing $\Delta M$ within several 10\% of our
standard value (varying between 0.027\,$M_\odot$ and 0.068\,$M_\odot$) and also tested the 
extremely small value of $\Delta M = 0.005\,M_\odot$, but could not find any relevant
$^{56}$Ni differences compared to our U$M_*$D simulations (a detailed
discussion of this aspect is provided in Section~\ref{sec:massvariations}).

The reason for the anti-correlation of $^{56}$Ni yield and energy-injection timescale can be
inferred from the top panel of Figure~\ref{fig:peakTtinj}, which displays the peak temperatures
as functions of enclosed mass for all investigated values of $t_\mathrm{inj}$ in the 
21\,$M_\odot$ CCSN runs. Efficient
$^{56}$Ni production requires the temperature in the expanding ejecta to reach the regime of 
NSE or complete silicon burning. Moreover, $Y_e$ has to exceed $\sim$0.48 considerably, which 
is obvious from the upper and lower right panels of Figure~\ref{fig:psn_closer}, where
$^{56}$Ni mass fractions above 0.1 occur only in regions where $Y_e \gtrsim 0.485$. Only when
these requirements are simultaneously fulfilled, freeze-out from NSE or explosive nuclear
burning are capable of contributing major fractions to the $^{56}$Ni yield. The top panel
of Figure~\ref{fig:peakTtinj} shows that for longer energy-injection times not only the 
maximum value of the peak temperature that can be reached in the heated matter drops, but 
also the total mass that is heated to the threshold temperature of complete Si burning 
(about 5\,GK) decreases. Therefore less $^{56}$Ni is nucleosynthesized when the 
energy injection of the thermal bomb (for a given value of the final explosion energy)
is stretched over a longer time interval.

This behavior is a consequence of the fact that the heated 
matter begins to expand as soon as the thermal bomb is switched on (see the upper panel of
Figure~\ref{fig:second_wave} for the uncollapsed 21.0\,$M_\odot$ model with 
$t_\mathrm{inj}=1.0$\,s). When the energy injection
is quasi-instantaneous, i.e., short compared to the hydrodynamical timescale for the 
expansion,\footnote{The hydrodynamical timescale, by its order of magnitude,
is given by the radial extension of the bomb-heated layer divided by the 
average sound speed in this layer. For the uncollapsed models it is roughly 
$\Delta R/\bar{c}_\mathrm{s}\sim 10^7\,\mathrm{cm}/(10^9\,\mathrm{cm\,s}^{-1}) = 10^{-2}$\,s.
Since the gravitational binding energy of the uncollapsed stellar structure at $r > R_\mathrm{ib}$ 
is low, this means that the outward expansion of the thermal-bomb-heated layer gains momentum 
within several 10\,ms at the longest.} 
the thermal energy deposition leads to an abrupt and strong increase of the temperature before 
the matter can react by its expansion. If, in contrast, the energy release by the thermal bomb
for the same final explosion energy is spread over a long time interval, i.e., longer than the 
hydrodynamical timescale, the expansion occurring during this energy injection has two effects
that reduce the temperature increase, in its maximum peak value as well as in the volume
that gets heated to high temperatures: First, cooling by expansion ($p\mathrm{d}V$) 
work limits the temperature
rise and, second, the thermal energy dumped by the bomb is distributed over a wider volume 
because the fixed mass $\Delta M$, into which the energy is injected, expands continuously.
This is visible in the mass-shell plots of Figure~\ref{fig:second_wave} by the outward motion
of the red line, which corresponds to the outer boundary radius, $R_\mathrm{OBED}$, of the 
energy-deposition layer. Because the gravitational binding energy of the uncollapsed stellar
profile is comparatively low, the expansion of the energy-injection layer sets in basically
promptly when the thermal bomb starts releasing its energy at $t = 0$. This holds true 
even if the specific energy-deposition rate $\dot e_\mathrm{inj, M}$ is relatively low 
because of a long injection timescale of $t_\mathrm{inj} = 1.0$\,s, for example
(top panel of Figure~\ref{fig:second_wave}).

Comparing the results for the four progenitors in the top panel of Figure~\ref{fig:mni0},
we notice three different aspects: (i) The absolute amount of the produced $^{56}$Ni and its 
steep variation with $t_\mathrm{inj}$ are quite similar for the 19.7\,$M_\odot$ and 21\,$M_\odot$
progenitors; (ii) these progenitors yield considerably less $^{56}$Ni for all energy-injection
timescales than the 26.6\,$M_\odot$ case; (iii) the 12.3\,$M_\odot$ progenitor exhibits the 
weakest variation of the ejected $^{56}$Ni mass with $t_\mathrm{inj}$ among all of the four 
considered stars.

\begin{figure}
\centering

\includegraphics[width=\columnwidth]{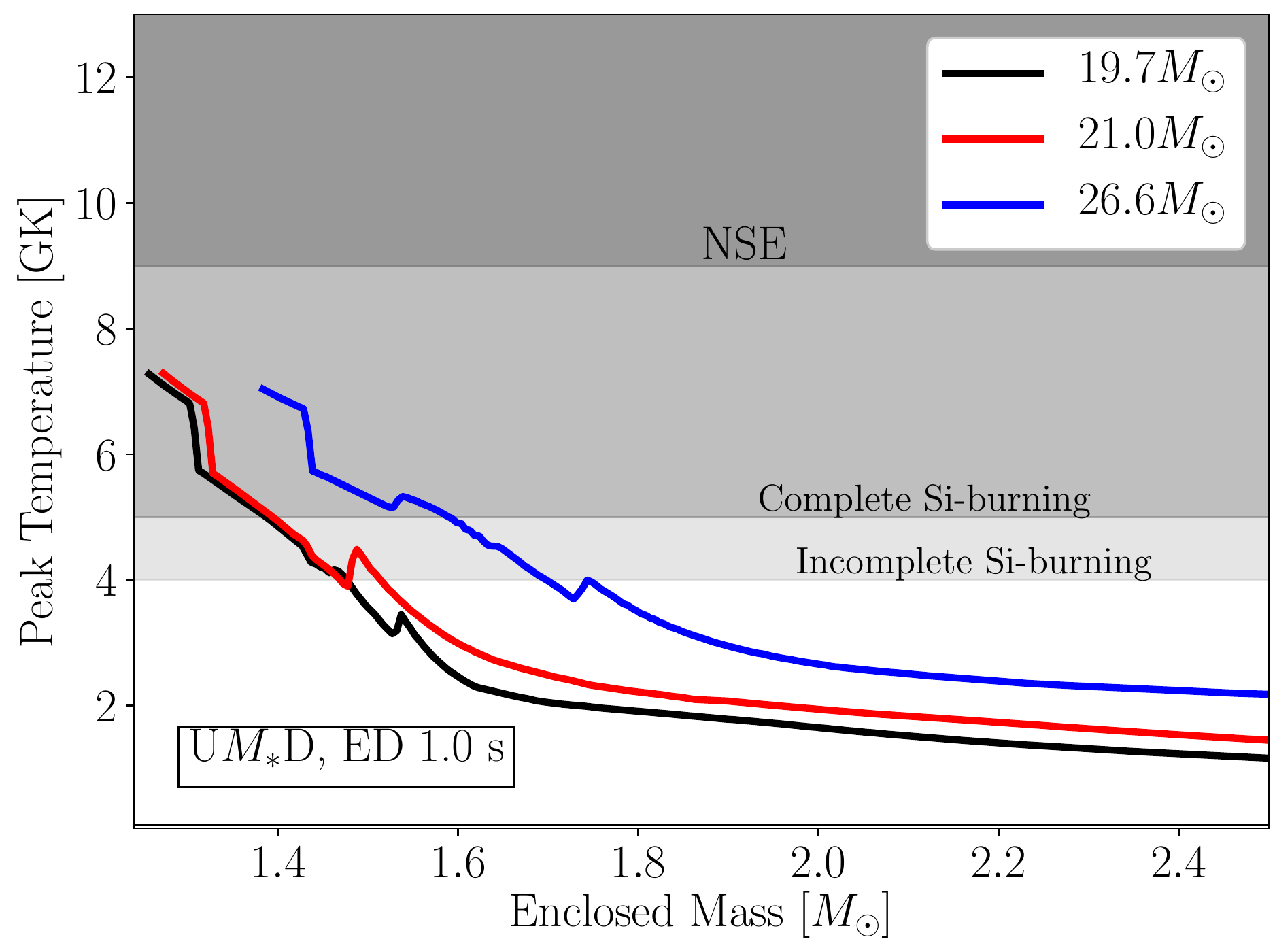}
\includegraphics[width=\columnwidth]{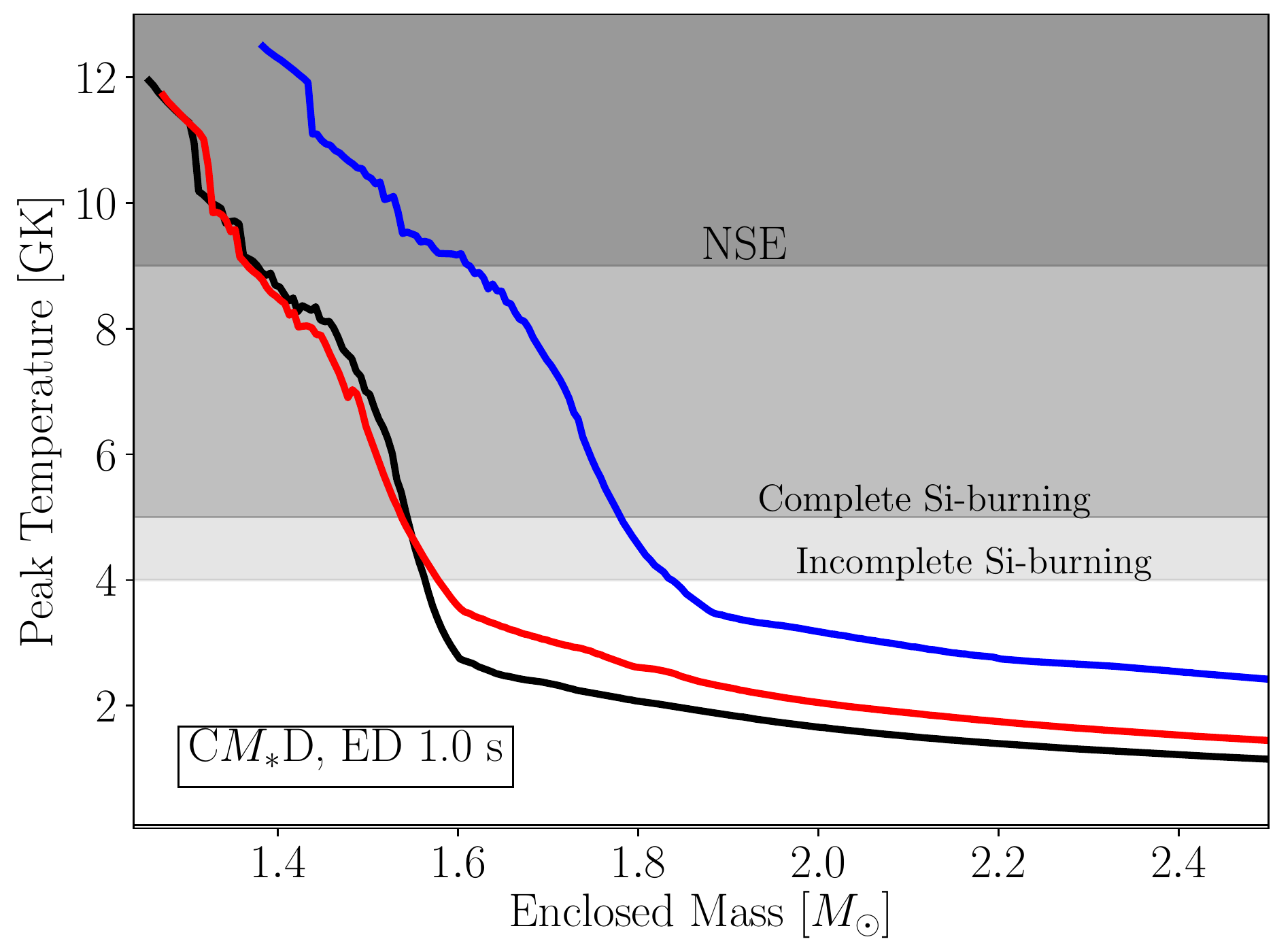}
\includegraphics[width=\columnwidth]{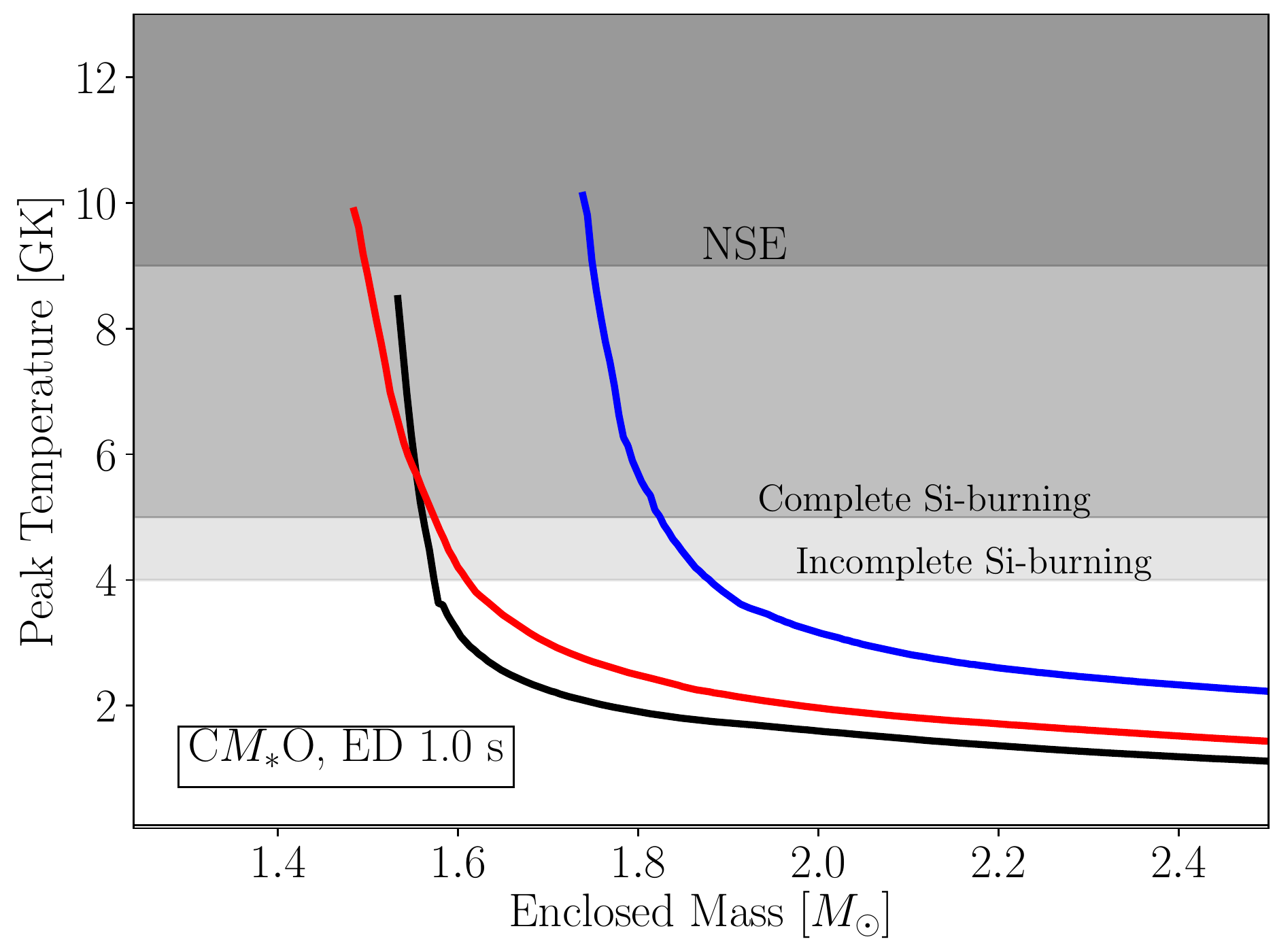}

        \caption{Peak temperatures as functions of enclosed mass for CCSN models for different progenitors using the standard value of $\Delta M = 0.05\,M_\odot$ for the fixed mass in the energy-injection layer and a representative energy-deposition timescale of 1.0\,s: uncollapsed (top) and collapsed (middle) with deep inner grid boundary, and collapsed with inner grid boundary shifted farther out (bottom). Grey shading again indicates different regimes of explosive nucleosynthesis as in Figure~\ref{fig:peakTtinj}. Note that the peak temperatures are displayed only for the runs with our default choice of $\Delta M = 0.05\,M_\odot$, because the differences compared to the other choices of $\Delta M$ are effectively indistinguishable.}
\label{fig:peakTC}

\end{figure}

\begin{figure}

\centering
\includegraphics[width=\columnwidth]{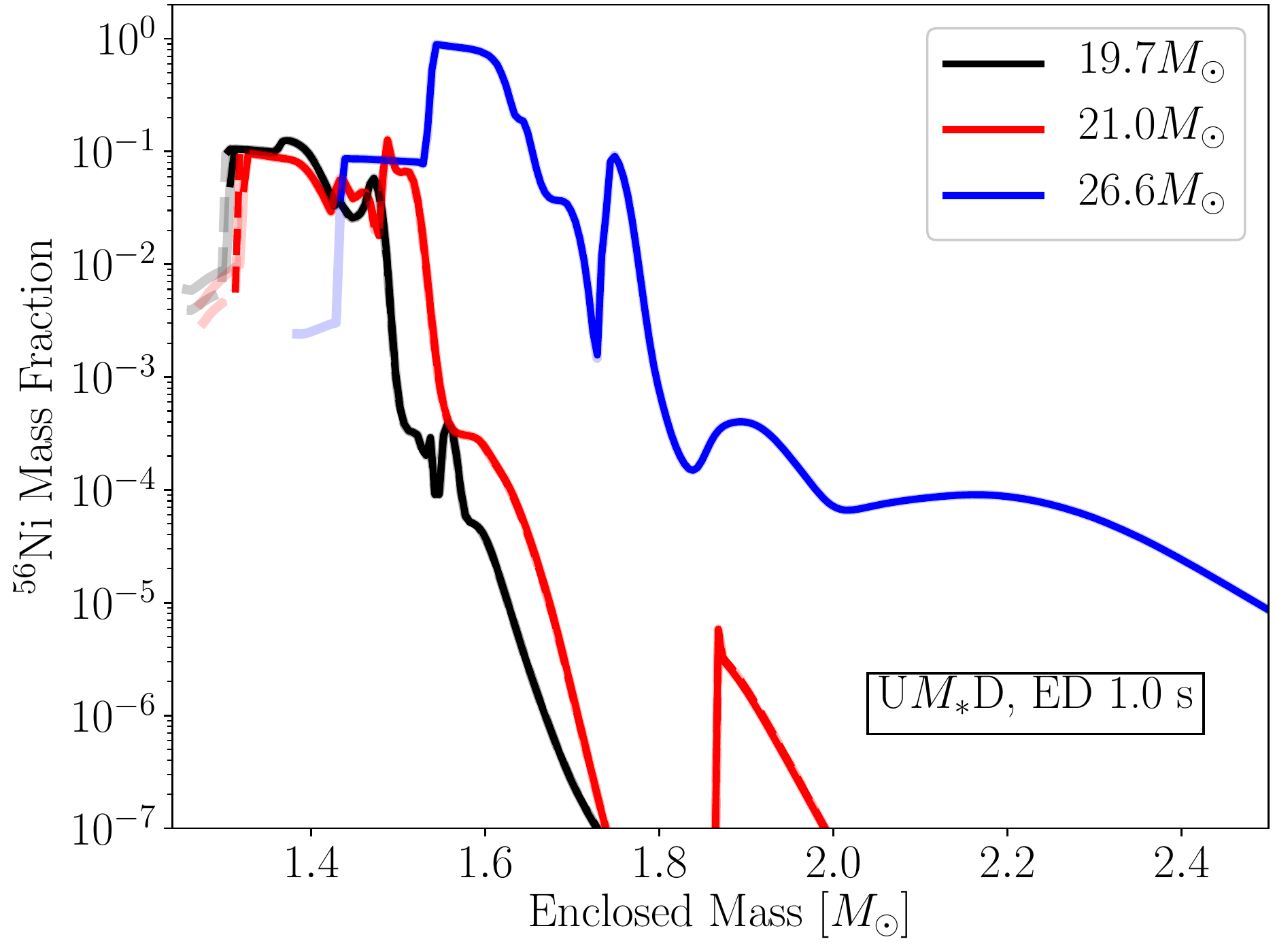}

\includegraphics[width=\columnwidth]{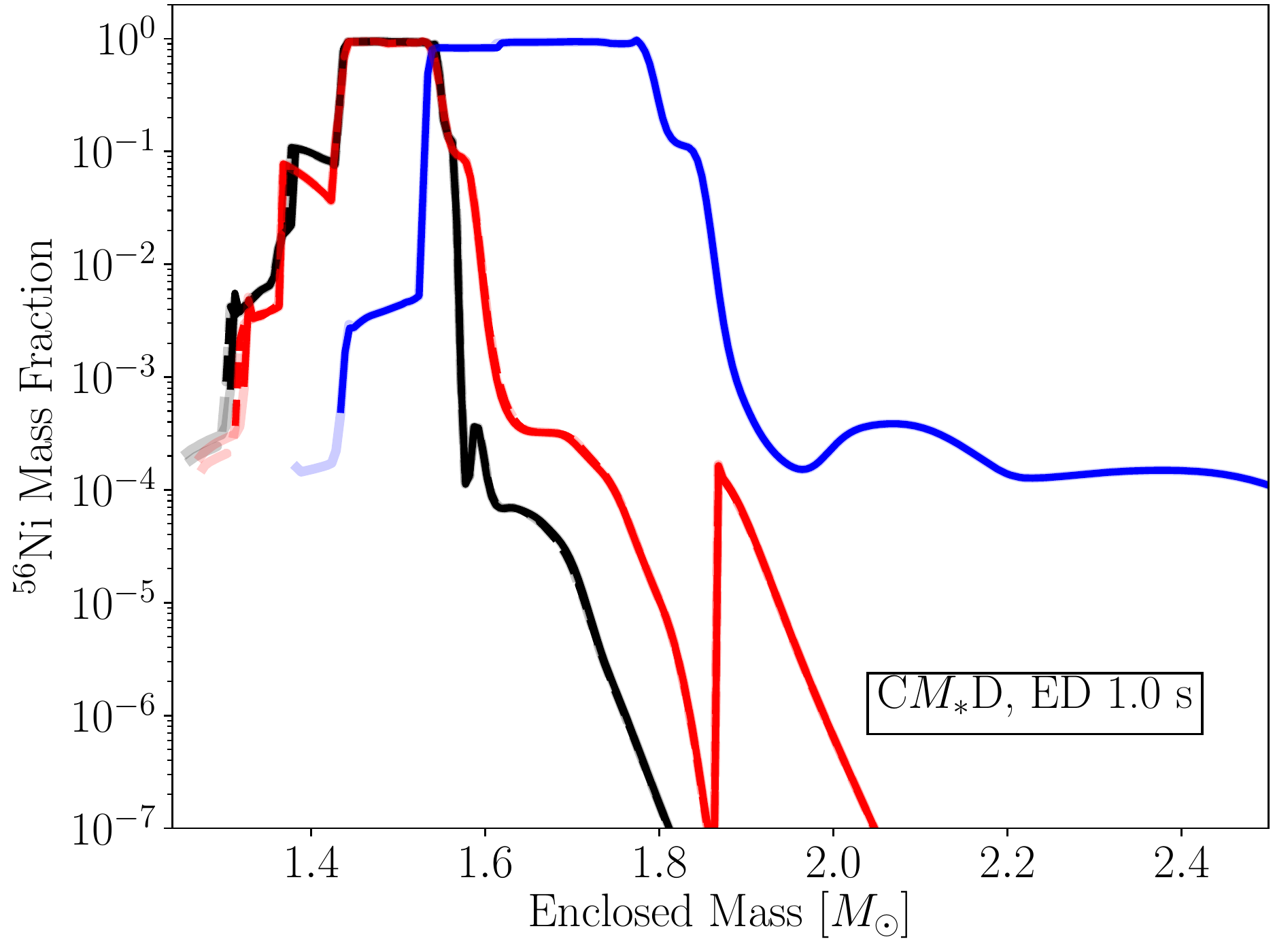}

\includegraphics[width=\columnwidth]{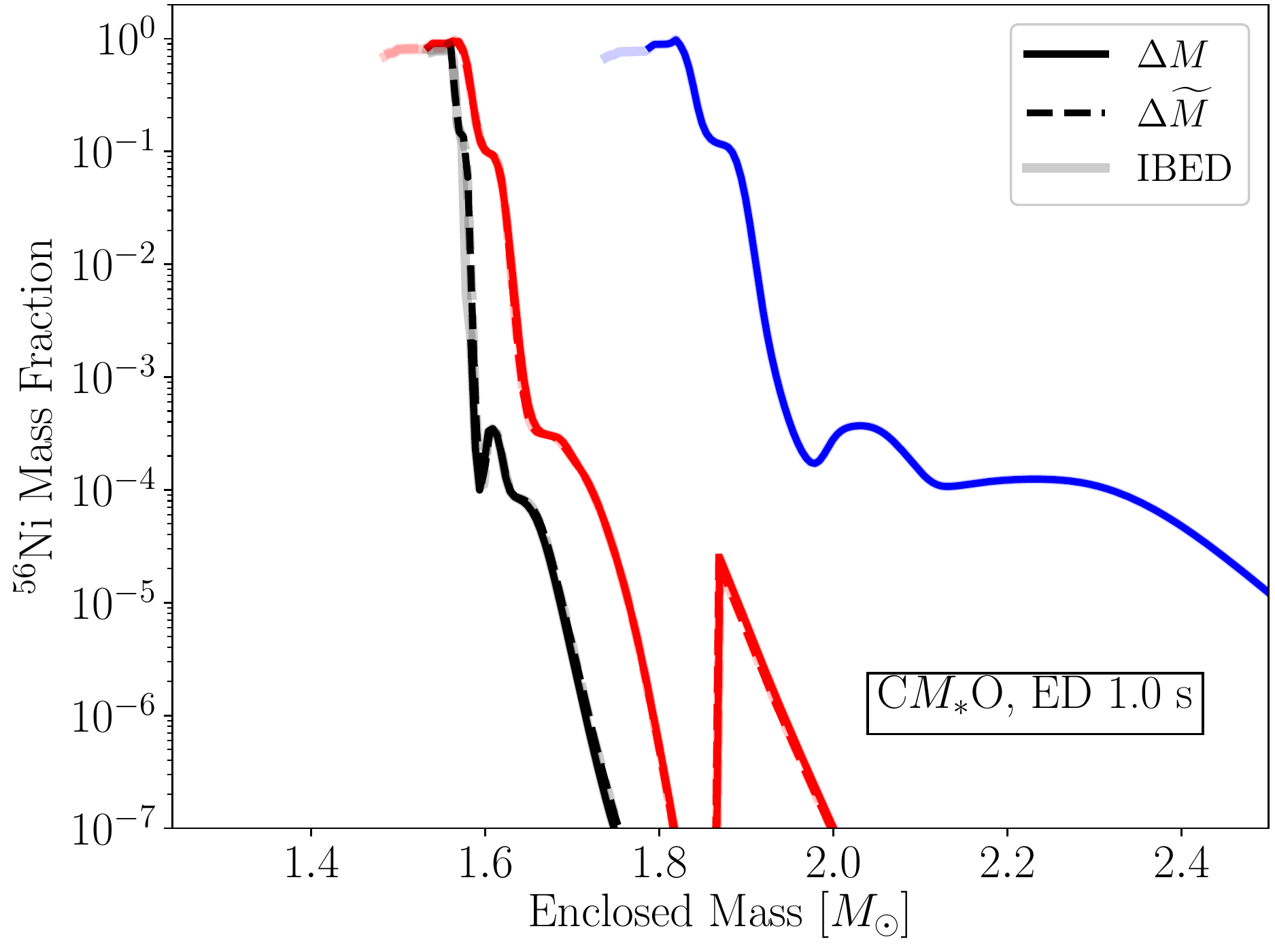}

     \caption{$^{56}$Ni mass fractions as functions of enclosed mass as produced in the CCSN models shown in Figure~\ref{fig:peakTC}. Here, we plot the results for our standard value of $\Delta M = 0.05\,M_\odot$ for the fixed mass in the energy-injection layer (solid lines) and for the cases with varied mass values $\Delta \widetilde{M}$ (models with unprimed M in their names, see Table~\ref{tab:explosions}; dashed lines). Note that the solid and dashed lines mostly overlap and therefore are hardly distinguishable. Moreover, we highlight the contribution to the $^{56}$Ni production from the mass in the energy-injection layer, which is included in our definition of the ejecta (indicated by light-colored parts of the solid and dashed lines).}
\label{fig:xni}
\end{figure}

These differences can be traced back to the progenitor structures plotted in 
Figure~\ref{fig:psn_closer} and to the peak temperature profiles in the ejecta caused by
the thermal bomb (see top panel in Figure~\ref{fig:peakTC}). Because of the shallow 
density profile at $r > R_\mathrm{ib}$ in the 26.6\,$M_\odot$ progenitor, the outward 
going shock wave that is generated by a thermal bomb with final explosion energy of
$10^{51}$\,erg heats much more mass to the temperatures required for strong $^{56}$Ni
production. The $^{56}$Ni nucleosynthesis is actually hampered in the 26.6\,$M_\odot$ 
progenitor by the fact that its innermost layer of $\sim$0.15\,$M_\odot$ possesses 
$Y_e$ values below 0.485 (Figure~\ref{fig:psn_closer}, upper right panel). In such conditions
the mass fraction of $^{56}$Ni does not exceed a few percent, see Figure~\ref{fig:psn_closer}, 
lower right panel, and Figure~\ref{fig:xni}, top panel, for $t_\mathrm{inj} = 0.01$\,s and
$t_\mathrm{inj} = 1.0$\,s, respectively. Nevertheless, the 26.6\,$M_\odot$ runs
produce a lot of $^{56}$Ni because considerable abundances of this isotope can
be nucleosynthesized even beyond an enclosed mass of $\sim$1.8\,$M_\odot$, in 
particular for short energy-injection times. 

In contrast, the 12.3\,$M_\odot$ 
progenitor possesses only a narrow layer of less than $\sim$0.07\,$M_\odot$ with
$Y_e\lesssim 0.485$ around $R_\mathrm{ib}$. This enables a relatively abundant 
production of $^{56}$Ni in the thermal-bomb models with this star for all 
energy-injection times and in spite of the steeper density profile compared to the
26.6\,$M_\odot$ progenitor. Finally, the two stellar models with 19.7\,$M_\odot$ and 
21\,$M_\odot$ exhibit very similar $Y_e$ profiles and also their density profiles
are close to each other up to the base of the oxygen shell, which is at roughly 
1.48\,$M_\odot$ in the 21\,$M_\odot$ model, but at about 1.53\,$M_\odot$ in the 
19.7\,$M_\odot$ case (see Table~\ref{tab:psn}). This difference, however, is 
located quite far away from the inner grid boundaries (which are at 1.256\,$M_\odot$ 
and 1.272\,$M_\odot$ for 19.7\,$M_\odot$ and 21\,$M_\odot$, respectively; see
Table~\ref{tab:explosions}) and its consequence (i.e., higher $^{56}$Ni mass 
fractions up to larger mass coordinates in the 21.0\,$M_\odot$ runs; 
Figure~\ref{fig:xni}) is partly compensated by more efficient $^{56}$Ni production
in the layers just exterior to the energy-injection domain in the 19.7\,$M_\odot$
runs (Figure~\ref{fig:psn_closer}, lower right panel, and Figure~\ref{fig:xni},
top panel). The overall effect is that both progenitors resemble each other closely 
in their $^{56}$Ni outputs for all values of $t_\mathrm{inj}$, at least when 
uncollapsed thermal-bomb models with deep inner boundary are considered.

In the following we will not use the 12.3\,$M_\odot$ runs any further, because 
they exhibit the weakest variation of the produced $^{56}$Ni mass with $t_\mathrm{inj}$, 
whereas our main focus is on how this variation is affected when an initial collapse 
phase is included in the thermal-bomb treatment.



\subsection{Collapsed models}
\label{sec:collapsedmodels}

The picture changes radically when a collapse phase is introduced into the explosion modelling
before the energy injection by the thermal bomb is switched on Figure~\ref{fig:mni0}, middle
panel, displays the $^{56}$Ni yields for the corresponding models with deep inner boundary
(our C$M_*$D simulations). 
For short energy-injection timescales ($t_\mathrm{inj}\lesssim 0.05$\,s) we find amounts of 
$^{56}$Ni very similar to those obtained in the uncollapsed models, but now also the explosion 
simulations with longer $t_\mathrm{inj}$ are efficient in producing $^{56}$Ni. In fact, there 
is little variation of the $^{56}$Ni yields when $t_\mathrm{inj}$ increases from 0.01\,s to 2\,s.
The anti-correlation of the $^{56}$Ni production with $t_\mathrm{inj}$ observed for the 
U$M_*$D models is gone and instead the C$M_*$D models exhibit a $^{56}$Ni nucleosynthesis that 
varies much less with the duration of the energy release by the thermal bomb.

Inspecting the peak temperature profiles versus enclosed mass (Figure~\ref{fig:peakTtinj}, 
middle panel),
one recognizes three main differences compared to the uncollapsed cases in the top panel of 
this figure. First, the maximum peak temperatures for all energy-injection times reach higher values
in the C-models and extend well into the NSE regime. Second, the peak temperature profiles are more similar
to each other than in the U-models when $t_\mathrm{inj}$ is varied. And third, this implies that
for all values of $t_\mathrm{inj}$ a wider mass layer is heated to the temperatures required
for complete Si burning or NSE.

These differences in the collapsed models compared to the uncollapsed ones have several reasons, 
whose relative importance varies with the energy-injection timescale.
Because of the compression heating during the collapse, the temperatures at the onset of the 
energy injection by the thermal bomb are already higher. A more important effect, however, 
is connected to the fact that the shock expands into stellar layers that have collapsed for 
$\sim$0.5\,s or longer and over 
radial distances between several 100\,km and more than 1000\,km. The growing kinetic energy 
of the infalling gas is converted to thermal energy in the shock. Moreover, the energy input 
into the collapsed mass layer of $\Delta M = 0.05\,M_\odot$ means that the 
energy is injected into a much smaller volume than in the uncollapsed models (Table~\ref{tab:dR}), 
implying considerably higher heating rates per 
unit volume. For the uncollapsed 26.6\,$M_\odot$ model with deep inner boundary, for example, 
the initial radii bounding the heating layer are $R_\mathrm{IBED}\approx 1280$\,km and 
$R_\mathrm{OBED}\approx 1380$\,km, i.e., the layer has a width of $\sim$100\,km, whereas in 
the corresponding collapsed model the initial radial extension of the heating layer is only 
40\,km between 500\,km and 540\,km (see Table~\ref{tab:dR}). In addition, the expansion of
the heated matter sets in much more slowly in the collapsed models, where the energy-injection
layer sits deeper in the gravitational potential and the overlying, infalling mass shells
provide external pressure, hampering the outward acceleration. One can clearly see this
effect when comparing the top and middle panels of Figure~\ref{fig:second_wave}. This inertia 
of the matter in the wake of the outgoing shock permits the energy injection to boost the 
temperature and thus the postshock pressure to high values even when the energy-deposition 
timescales are long. As a
consequence, the shock is pushed strongly into the infalling, overlying shells, and the 
peak-temperature profiles (Figure~\ref{fig:peakTtinj}) as well as the mass that is heated 
sufficiently to enable abundant $^{56}$Ni production become quite similar for different 
$t_\mathrm{inj}$. 

Again, as for the U-models, the thermal-bomb runs for the collapsed 26.6\,$M_\odot$ models
lead to the highest  yields when the final explosion energy is fixed to 
$\sim$\,$10^{51}$\,erg for all progenitors. Once again, this is connected to the more shallow 
pre-collapse density profile of the 26.6\,$M_\odot$ star, for which reason more mass is heated to $^{56}$Ni-production temperatures (Figure~\ref{fig:peakTC}, middle panel). Correspondingly, 
the mass layer with a high mass fraction of this isotope is much more extended in the C26.6 
models (see Figure~\ref{fig:xni}, middle panel). More energy input by the thermal bomb is
needed and, accordingly, a stronger shock wave is created to lift the ejecta out of the 
deeper gravitational potential of the central mass of the new-born neutron star
($M_\mathrm{ib} = 1.383\,M_\odot$ in model C26.6D compared to 1.256\,$M_\odot$ and
1.272\,$M_\odot$ in models C19.7D and C21.0D, respectively).

The $^{56}$Ni yields of the 19.7\,$M_\odot$ and 21.0\,$M_\odot$ models are somewhat more 
different in the simulations with initial collapse than in the runs without collapse, 
especially for energy-injection times shorter than 0.5\,s (Figure~\ref{fig:mni0}, middle panel),
despite the similar density profiles of the two stars up to the base of the oxygen shell and 
despite their steep increase from $Y_e < 0.485$ to $Y_e > 0.495$ happening at the same 
mass coordinate (Figure~\ref{fig:psn_closer}, upper two panels). The C21.0D models 
nevertheless produce more $^{56}$Ni because the interface to the O-layer with decreasing
density and increasing entropy lies at a lower enclosed mass, permitting stronger shock
heating and more $^{56}$Ni nucleosynthesis in the oxygen shell (Figure~\ref{fig:xni},
middle panel). For long energy-injection times, however, this effect is again 
compensated by slightly more $^{56}$Ni production in the innermost layers of the 
C19.7D runs.

A special feature requires brief discussion: At intermediate energy-deposition timescales
the C21.0D and C26.6D models exhibit local maxima of their $^{56}$Ni yields, more
prominently in the 26.6\,$M_\odot$ cases and only shallow in the 21.0\,$M_\odot$ runs.
This phenomenon is caused by the thermal-bomb prescription of energy-injection into
a fixed mass shell $\Delta M$ that starts expanding when the energy deposition sets
in. This creates a compression wave when the energy deposition takes place on a shorter
timescale than the expansion, which leads to peak temperatures in the ejecta that are 
reached not exactly right behind the outgoing shock wave but at some distance 
behind the shock, thus causing high temperatures for a longer period in a wider layer 
of mass and therefore more $^{56}$Ni production. This effect can be seen in a weak
variant in the middle panel of Figure~\ref{fig:second_wave}, where between $t\sim 1.1$\,s 
and $t\sim 1.3$\,s the peak temperatures of the expelled mass shells (marked by crosses)
appear detached from the shock. In this 21.0\,$M_\odot$ model with 
$t_\mathrm{inj} = 1.0$\,s, however, the effect is mild and has no relevant impact on the 
$^{56}$Ni nucleosynthesis. For simulations with very short $t_\mathrm{inj}$ the energy 
deposition is so fast that the compression wave quickly merges with the shock, whereas
for very long timescales $t_\mathrm{inj}$ the energy injection is gentle and keeps pace with
the outward acceleration of the mass shells, for which reason a strong compression wave
is absent. Only at intermediate values of $t_\mathrm{inj}\sim 0.2$\,s this compression 
wave has a significant influence on the temperature evolution of the ejected mass shells 
in the postshock domain and thus a noticeable effect on enhanced $^{56}$Ni production.


\subsection{Shifted inner boundary}
\label{sec:shiftedIBED}

In a next test we moved the inner grid boundary from the deep location to the position at the
base of the O-shell (where $s/k_\mathrm{B} = 4$). This choice for the C$M_*$O models is more realistic 
than the deep inner boundary, because it is 
better compatible with our current understanding of the neutrino-driven explosion mechanism of CCSNe 
\citep[e.g.,][]{2016ApJ...818..124E,2016ApJ...821...38S}. The corresponding $^{56}$Ni yields of the 
thermal-bomb simulations with our standard setting of $\Delta M = 0.05\,M_\odot$ for the energy-injection
layer and different values of $t_\mathrm{inj}$ are displayed by solid lines in the bottom panel of 
Figure~\ref{fig:mni0}.

First, we notice that the $^{56}$Ni yields of C$M_*$O models are much lower for all $t_\mathrm{inj}$ than in the 
C$M_*$D models in the panel above. In absolute numbers these yields are closer to the typical values of $\sim$0.05--$0.1\,M_\odot$ for the $^{56}$Ni production in CCSNe with explosion energies around 
(1--2)$\times 10^{51}$\,erg \citep[see, e.g.,][]{1989ARA&A..27..629A,1994ApJ...437L.115I,2017ApJ...841..127M}.
While models C26.6O and C21.0O eject similar amounts of $^{56}$Ni, model C19.7O, in contrast,
produces considerably less $^{56}$Ni.

Several important aspects in the C-models with the O-boundary are different from those with the D-boundary: The densities and therefore the ram pressure in the pre-shock matter are significantly lower, for which reason the expansion of the shock and thus also of the matter in the energy-injection layer and above occurs much faster. This can be seen by comparing the middle and bottom panels of Figure~\ref{fig:second_wave}. Moreover, since the density is low, the energy injected into a given mass layer $\Delta M$ is distributed over a considerably wider volume, which can be concluded from the values of $R_\mathrm{OBED}$ given for the C$M_*$O and C$M_*$D models in Table~\ref{tab:dR} ($1.76\cdot 10^8$\,cm and $5.4\cdot 10^7$\,cm, respectively). The effect, however, is not quite as dramatic as the different $R_\mathrm{OBED}$ might suggest, because the density gradient is steep and most of the heated mass $\Delta M$ is still located relatively close to $R_\mathrm{IBED} = 5\cdot 10^7$\,cm. Overall, however, these differences lead to steeper declines of the peak temperatures with enclosed mass than in the models with D-boundary (compare the bottom panels of Figures~\ref{fig:peakTtinj} and \ref{fig:peakTC} with the top and middle panels of these figures). This explains why in the CCSN models with O-boundary less mass is heated to $^{56}$Ni production temperatures. As a consequence, the layer of abundant $^{56}$Ni nucleosynthesis is much narrower in mass and very close to the inner grid boundary (Figure~\ref{fig:xni}), and the total $^{56}$Ni yields are considerably lower than in the CCSN models with deep boundary, even when the final explosion energy is tuned to the same value. 

In the C$M_*$O models the peak temperature profiles are quite similar for different energy-injection timescales (Figure~\ref{fig:peakTtinj}, bottom panel), for which reason the $^{56}$Ni outputs of the 21.0\,$M_\odot$ and 26.6\,$M_\odot$ models are relatively similar with a moderate decrease for longer $t_\mathrm{inj}$. In the case of the 19.7\,$M_\odot$ simulations, however, the peak temperature declines extremely steeply as function of enclosed mass (Figure~\ref{fig:peakTC}) because of the very low densities of the heated mass layer (due to the low densities in the oxygen layer of the progenitor; Figure~\ref{fig:psn_closer}). Therefore the expansion of this layer proceeds extremely quickly and the expansion cooling as well as the dilution of the energy deposition over a quickly growing volume do not permit high peak temperatures in a large mass interval. This leads to the result that the $^{56}$Ni yields in the C19.7O models are the lowest of all of the three considered progenitors.

Another difference between C-models with D-boundary and O-boundary is the fact that in the latter
the inclusion of the heated mass $\Delta M$ in the ejecta or its exclusion can make a sizable difference 
in the $^{56}$Ni yields. In contrast to the U$M_*$D and C$M_*$D models, the 
simulations with collapse and O-boundary produce considerably less $^{56}$Ni when the matter in the
energy-injection layer is not taken into account in the ejecta (see the light-colored solid lines in
the bottom panel of Figure~\ref{fig:mni0}). In particular, C19.7O underproduces
$^{56}$Ni massively in this case, and for the models with the 21.0\,$M_\odot$ and 26.6\,$M_\odot$
progenitors we witness again a strong trend of decreasing $^{56}$Ni yields with longer energy-injection
timescales when only material exterior to $R_\mathrm{OBED}$ is counted as ejecta.

Such a trend, however, disappears essentially entirely when the $^{56}$Ni nucleosynthesized 
in the energy-deposition layer is included in the ejecta (heavy solid lines compared to 
light-colored solid lines in the bottom panel of Figure~\ref{fig:mni0}). We recall that the exclusion
of the heated mass from the ejecta or its inclusion does not have any relevant influence on the total
$^{56}$Ni yields of our U- and C-models with deep inner boundary, because the low $Y_e$ in the vicinity of this boundary location (see Figure~\ref{fig:psn_closer}) prevents abundant production of $^{56}$Ni in the heated mass layer (Figure~\ref{fig:xni}, top and middle panels). The situation is different now for the O-models, because $Y_e$ is close to 0.5 near the inner grid boundary in this case
(Figure~\ref{fig:psn_closer}). Much of the $^{56}$Ni is then produced in the mass layers just 
exterior to $R_\mathrm{ib}$ in addition to the fact that the total $^{56}$Ni yields are much smaller
(Figure~\ref{fig:xni}, bottom panel). Therefore the $^{56}$Ni assembled in the heated mass can make a significant or even dominant contribution to the total yield of this isotope. The C19.7O models are the most extreme cases in this respect. Their $^{56}$Ni yields are extremely low when only matter exterior to the heated layer is considered as
ejecta. This is especially problematic since our default value of 0.05\,$M_\odot$ for the
energy-injection mass $\Delta M$ is fairly large. This fact is further illuminated in the following section, where we will discuss the results for variations of $\Delta M$.


\begin{figure*}
\centering

\includegraphics[width=\columnwidth]{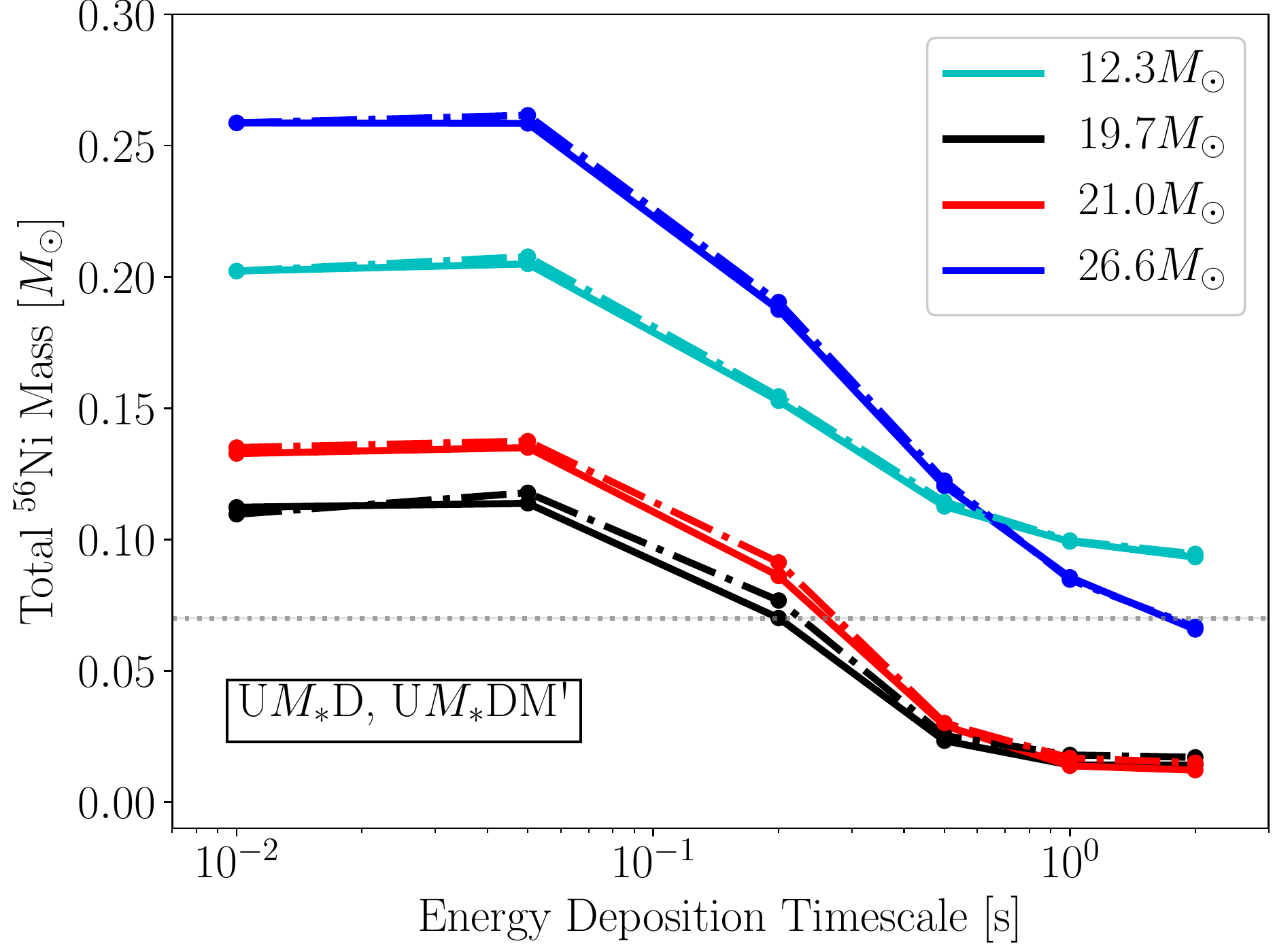}
\includegraphics[width=\columnwidth]{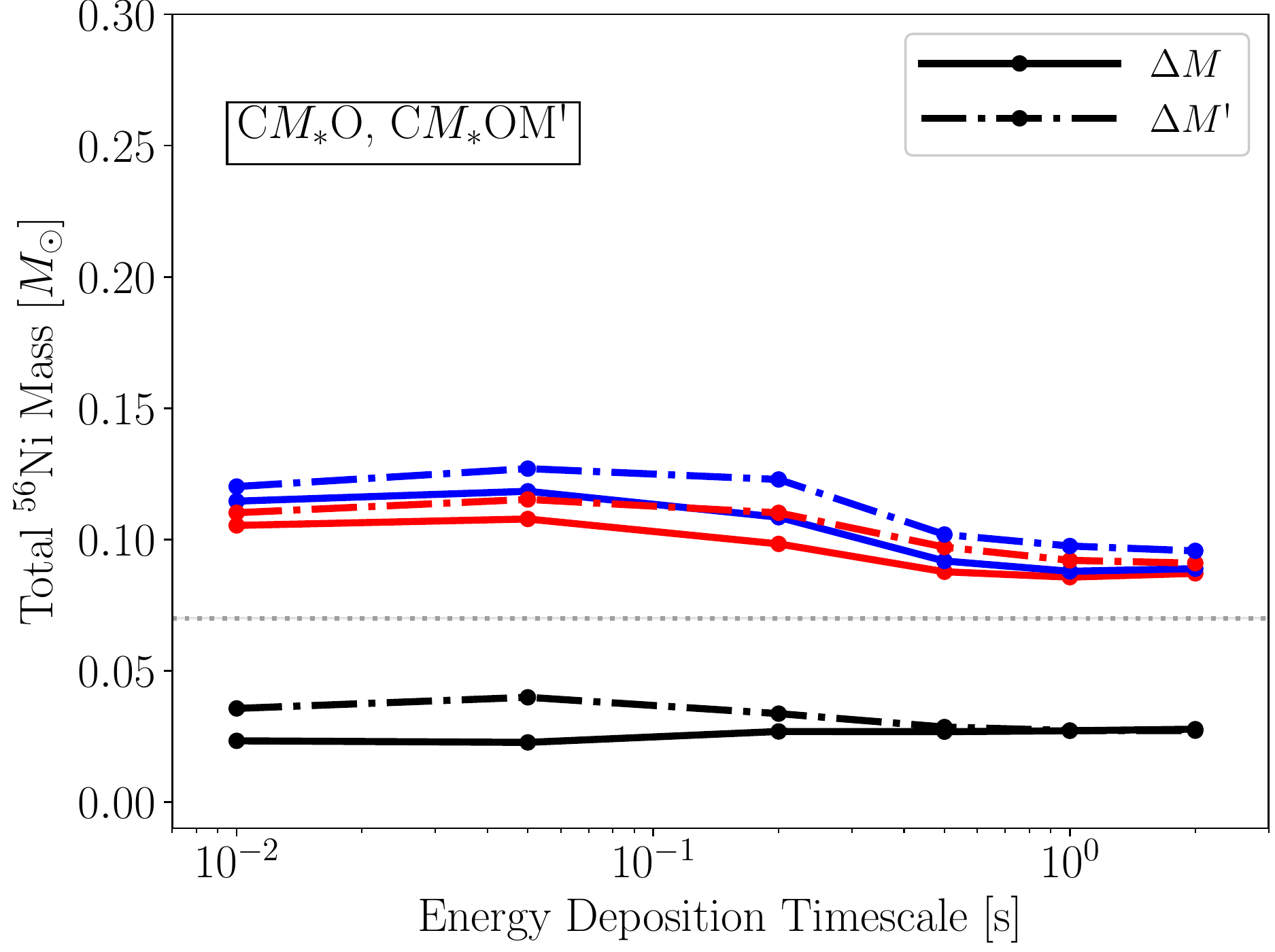}
   
      \caption{ 
      $^{56}$Ni yields as functions of energy-injection timescale for uncollapsed CCSN models (left panel) and collapsed CCSN models with inner grid boundary shifted farther out (right panel). The different colors correspond to the different progenitors as labelled in the left panel. Solid lines belong to our standard choice of $\Delta M = 0.05\,M_\odot$ for the fixed mass in the energy-deposition layer and dash-dotted lines refer to the values of $\Delta M$\textquotesingle$=0.005\,M_\odot$ (see Table~\ref{tab:explosions}). The horizontal grey dotted line indicates the $^{56}$Ni yield of 0.07\,$M_\odot$ for a $\sim$\,$10^{51}$\,erg explosion, e.g., SN~1987A \citep{1989ARA&A..27..629A}.}
     \label{fig:mni005}   
\end{figure*}

\subsection{Variations of mass in energy-injection layer}
\label{sec:massvariations}

We also simulated some test cases of U-models and C-models using moderately different values of 
the fixed heated mass $\Delta M$, varied within plausible ranges such that the initial volumes 
of the heated masses are the same for the C-models of all progenitors (see Table~\ref{tab:dR}
and Section~\ref{sec:bombvariations}). These models are denoted by U$M_*$DM, C$M_*$DM, and
C$M_*$OM, represented by dashed lines in the panels of Figure~\ref{fig:mni0}. 

There are no relevant effects with respect to the $^{56}$Ni production, neither in U-models nor C-models, in the cases with deep inner boundary when $\Delta M \approx 0.04\,M_\odot$ is used instead of $\Delta M = 0.05\,M_\odot$; the dashed lines are mostly indistinguishable from the solid lines in the top and middle panels of Figure~\ref{fig:mni0}. However, slightly more sensitivity of the $^{56}$Ni yields to the choice of $\Delta M$ is obtained in the cases of the C$M_*$O models (bottom panel of Figure~\ref{fig:mni0}). Changing to $\Delta M \approx 0.03\,M_\odot$ (C19.7OM models) increases the nickel production for $t_\mathrm{inj} \lesssim 0.2$\,s, whereas a change to $\Delta M \approx 0.07\,M_\odot$ decreases the $^{56}$Ni yield (C21.0OM models), displayed by heavy dashed lines in the bottom panel of Figure~\ref{fig:mni0}. In both cases the relative difference in the $^{56}$Ni yields compared to the standard setup with $\Delta M = 0.05\,M_\odot$ depends on $t_\mathrm{inj}$ and is largest
for short $t_\mathrm{inj}$ and low $^{56}$Ni production with the standard value of $\Delta M$.

We notice again that this effect is considerably stronger if the nucleosynthesis in the heated mass $\Delta M$ itself is excluded from the $^{56}$Ni budget (light-colored dashed lines in the bottom panel of Figure~\ref{fig:mni0}) instead of counting unbound matter in the energy-deposition layer also as ejecta (heavy dashed lines in the bottom panel of Figure~\ref{fig:mni0}). When $\Delta M$ is excluded from the ejecta, the $^{56}$Ni yields in the C$M_*$O (light-colored solid lines) and the C$M_*$OM models (light-colored dashed lines) do not only become significantly lower but also very sensitive to the energy-injection timescale, as already mentioned in Section~\ref{sec:shiftedIBED}. This strong variation with $t_\mathrm{inj}$ in the case of our O-boundary models reminds one of the SM19 results with D-boundary, but the effect vanishes almost entirely for all O-models when the $^{56}$Ni production within the heated mass layer is added to the ejecta.

For completeness, we also tested a radical reduction of $\Delta M$ from our default of 0.05\,$M_\odot$ to the value of 0.005\,$M_\odot$ adopted by SM19 for the fixed mass in the
energy-deposition layer (U- and C-models in Table~\ref{tab:explosions} with M\textquotesingle\ as
endings of their names). These simulations reproduce the trend witnessed for the C19.7OM
models compared to the C19.7O models in the bottom panel of Figure~\ref{fig:mni0}, 
namely that a reduced $\Delta M$ tends to increase the $^{56}$Ni production 
(see Figure~\ref{fig:mni005}). While the difference is small and thus
has no relevant effect in the uncollapsed (and collapsed) models with the D-boundary (left
panel of Figure~\ref{fig:mni005}) the increase is more significant in the simulations with 
O-boundary (right panel). However, considering all the results provided by Figures~\ref{fig:mni0} 
and \ref{fig:mni005}, one must conclude that, overall, the $^{56}$Ni yields are not
overly sensitive to the exact value chosen for $\Delta M$, and that the corresponding 
variations are certainly secondary compared to the differences obtained between 
collapsed and uncollapsed models and between changing from D-boundary to O-boundary.
 
These findings shed light on the many ambiguities and the somewhat arbitrary choices that can be made in the treatments of artificial explosions with parametric methods. In any case, it is advisable to include also the mass of the energy-injection layer in the ejecta of the thermal bomb, if this matter gets ultimately expelled during the explosion. This is particularly relevant when the initial mass cut is assumed to be located at the more realistic $s/k_\mathrm{B} = 4$ position and the thermal energy is dumped into an extended layer with mass $\Delta M$, whose choice is inspired (roughly) by the mass heated by neutrinos in CCSNe. If otherwise the mass of $\Delta M$ is excluded from the ejecta, the $^{56}$Ni production can become highly sensitive to the exact values of both $\Delta M$ and $t_\mathrm{inj}$, depending on the density structure of the progenitor star.


\begin{figure*}
\includegraphics[width=\columnwidth]{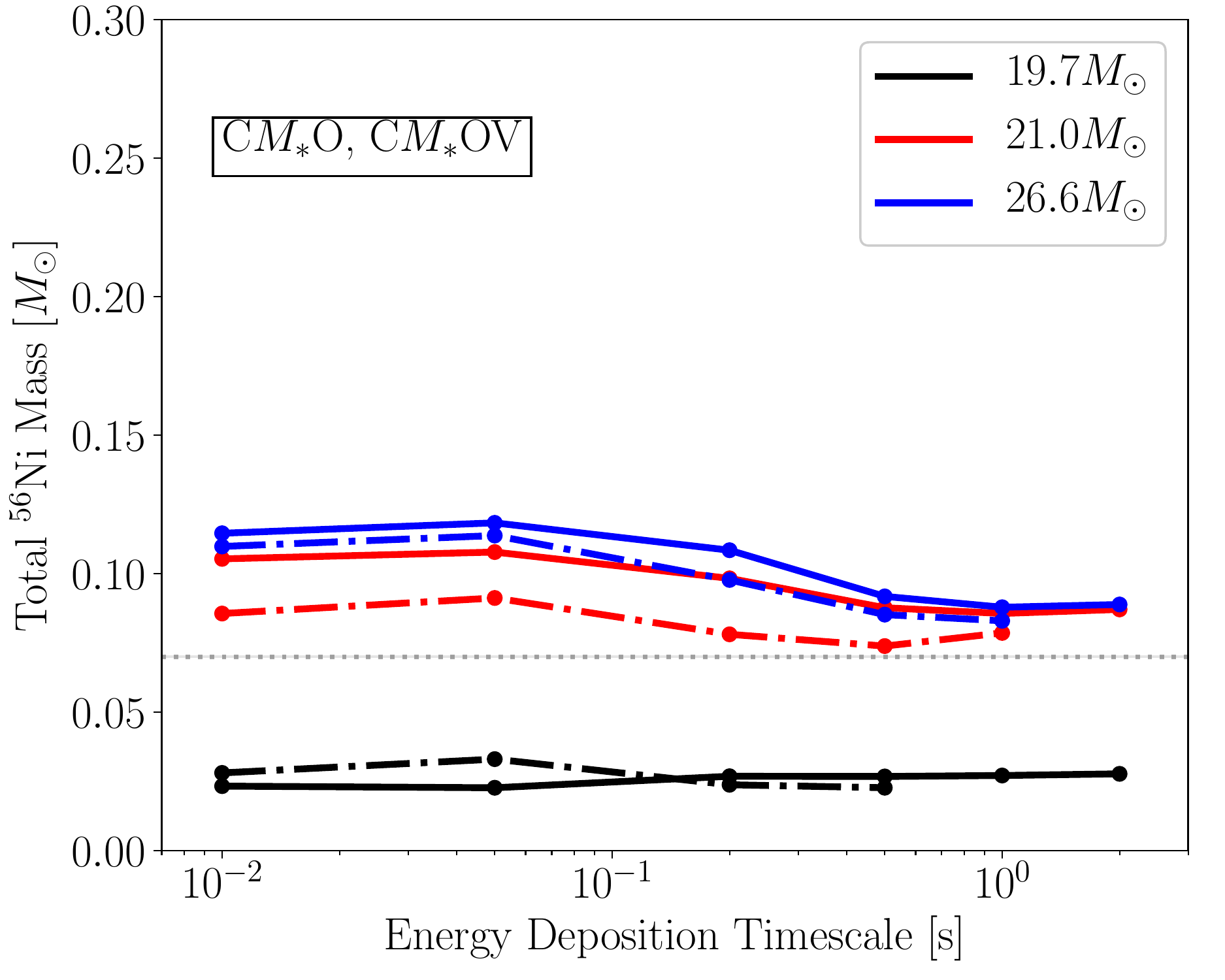}
\includegraphics[width=\columnwidth]{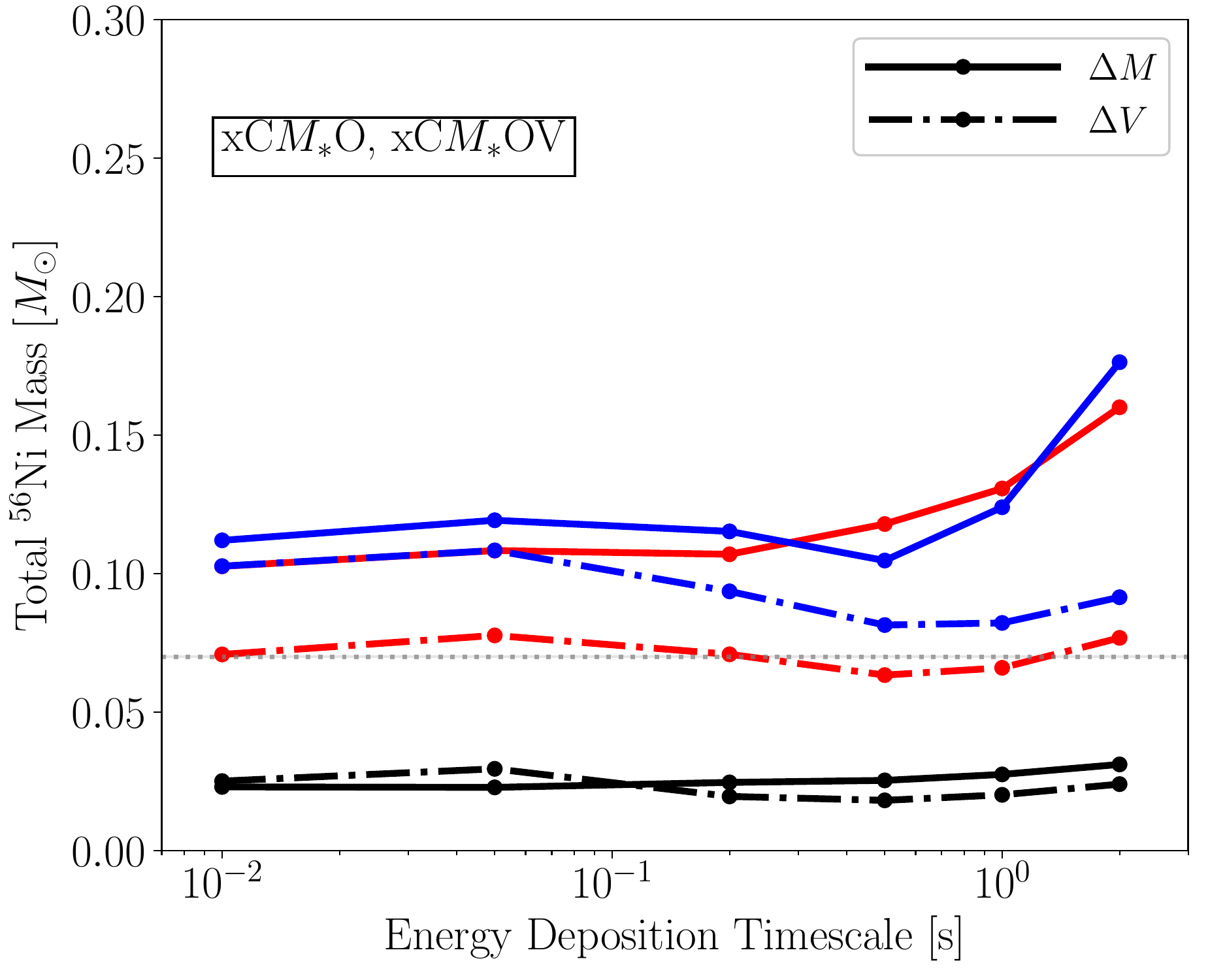}
\caption{$^{56}$Ni yields as functions of energy-injection timescale for collapsed CCSN models with fixed mass $\Delta M = 0.05\,M_\odot$ (solid lines) and fixed volume (dash-dotted lines) of the energy-deposition layer. The left panel displays the results for our standard collapse to $r_\mathrm{min}=500$\,km, the right panel the cases with extreme collapse to $r_\mathrm{min}=150$\,km. The different colors correspond to the different progenitors as labelled in the left panel. Note that the models with fixed volume for the longest energy-deposition timescales in the left panel could not be finished because of the computational demands connected to small time steps. The horizontal grey dotted line indicates the $^{56}$Ni yield of 0.07\,$M_\odot$ for a $\sim$\,$10^{51}$\,erg explosion, e.g., SN~1987A \citep{1989ARA&A..27..629A}.}
\label{fig:final_ni_150}
\end{figure*}

\begin{figure*}
\includegraphics[width=\columnwidth]{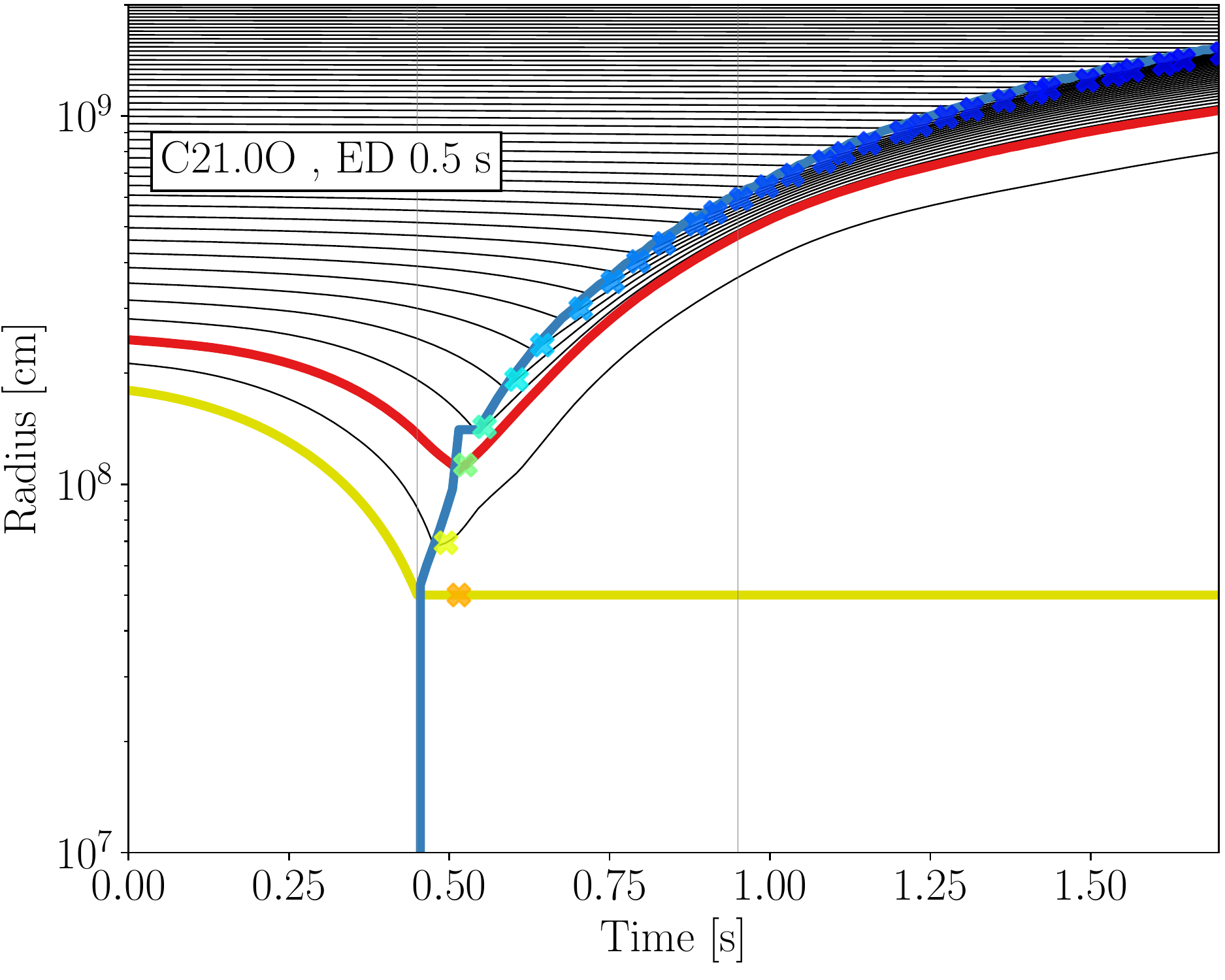}
\includegraphics[width=\columnwidth]{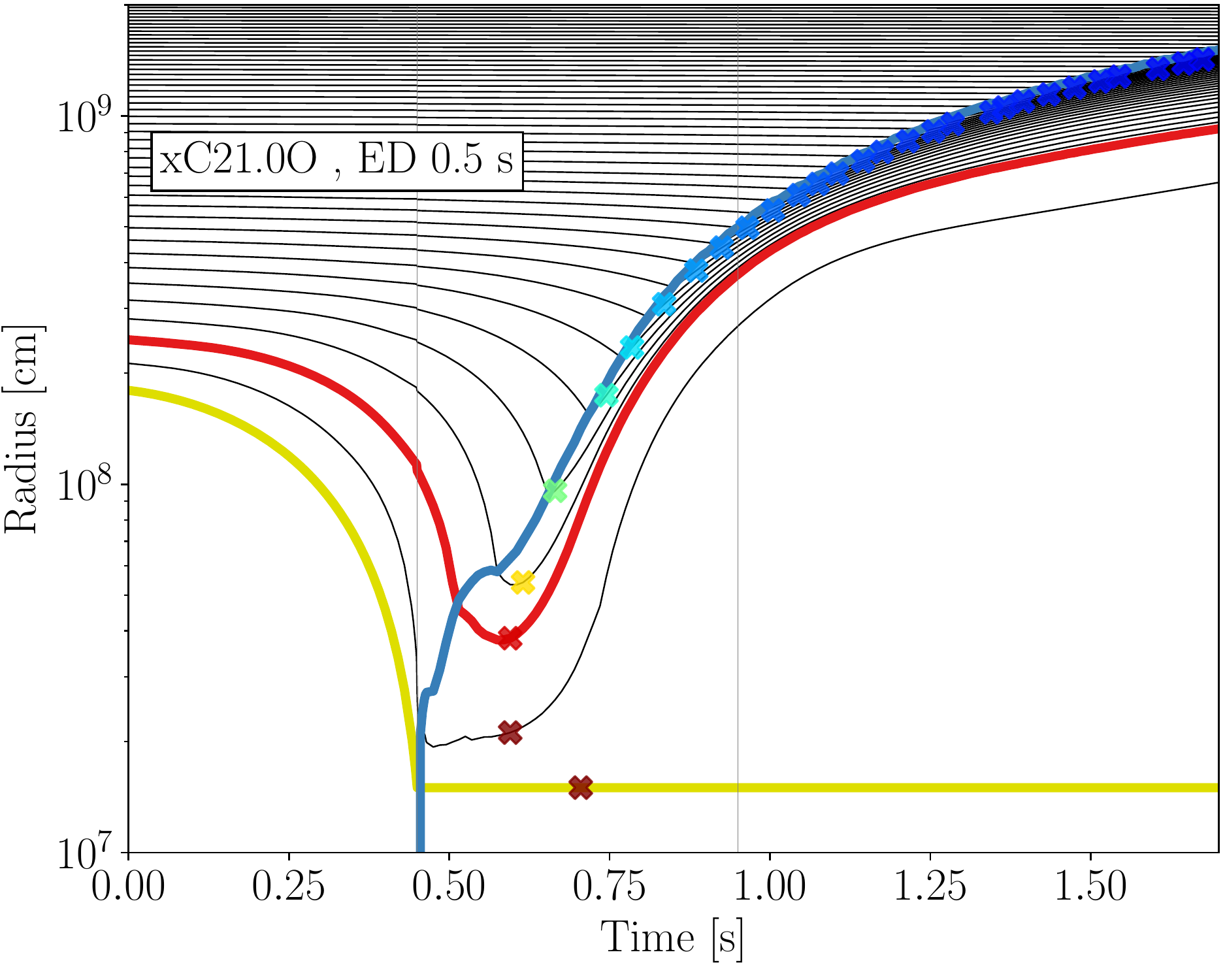}

\includegraphics[width=\columnwidth]{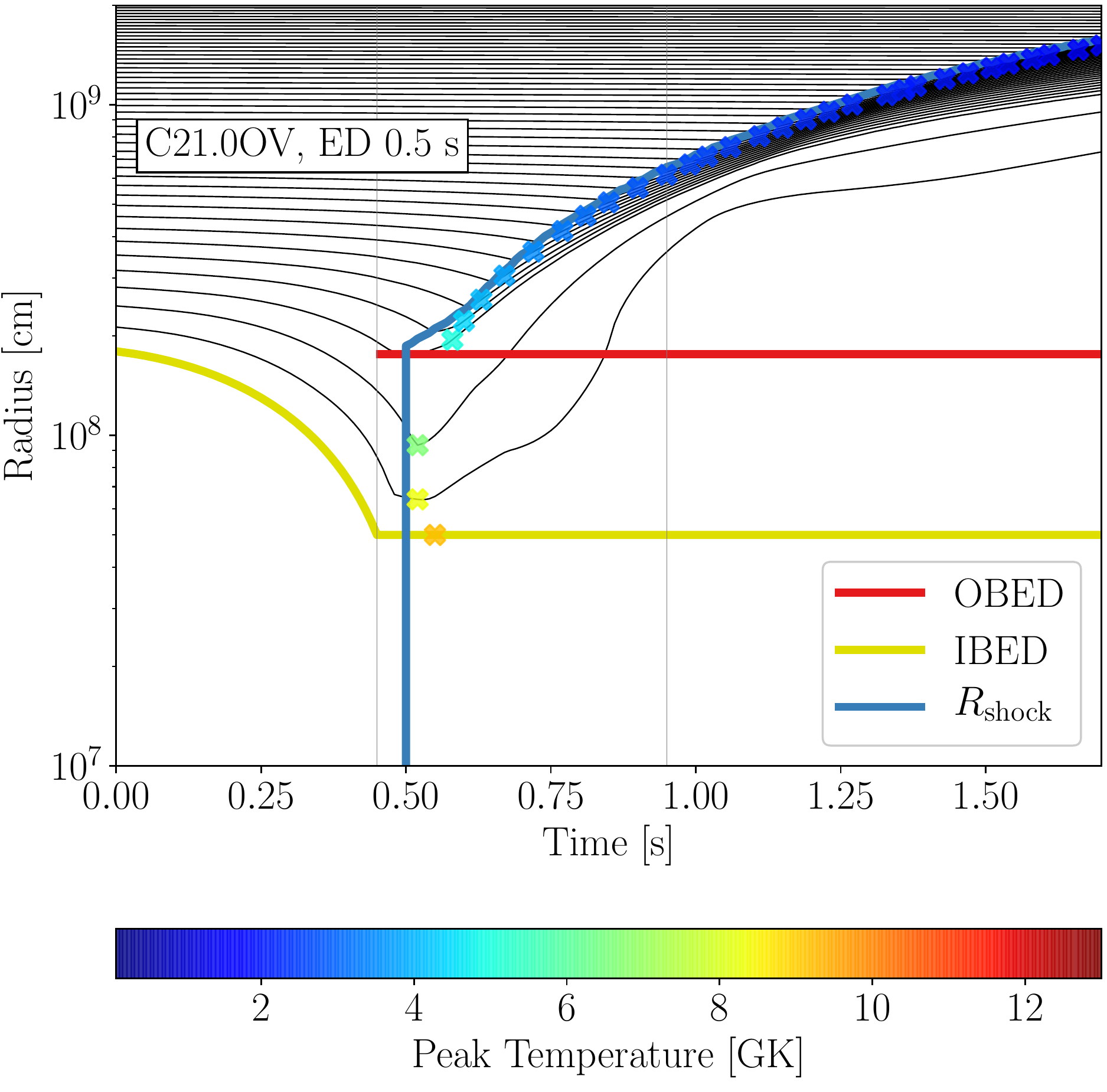}
\includegraphics[width=\columnwidth]{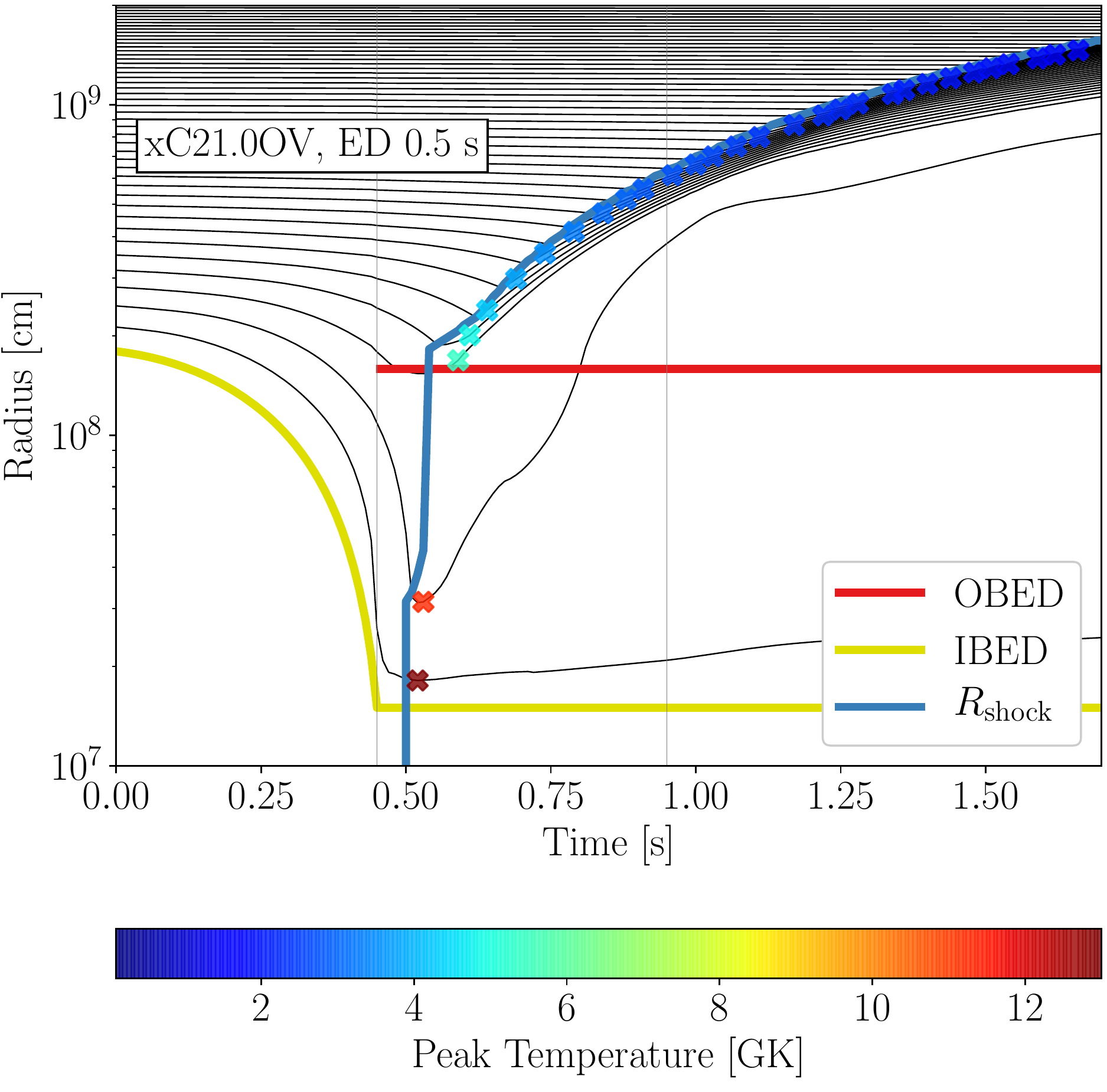}
\caption{Radius evolution of Lagrangian mass shells versus time for CCSN runs of the 21\,$M_\odot$ progenitor with collapse phase and a representative energy-deposition timescale of 0.5\,s; top left: for fixed mass of $\Delta M = 0.05\,M_\odot$ in the energy-deposition layer and collapse to our default value for the minimum radius of $r_\mathrm{min}=500$\,km; top right: for the same fixed mass in the energy-deposition layer but collapse to $r_\mathrm{min}=150$\,km; bottom left: for fixed volume of the energy deposition and collapse to $r_\mathrm{min}=500$\,km; bottom right: for fixed energy-deposition volume and collapse to $r_\mathrm{min}=150$\,km. The thin black solid lines are the mass shells, spaced in steps of 0.025\,$M_\odot$, the blue line marks the shock radius, the yellow line the inner grid boundary, which is also the lower boundary of the energy-deposition layer, and the red line indicates the outer boundary of the energy-deposition layer, either at a fixed mass interval of 0.05\,$M_\odot$ above the inner boundary or at a fixed radius. Crosses indicate the instants when the peak temperature of each mass shell is reached; their colors correspond to temperature values as given by the color bars. Vertical lines mark the beginning and the end of the energy deposition.}
\label{fig:shells_150}
\end{figure*}

\begin{figure*}
\includegraphics[width=\columnwidth]{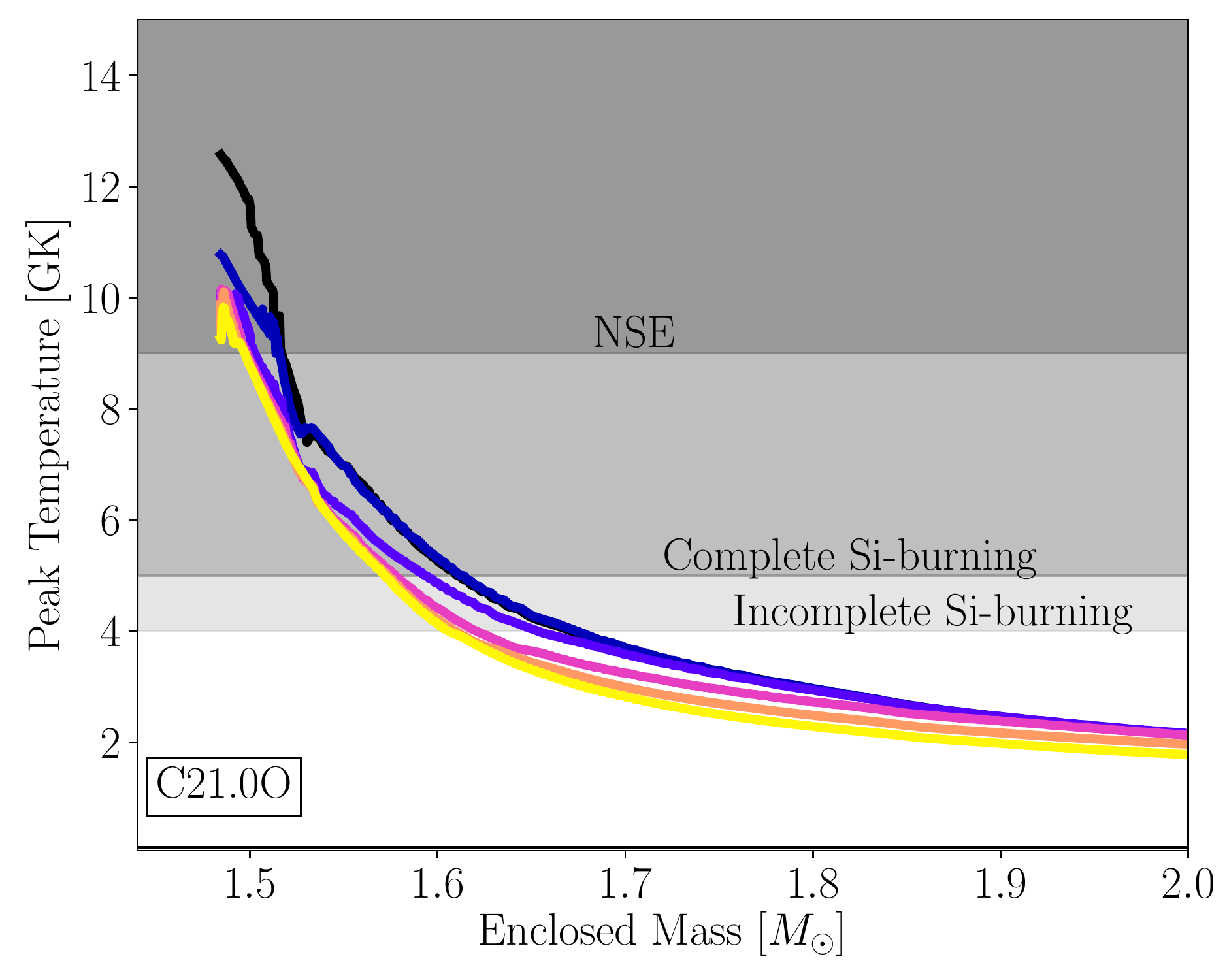}
\includegraphics[width=\columnwidth]{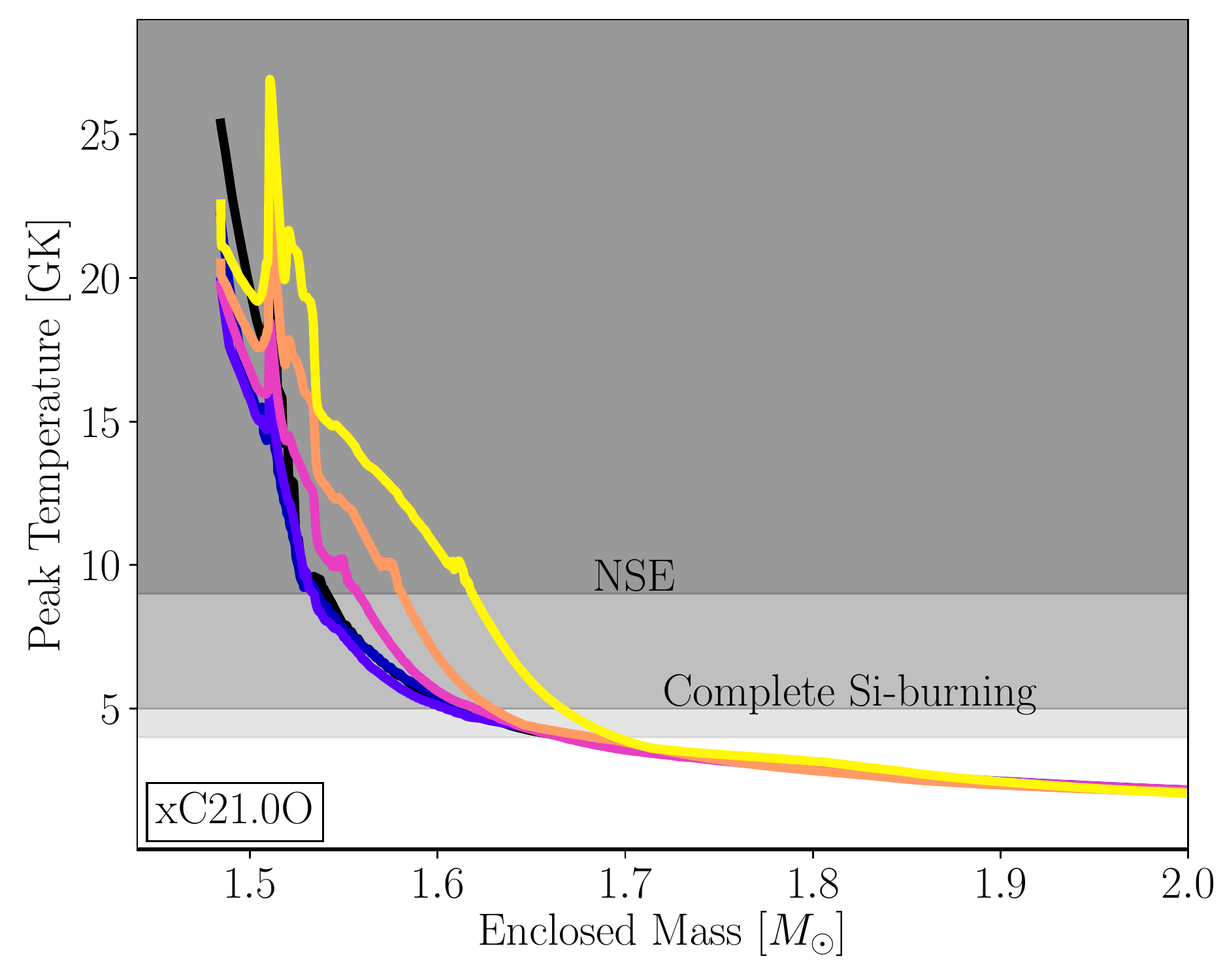}
\includegraphics[width=\columnwidth]{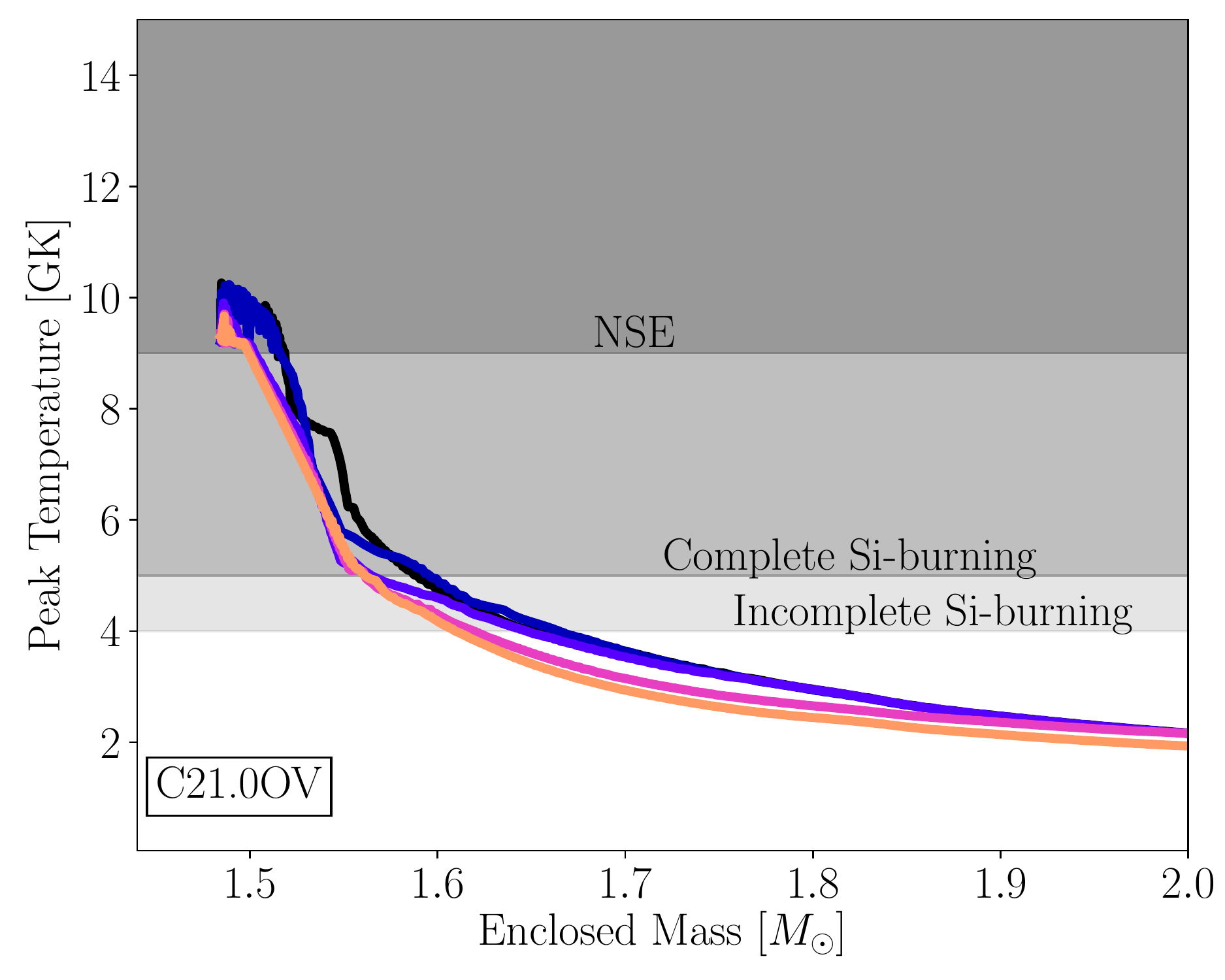}
\includegraphics[width=\columnwidth]{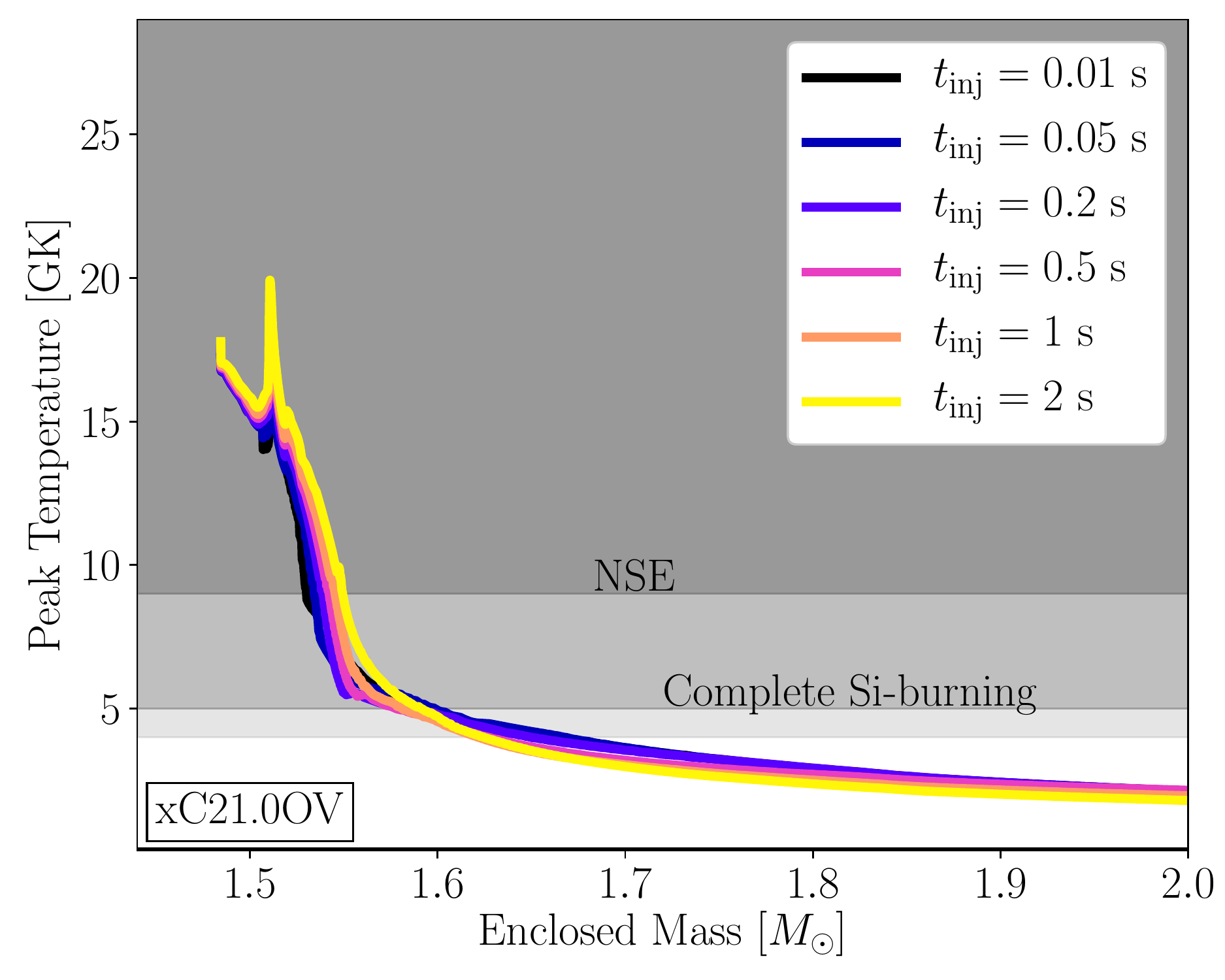}
\caption{Peak temperatures as functions of enclosed mass for the CCSN runs with the 21\,$M_\odot$ progenitor and different energy-injection timescales for the same modelling setups shown in Figure~\ref{fig:shells_150}; top left: for fixed mass of $\Delta M = 0.05\,M_\odot$ in the energy-deposition layer and collapse to our default value for the minimum radius of $r_\mathrm{min}=500$\,km; top right: for the same fixed mass in the energy-deposition layer but collapse to $r_\mathrm{min}=150$\,km; bottom left: for fixed volume of the energy deposition and collapse to $r_\mathrm{min}=500$\,km; bottom right: for fixed energy-deposition volume and collapse to $r_\mathrm{min}=150$\,km. Different intensities of grey shading indicate different regimes of explosive nucleosynthesis as labelled.}
\label{fig:peak_temp_150}
\end{figure*}

\subsection{Fixed volume for energy-injection layer}
\label{sec:fixedvolume}

In another variation of the thermal-bomb modelling we also performed runs with fixed volume
$\Delta V$ for the energy deposition, constrained to simulations including the
collapse phase and applying the O-boundary (models C$M_*$OV in Table~\ref{tab:explosions}). 
These simulations used the same volume for all of the three considered progenitors, and 
correspondingly the initial masses in the energy-injection volume were slightly different
between these progenitors (Table~\ref{tab:dR}). Moreover, these initial mass values were also 
different from the fixed masses $\Delta M$ in the heating layer of the C$M_*$O models (except
for the 26.6\,$M_\odot$ case), which we will compare the C$M_*$OV models to. Although we found
only a modest influence by variations of the fixed mass in the energy-deposition layer in
Section~\ref{sec:massvariations},
we will see that the moderate differences in the initial mass contained by the fixed heated 
volume can cause some subtle relative differences in the behavior of the simulations for 
different progenitor masses.

Our CCSN models with fixed volume for the energy-injection behave, overall, quite similarly
to the models with fixed mass. This holds concerning the $^{56}$Ni yields (left panel of
Figure~\ref{fig:final_ni_150}) as well as the explosion dynamics (left panels of 
Figure~\ref{fig:shells_150}) and the peak-temperature distribution (left panels of
Figure~\ref{fig:peak_temp_150}). However, the computation of the fixed $\Delta V$-models is 
partly more difficult and more time consuming, because the time steps become small when 
the mass in the energy-deposition volume decreases and therefore the entropy per nucleon $s$ 
increases. This implies a growth of the sound speed, because $c_\mathrm{s} \approx 
\sqrt{(4/3)\cdot P/\rho} \propto \sqrt{(4/3)\cdot s\,T}$ for the radiation-dominated conditions
in the heated volume, and therefore it leads to a corresponding reduction of the Courant-Friedrichs-Lewy 
limit for the length of the time steps. For this reason our C$M_*$OV simulations with the longest
energy-deposition timescales could partly not be finished due to their computational demands.
Nevertheless, the available runs are sufficient to draw the essential conclusions.

In Figure~\ref{fig:final_ni_150}, left panel, only minor differences in the $^{56}$Ni production 
are visible between the C$M_*$O models and the C$M_*$OV models. Only the 21.0\,$M_\odot$ 
runs exhibit more sizable differences, i.e., the C21.0OV models eject systematically
lower $^{56}$Ni yields than the C21.0O simulations, especially for short energy-injection times. 
The special role of the C21.0OV models among the CCSN simulations for
the three progenitors is explained by the fact that the initially heated mass in the 21.0\,$M_\odot$
models is the largest of all of the constant-volume models (see Table~\ref{tab:dR}), 
whereas the heated volumes are the same for all cases. This implies that the heating rate
per unit mass is smallest in the C21.0OV models of the 21.0\,$M_\odot$ progenitor.
In addition, the initial mass in 
the heated volume of the C21.0OV models is also larger than the mass in the heating layer
of the C21.0O simulations ($0.068\,M_\odot$ instead of $0.05\,M_\odot$). For this reason
the volume over which the heating is spread is greater in the C21.0OV models, reducing
the heating rate per volume in the innermost ejecta. 

These differences have consequences for the shock strength. The shock in the C21.0OV 
simulations is weaker and the peak temperatures remain lower than in the C21.0O models
(Figure~\ref{fig:peak_temp_150}, left panels), where the heated mass is not only smaller 
but the energy injection also occurs into a fixed mass and thus follows the expanding
gas. In contrast, in the C21.0OV simulations the heated gas expands out of the heated
volume. For long heating timescales the energy injection into a fixed mass or a
fixed volume makes little difference because the gas expands only slowly, allowing the
infall of the preshock gas to proceed for a longer time, leading to higher kinetic 
energies and thus to stronger shock heating. Therefore the
solid and dash-dotted lines in the left panel of Figure~\ref{fig:final_ni_150} approach
each other for all progenitors when the heating timescales are long, consistent with
the observation that the peak temperatures in the left panels of Figure~\ref{fig:peak_temp_150} 
become very similar for the higher values of $t_\mathrm{inj}$. Instead, if the heating 
timescale is short, the heated gas in the 21.0\,$M_\odot$ models with fixed 
energy-deposition volume 
experiences lower heating rates per unit volume and moves out of the heated volume 
rather than receiving continuous energy input as in the C21.0O models, where the heating shifts
outward with the expanding matter. Therefore the shock becomes weaker and the peak 
temperatures in particular of the
innermost ejecta in the C21.0OV simulations with short $t_\mathrm{inj}$ remain lower 
than in the C21.0O models. Since the initially heated mass in the C21.0OV models 
is larger than in the fixed $\Delta V$-simulations for the other progenitors, this temperature
effect and the correspondingly lower $^{56}$Ni production are most pronounced in
the C21.0OV runs. A moderate opposite trend is visible for the C19.7OV models with short
$t_\mathrm{inj}$ because of the smallest value of the initial mass in the fixed heated volume in simulations with the 19\,$M_\odot$ progenitor (Table~~\ref{tab:dR}).

\subsection{Effects of minimum radius for collapse}
\label{sec:minradius}

Finally, we also tested the influence of the minimum radius $r_\mathrm{min}$ in the 
prescription of the initial collapse phase of the C-models by running thermal-bomb
models with $r_\mathrm{min} = 150$\,km, which is close to the radial location 
of the neutrino-heating layer in neutrino-driven explosion models, instead of our 
canonical choice of $r_\mathrm{min} = 500$\,km. For doing these tests we constrained
ourselves to the models with O-boundary for fixed mass layer $\Delta M$ 
(models xC$M_*$O in Table~\ref{tab:explosions}) and fixed volume $\Delta V$
(models xC$M_*$OV in Table~\ref{tab:explosions}) for the energy injection, 
and we will compare them with the 
default-collapse models of C$M_*$O and C$M_*$OV. Here one has to keep in mind that
all C$M_*$O and xC$M_*$O models, for all progenitors, were computed with exactly the 
same fixed mass of $\Delta M = 0.05\,M_\odot$ for the energy-injection layer. The
C$M_*$OV and xC$M_*$OV models for a given progenitor had effectively the same initial
mass (up to the third digit) and nearly the same volume of the heated layer
(Table~\ref{tab:dR}). However, while the heated volume is the same in the CCSN runs
for all progenitors, the initial masses in this volume differ between the three
progenitors (Table~\ref{tab:dR}).

Comparing the left and right panels of Figure~\ref{fig:final_ni_150}, we witness only 
small differences in the $^{56}$Ni production for short heating timescales between the 
xC$M_*$O and the C$M_*$O simulations, and also between the xC$M_*$OV and the C$M_*$OV 
simulations there are only relatively modest differences. The most prominent effect
is a spreading between the $^{56}$Ni yields of the xC21.0O and xC21.0OV models that is about 
twice as big as it is between the C21.0O and C21.0OV cases (right panel of Figure~\ref{fig:final_ni_150}). There is also a slightly greater 
gap between the yields of the xC26.6O and xC26.6OV simulations; this difference is again 
about double the size of that between the C26.6O and C26.6OV models, where it is effectively insignificant. 
The reasons for the somewhat lower production of $^{56}$Ni in the fixed-volume models with 
short energy-injection times were discussed in Section~\ref{sec:fixedvolume}, and they lead
to stronger effects in simulations with more extreme collapse.

For long heating timescales we observe an interesting, new phenomenon in the 
extreme-collapse models that is exactly opposite to the pronounced decrease of the 
$^{56}$Ni yields for longer $t_\mathrm{inj}$ in U-models reported by SM19 and
reproduced by our calculations, and the similar but much weaker trends that one can spot
in most of our C-models, too. Allowing for a deep collapse to $r_\mathrm{min} = 150$\,km 
we obtain increasing $^{56}$Ni yields for longer energy-injection timescales
in particular for the fixed-$\Delta M$ cases, but also, though less drastic, for the 
fixed-$\Delta V$ models (Figure~\ref{fig:final_ni_150}, right panel). (It is possible that a
mild version of this trend is also present in our default-collapse models with
fixed heating volume, but unfortunately the corresponding simulations for long $t_\mathrm{inj}$
could not be finished.) The increase of the $^{56}$Ni production for $t_\mathrm{inj} = 1$\,s and
2\,s reverses the shallow decline that can be seen between $t_\mathrm{inj} = 0.05$\,s and 0.5\,s.

The reason for this new effect can be inferred from the right panels of Figure~\ref{fig:peak_temp_150}.
In stark contrast to all the other model sets plotted in Figure~\ref{fig:peakTtinj} and in the left 
panels of Figure~\ref{fig:peak_temp_150}, the extreme-collapse models with the longest 
energy-injection times tend to reach higher peak temperatures in a wider mass range than 
the corresponding simulations with short $t_\mathrm{inj}$. This effect is particularly 
strong for the xC-models with fixed mass $\Delta M$ of the heating layer (upper right panel 
of Figure~\ref{fig:peak_temp_150} for the CCSN runs with the 21.0\,$M_\odot$ progenitor).
The mass-shell plots of Figure~\ref{fig:shells_150}, right panels compared to the left
panels, provide an explanation of this phenomenon. In the deep collapse cases, the 
matter is much more strongly compression-heated during the infall, and it also expands 
more slowly behind the shock than in the standard C-models. This effect is especially 
relevant when the heating timescales are long, because in such cases the shock accelerates outward less
quickly, thus the gas ahead of the shock has more time to fall deeper into the
gravitational potential of the newly formed neutron star, and when the outward moving shock
sweeps up the infalling matter, the higher gas velocities lead to much stronger shock
heating. 

In the xC21.0OV and xC21.0O models there is an additional effect. 
In the fixed-$\Delta M$ models of the 21.0\,$M_\odot$ progenitor, the energy injection
is initially constrained to a more narrow volume containing 0.05\,$M_\odot$, and it tracks 
the ejected matter. This leads to maximum peak temperatures in the mass shells well 
behind the shock (see upper right panel of Figure~\ref{fig:shells_150}). In contrast, in the fixed-$\Delta V$ models of the same progenitor, the heated volume (initially containing 0.068\,$M_\odot$) is considerably larger than the initial
heating volume in the corresponding fixed-$\Delta M$ models. Therefore the shock expansion reaches
a larger radius within a shorter period of time, preventing the deep infall of the 
preshock material in the xC21.0-cases with fixed $\Delta V$ (compare upper and lower
right panels of Figure~\ref{fig:shells_150}). 
Consequently, the postshock heating is less extreme
in the simulations with fixed energy-injection volume than in the 
models with fixed mass (see the upper and lower right panels of 
Figure~\ref{fig:peak_temp_150}). 

In the extreme-collapse cases with fixed $\Delta V$
the heated volume is somewhat smaller than in the corresponding models with standard 
collapse because of smaller values of $R_\mathrm{IBED}$ and $R_\mathrm{OBED}$ 
(Table~\ref{tab:dR}). Therefore the energy-deposition rate per volume in these xC-models 
is higher than in the C-models, and the innermost ejecta come from regions with stronger 
heating, for which reason also the xC-models with fixed $\Delta V$ exhibit a mild
trend to higher postshock temperatures for long energy-injection timescales. Of course, 
the combined heating effect (compression by infall and shock, plus energy injection) is 
significantly stronger when the heating follows the ejected mass in the xC$M_*$O models, 
for which reason these models show a considerably steeper increase of the $^{56}$Ni 
production with longer $t_\mathrm{inj}$.

In contrast, for short heating timescales the explosion dynamics of models with default 
collapse and extreme collapse are quite similar and the differences in the peak-temperature
distributions are mostly connected to the initially stronger compression heating in the
xC-models. However, in both prescriptions of the collapse phase, similar amounts of mass are
heated to NSE and complete Si-burning temperatures (compare the upper left with the upper
right panel and the lower left with the lower right panel in Figure~\ref{fig:peak_temp_150}).
Therefore the $^{56}$Ni yields for short $t_\mathrm{inj}$ are similar between the C-models 
and the xC-models of each progenitor and both for fixed $\Delta M$ and for fixed $\Delta V$,
except for the effect that we already mentioned above, namely that the $^{56}$Ni production 
in the xC21.0OV
and xC26.6OV models compared to the xC21.0O and xC26.6O models is somewhat more reduced
than in the C21.0OV and C26.6OV models relative to the C21.0O and C26.6O models (see the
left and right panels of Figure~\ref{fig:final_ni_150}).

By default our $^{56}$Ni yields include nickel produced in the energy-deposition layer 
(see Section~\ref{sec:compSM19models}). In principle, one has to consider that some of this
innermost matter may be unable to achieve escape conditions and thus may stay gravitationally
bound, thus not contributing to the CCSN ejecta. From our model sets this issue affects 
especially the extreme-collapse cases with fixed volume for the energy injection, where the
heated gas resides deep in the gravitational potential of the newly formed neutron star
and the energy deposition does not follow the outward moving matter. Among these 
xC-models mainly the 21.0\,$M_\odot$ simulations are concerned, since the initial mass 
in the heated volume of these models is largest (see Table~\ref{tab:dR}). One can
see this in the lower right panel of Figure~\ref{fig:shells_150}, because the innermost 
displayed mass shell exterior to $R_\mathrm{IBED}$ expands only very slowly there. The 
radial velocities of this shell over 30\,s in the xC21.0OV simulations are only around
100\,km\,s$^{-1}$ and therefore considerably lower than the escape velocity, which is 
on the order of 1000\,km\,s$^{-1}$ at a radius
of some 1000\,km. Consequently, this matter might not become unbound despite its 
continuous, slow expansion until the end of our simulations. Subtracting the $^{56}$Ni 
contained in this innermost material would somewhat reduce the nickel production, but 
such a correction would not mean a dominant effect for the xC21.0OV models. Nevertheless, 
it might damp the increase of the $^{56}$Ni yields in these model runs for long
energy-injection times seen in Figure~\ref{fig:final_ni_150}, right panel.


\section{Summary and discussion}
\label{section:conclusions}

The thermal bomb method is a widely used modelling approach to trigger CCSN explosions
artificially by releasing energy into a chosen mass layer or chosen volume around a 
chosen location of the (initial, i.e. before fallback) mass cut, which usually 
coincides with the inner boundary of the computational grid. In the present paper
we explored various dependencies of the thermal-bomb parameterization, in particular 
we considered models with and without an initial collapse phase, different timescales
for the energy release, different radial positions of the mass cut, energy deposition
in a fixed mass layer or fixed volume, different masses for this layer, and different
minimum radii for the contraction during the collapse phase. For this purpose
we performed 1D CCSN simulations with the thermal-bomb method, using the 
\textsc{Prometheus-HOTB} code, and we post-processed the ejecta for nucleosynthesis
with the SkyNet open-source network. We focused here on the production of $^{56}$Ni
because of its pivotal importance for observational SN diagnostics. Moreover,
the production of this dominant radioactive isotope can be considered as 
representative of the total output in iron-group and intermediate-mass nuclei
without entering the discussion of yields of other isotopes, whose 
relative amounts are highly sensitive to the exact distribution of $Y_e$ in the 
ejecta.

Our work was motivated by the recent finding of SM19, deduced from thermal-bomb 
simulations for three progenitors with different masses, that the production of
$^{56,57}$Ni and $^{44}$Ti decreases dramatically for energy-injection timescales
longer than about 100\,ms. SM19 concluded that the production of these nuclear 
species and other elements is best compatible with observational constraints 
for nearly instantaneous explosions, i.e.,
for energy-release timescales of the thermal bomb as short as $\lesssim$50\,ms.
If correct, this result would be a strong argument against the neutrino-driven 
explosion mechanism for CCSNe, because self-consistent ab initio simulations show
that this mechanism provides the energy of the explosion only over timescales of 
seconds \citep[see, e.g.,][]{2021ApJ...915...28B}. 

In our simulations, mainly considering 19.7, 21.0, and 26.6\,$M_\odot$ progenitors
with significantly different pre-collapse structures, we confirmed the results 
obtained by SM19, namely a strong anti-correlation between $^{56}$Ni yields and 
energy-injection timescale. However, we obtained these results only when the 
thermal bomb was assumed to release its energy in the uncollapsed progenitor
models. Including an initial collapse phase, which is the more realistic approach
when stellar core collapse, neutron star formation, and CCSN explosions 
are supposed to be simulated, the trend
witnessed by SM19 effectively disappears and the $^{56}$Ni production becomes
almost independent of the timescale for the energy release. Allowing for an 
initial collapse to a minimal radius of 150\,km instead of our default value 
of 500\,km, thus more closely adopting conditions similar to those in 
neutrino-driven explosions, we even obtained a reversal of the trend seen in 
uncollapsed models. In such calculations with the more extreme collapse, we 
found that long energy-injection timescales, especially when longer 
than $\sim$1\,s, lead to a higher production of $^{56}$Ni than the shorter 
energy-deposition times, which trigger more rapid explosions. 

Therefore there is no 
reason to conclude on grounds of thermal-bomb simulations that the $^{56}$Ni 
production in slow explosions as expected for the neutrino-driven mechanism
is in conflict with observational data. The result reported by SM19 for their
thermal-bomb explosions of uncollapsed progenitor models was caused by the 
energy injection into the low-density, hydrostatic stellar profiles, which 
permits easy expansion of the ejecta with corresponding expansion-cooling 
as soon as the energy release is switched on. Therefore only small amounts 
of matter close to the heated mass shell (i.e., the defined mass cut) can 
reach temperatures that are sufficiently high for NSE and Si-burning. The 
conditions for such temperatures are strongly disfavored for longer 
energy-injection timescales. In contrast, when an
initial collapse phase is included in the thermal-bomb modelling, the 
energy deposition occurs in infalling matter, which expands much less
readily, because the SN shock wave needs to propagate outward against the
ram pressure of infalling stellar layers. In this case it has to receive 
more energy input for a predefined value of the final explosion energy, 
and the correspondingly stronger explosion shock can heat more mass to 
NSE and Si-burning conditions. 

Varying the different inputs for the parametric description of the thermal 
bombs for a fixed value of the explosion energy, we found that the most sensitive 
aspects for the production of $^{56}$Ni are the inclusion of the initial
collapse instead of releasing the energy into the uncollapsed progenitor, and
the location of the initial mass cut at the radius where the entropy per nucleon reaches
$s/k_\mathrm{B} = 4$ instead of the position where $Y_e = 0.48$. There is only
a relatively modest influence of the exact value of the fixed mass $\Delta M$ 
in the energy-deposition layer. Also the choice of a fixed volume for the 
energy release instead of a fixed mass causes only secondary differences.
Once the initial collapse is included, also the timescale of the energy release
by the thermal bomb leads to variations only on a secondary level. For the 
more realistic choice of the initial mass cut at $s/k_\mathrm{B} = 4$, which can
be better motivated by neutrino-driven explosion models, it is crucial to also
include matter in the heated layer in the ejecta, if this matter becomes unbound
during the explosion.

Because of their numerous degrees of freedom, thermal-bomb models can certainly
not be employed to assess the viability of any kind of physical explosion 
mechanism. For example, artificial explosion methods like the thermal bombs can 
hardly be expected to reproduce the dynamics of neutrino-driven explosions in 
a physically correct and reliable way. In particular, fixing the mass layer for 
the energy injection means that the energy input follows the expanding matter, 
which is unrealistic. Fixing instead the volume for the energy release either 
overestimates the heated volume or underestimates the heated mass in this 
heated volume, where in addition the mass decreases with time, which again is 
not a realistic description of the neutrino-driven mechanism. Fortunately, the 
$^{56}$Ni production of thermal bomb simulations that include a collapse 
phase turned out not to be overly sensitive to such alternative choices.

Thermal bombs are a numerical recipe that depends on a variety of parameterized 
inputs that need to be defined. Nevertheless, even with the best choice of these 
inputs, their usefulness for quantitative predictions of iron-group and 
intermediate-mass-element nucleosynthesis will always be hampered by 
the unknown value of the explosion energy and, in principle, also of the initial 
mass cut. Moreover, iron-group species such as the isotopes of $^{56,57}$Ni and
of $^{44}$Ti are formed in ejecta whose $Y_e$ evolves due to weak-force 
interactions of neutrinos and where multi-dimensional flows play a crucial role. 
None of these are taken into account in a simple 
thermal-bomb treatment. Therefore the best one can 
expect of any artificial explosion trigger is that the method 
should be set up such that it does not massively overproduce or underproduce 
nickel and it should also be set up such that the correct trends of the 
$^{56}$Ni production with explosion energy, explosion time scale, and progenitor 
structure can be maintained.

Since thermal bombs provide an easy-to-apply recipe to trigger explosions, it is 
very likely that they will remain in use as a method of choice for the 
exploration of CCSN nucleosynthesis, for example in large sets of progenitor 
models, despite all the mentioned caveats \citep[e.g.][]{Farmer+2021}. 
In view of the results of our study, we recommend the following prescriptions:
\begin{enumerate}
    \item Include a collapse phase before the energy release of the thermal 
    bomb is started. A minimum collapse radius near 500\,km seems to be 
    sufficient and is computationally less demanding than a smaller radius.
    \item Since self-consistent simulations of neutrino-driven CCSNe show that
    the explosion sets in when the infalling Si/O interface reaches the stagnant
    bounce shock, the initial mass cut should be chosen near the 
    $s/k_\mathrm{B} = 4$ location instead of putting it close to the edge 
    of the iron core. Therefore $Y_e$ in the layer of energy injection by 
    the thermal bomb is very close to 0.5 (typically higher than 0.497).
    \item For this reason $^{56}$Ni will be efficiently produced in the 
    energy-injection layer and the matter in this layer should be included in
    the ejecta, if it becomes gravitationally unbound by the explosion.
    \item Using a fixed mass layer $\Delta M$ for the energy injection is 
    numerically easier than a fixed volume, and both choices do not cause 
    any major differences. The exact value of $\Delta M$ is not crucial. 
    We suggest $0.05\,M_\odot$, but smaller masses lead to very similar 
    nickel yields.
    \item With the recommended setup the $^{56}$Ni production is basically 
    insensitive to the timescale chosen for the energy injection by the thermal
    bomb.
\end{enumerate}
Of course, these recommendations are based on a small set of simulations for 
only three progenitors and a defined explosion energy of $10^{51}$\,erg 
in all of our thermal-bomb calculations. A wider exploration is desirable
to test the more general reliability of our proposed parameter settings.  
Beyond the prescriptions listed above, the value of the explosion 
energy is another crucial input into the thermal-bomb modelling. Its specification
has to be guided by our first-principle understanding of the physics of the 
CCSN mechanism in stars of different masses. In future work we plan to 
compare thermal-bomb models and direct simulations of neutrino-driven CCSN 
explosions with respect to the progenitor and explosion energy dependent 
production of $^{56}$Ni and other iron-group and intermediate-mass elements.

\section*{Acknowledgements}
We are grateful to Thomas Ertl for his assistance in the starting phase of the project and
thank Ewald M\"uller and Johannes Ringler for discussions.
Support by the Deutsche Forschungsgemeinschaft (DFG, German
Research Foundation) through Sonderforschungsbereich (Collaborative Research
Center) SFB-1258 ``Neutrinos and Dark Matter in Astro- and Particle Physics (NDM)'' and under Germany's Excellence Strategy through Cluster of Excellence ORIGINS (EXC-2094)-390783311 is acknowledged.


\section*{Data availability}

The data of our calculations will be made available upon reasonable request.

\section*{Software}
\textsc{Prometheus-HOTB}  \citep{1996A&A...306..167J,2003A&A...408..621K,2006A&A...457..963S,2007A&A...467.1227A,2012ApJ...757...69U,2016ApJ...818..124E};
KEPLER \citep{1978ApJ...225.1021W}; SkyNet \citep{2017ApJS..233...18L}; Matplotlib \citep{2007CSE.....9...90H}; NumPy \citep{2011CSE....13b..22V}.



\bibliographystyle{mnras}
\bibliography{imasheva_references}

\begin{thebibliography}{}
\makeatletter
\relax
\def\mn@urlcharsother{\let\do\@makeother \do\$\do\&\do\#\do\^\do\_\do\%\do\~}
\def\mn@doi{\begingroup\mn@urlcharsother \@ifnextchar [ {\mn@doi@}
  {\mn@doi@[]}}
\def\mn@doi@[#1]#2{\def\@tempa{#1}\ifx\@tempa\@empty \href
  {http://dx.doi.org/#2} {doi:#2}\else \href {http://dx.doi.org/#2} {#1}\fi
  \endgroup}
\def\mn@eprint#1#2{\mn@eprint@#1:#2::\@nil}
\def\mn@eprint@arXiv#1{\href {http://arxiv.org/abs/#1} {{\tt arXiv:#1}}}
\def\mn@eprint@dblp#1{\href {http://dblp.uni-trier.de/rec/bibtex/#1.xml}
  {dblp:#1}}
\def\mn@eprint@#1:#2:#3:#4\@nil{\def\@tempa {#1}\def\@tempb {#2}\def\@tempc
  {#3}\ifx \@tempc \@empty \let \@tempc \@tempb \let \@tempb \@tempa \fi \ifx
  \@tempb \@empty \def\@tempb {arXiv}\fi \@ifundefined
  {mn@eprint@\@tempb}{\@tempb:\@tempc}{\expandafter \expandafter \csname
  mn@eprint@\@tempb\endcsname \expandafter{\@tempc}}}

\bibitem[\protect\citeauthoryear{{Abbott} et~al.,}{{Abbott}
  et~al.}{2021}]{2021ApJ...913L...7A}
{Abbott} R.,  et~al., 2021, \mn@doi [\apjl] {10.3847/2041-8213/abe949}, \href
  {https://ui.adsabs.harvard.edu/abs/2021ApJ...913L...7A} {913, L7}

\bibitem[\protect\citeauthoryear{{Aguilera-Dena}, {M{\"u}ller}, {Antoniadis},
  {Langer}, {Dessart}, {Vigna-G{\'o}mez}  \& {Yoon}}{{Aguilera-Dena}
  et~al.}{2022}]{2022arXiv220400025A}
{Aguilera-Dena} D.~R.,  {M{\"u}ller} B.,  {Antoniadis} J.,  {Langer} N.,
  {Dessart} L.,  {Vigna-G{\'o}mez} A.,   {Yoon} S.-C.,  2022, arXiv e-prints,
  \href {https://ui.adsabs.harvard.edu/abs/2022arXiv220400025A} {p.
  arXiv:2204.00025}

\bibitem[\protect\citeauthoryear{{Aloy}, {M{\"u}ller}, {Ib{\'a}{\~n}ez},
  {Mart{\'\i}}  \& {MacFadyen}}{{Aloy} et~al.}{2000}]{2000ApJ...531L.119A}
{Aloy} M.~A.,  {M{\"u}ller} E.,  {Ib{\'a}{\~n}ez} J.~M.,  {Mart{\'\i}} J.~M.,
  {MacFadyen} A.,  2000, \mn@doi [\apjl] {10.1086/312537}, \href
  {https://ui.adsabs.harvard.edu/abs/2000ApJ...531L.119A} {531, L119}

\bibitem[\protect\citeauthoryear{{Arcones}, {Janka}  \& {Scheck}}{{Arcones}
  et~al.}{2007}]{2007A&A...467.1227A}
{Arcones} A.,  {Janka} H.~T.,   {Scheck} L.,  2007, \mn@doi [\aap]
  {10.1051/0004-6361:20066983}, \href
  {https://ui.adsabs.harvard.edu/abs/2007A&A...467.1227A} {467, 1227}

\bibitem[\protect\citeauthoryear{{Arnett}, {Bahcall}, {Kirshner}  \&
  {Woosley}}{{Arnett} et~al.}{1989}]{1989ARA&A..27..629A}
{Arnett} W.~D.,  {Bahcall} J.~N.,  {Kirshner} R.~P.,   {Woosley} S.~E.,  1989,
  \mn@doi [\araa] {10.1146/annurev.aa.27.090189.003213}, \href
  {https://ui.adsabs.harvard.edu/abs/1989ARA&A..27..629A} {27, 629}

\bibitem[\protect\citeauthoryear{{Aufderheide}, {Baron}  \&
  {Thielemann}}{{Aufderheide} et~al.}{1991}]{1991ApJ...370..630A}
{Aufderheide} M.~B.,  {Baron} E.,   {Thielemann} F.~K.,  1991, \mn@doi [\apj]
  {10.1086/169849}, \href
  {https://ui.adsabs.harvard.edu/abs/1991ApJ...370..630A} {370, 630}

\bibitem[\protect\citeauthoryear{{Barker}, {Harris}, {Warren}, {O'Connor}  \&
  {Couch}}{{Barker} et~al.}{2022}]{2022ApJ...934...67B}
{Barker} B.~L.,  {Harris} C.~E.,  {Warren} M.~L.,  {O'Connor} E.~P.,   {Couch}
  S.~M.,  2022, \mn@doi [\apj] {10.3847/1538-4357/ac77f3}, \href
  {https://ui.adsabs.harvard.edu/abs/2022ApJ...934...67B} {934, 67}

\bibitem[\protect\citeauthoryear{{Bersten}, {Benvenuto}  \& {Hamuy}}{{Bersten}
  et~al.}{2011}]{2011ApJ...729...61B}
{Bersten} M.~C.,  {Benvenuto} O.,   {Hamuy} M.,  2011, \mn@doi [\apj]
  {10.1088/0004-637X/729/1/61}, \href
  {https://ui.adsabs.harvard.edu/abs/2011ApJ...729...61B} {729, 61}

\bibitem[\protect\citeauthoryear{{Bollig}, {Yadav}, {Kresse}, {Janka},
  {M{\"u}ller}  \& {Heger}}{{Bollig} et~al.}{2021}]{2021ApJ...915...28B}
{Bollig} R.,  {Yadav} N.,  {Kresse} D.,  {Janka} H.-T.,  {M{\"u}ller} B.,
  {Heger} A.,  2021, \mn@doi [\apj] {10.3847/1538-4357/abf82e}, \href
  {https://ui.adsabs.harvard.edu/abs/2021ApJ...915...28B} {915, 28}

\bibitem[\protect\citeauthoryear{{Bruenn} et~al.,}{{Bruenn}
  et~al.}{2016}]{2016ApJ...818..123B}
{Bruenn} S.~W.,  et~al., 2016, \mn@doi [\apj] {10.3847/0004-637X/818/2/123},
  \href {https://ui.adsabs.harvard.edu/abs/2016ApJ...818..123B} {818, 123}

\bibitem[\protect\citeauthoryear{{Burrows} \& {Vartanyan}}{{Burrows} \&
  {Vartanyan}}{2021}]{2021Natur.589...29B}
{Burrows} A.,  {Vartanyan} D.,  2021, \mn@doi [\nat]
  {10.1038/s41586-020-03059-w}, \href
  {https://ui.adsabs.harvard.edu/abs/2021Natur.589...29B} {589, 29}

\bibitem[\protect\citeauthoryear{{Chieffi} \& {Limongi}}{{Chieffi} \&
  {Limongi}}{2013}]{2013ApJ...764...21C}
{Chieffi} A.,  {Limongi} M.,  2013, \mn@doi [\apj]
  {10.1088/0004-637X/764/1/21}, \href
  {https://ui.adsabs.harvard.edu/abs/2013ApJ...764...21C} {764, 21}

\bibitem[\protect\citeauthoryear{{Couch}}{{Couch}}{2017}]{2017RSPTA.37560271C}
{Couch} S.~M.,  2017, \mn@doi [Philosophical Transactions of the Royal Society
  of London Series A] {10.1098/rsta.2016.0271}, \href
  {https://ui.adsabs.harvard.edu/abs/2017RSPTA.37560271C} {375, 20160271}

\bibitem[\protect\citeauthoryear{{Couch}, {Warren}  \& {O'Connor}}{{Couch}
  et~al.}{2020}]{2020ApJ...890..127C}
{Couch} S.~M.,  {Warren} M.~L.,   {O'Connor} E.~P.,  2020, \mn@doi [\apj]
  {10.3847/1538-4357/ab609e}, \href
  {https://ui.adsabs.harvard.edu/abs/2020ApJ...890..127C} {890, 127}

\bibitem[\protect\citeauthoryear{{Cowan}, {Sneden}, {Lawler}, {Aprahamian},
  {Wiescher}, {Langanke}, {Mart{\'\i}nez-Pinedo}  \& {Thielemann}}{{Cowan}
  et~al.}{2021}]{2021RvMP...93a5002C}
{Cowan} J.~J.,  {Sneden} C.,  {Lawler} J.~E.,  {Aprahamian} A.,  {Wiescher} M.,
   {Langanke} K.,  {Mart{\'\i}nez-Pinedo} G.,   {Thielemann} F.-K.,  2021,
  \mn@doi [Reviews of Modern Physics] {10.1103/RevModPhys.93.015002}, \href
  {https://ui.adsabs.harvard.edu/abs/2021RvMP...93a5002C} {93, 015002}

\bibitem[\protect\citeauthoryear{{Curtis}, {Ebinger}, {Fr{\"o}hlich}, {Hempel},
  {Perego}, {Liebend{\"o}rfer}  \& {Thielemann}}{{Curtis}
  et~al.}{2019}]{2019ApJ...870....2C}
{Curtis} S.,  {Ebinger} K.,  {Fr{\"o}hlich} C.,  {Hempel} M.,  {Perego} A.,
  {Liebend{\"o}rfer} M.,   {Thielemann} F.-K.,  2019, \mn@doi [\apj]
  {10.3847/1538-4357/aae7d2}, \href
  {https://ui.adsabs.harvard.edu/abs/2019ApJ...870....2C} {870, 2}

\bibitem[\protect\citeauthoryear{{Curtis}, {Wolfe}, {Fr{\"o}hlich}, {Miller},
  {Wollaeger}  \& {Ebinger}}{{Curtis} et~al.}{2021}]{2021ApJ...921..143C}
{Curtis} S.,  {Wolfe} N.,  {Fr{\"o}hlich} C.,  {Miller} J.~M.,  {Wollaeger} R.,
    {Ebinger} K.,  2021, \mn@doi [\apj] {10.3847/1538-4357/ac0dc5}, \href
  {https://ui.adsabs.harvard.edu/abs/2021ApJ...921..143C} {921, 143}

\bibitem[\protect\citeauthoryear{{Dessart}, {Hillier}, {Sukhbold}, {Woosley}
  \& {Janka}}{{Dessart} et~al.}{2021a}]{2021A&A...652A..64D}
{Dessart} L.,  {Hillier} D.~J.,  {Sukhbold} T.,  {Woosley} S.~E.,   {Janka}
  H.~T.,  2021a, \mn@doi [\aap] {10.1051/0004-6361/202140839}, \href
  {https://ui.adsabs.harvard.edu/abs/2021A&A...652A..64D} {652, A64}

\bibitem[\protect\citeauthoryear{{Dessart}, {Hillier}, {Sukhbold}, {Woosley}
  \& {Janka}}{{Dessart} et~al.}{2021b}]{2021A&A...656A..61D}
{Dessart} L.,  {Hillier} D.~J.,  {Sukhbold} T.,  {Woosley} S.~E.,   {Janka}
  H.~T.,  2021b, \mn@doi [\aap] {10.1051/0004-6361/202141927}, \href
  {https://ui.adsabs.harvard.edu/abs/2021A&A...656A..61D} {656, A61}

\bibitem[\protect\citeauthoryear{{Diehl} et~al.,}{{Diehl}
  et~al.}{2021}]{2021PASA...38...62D}
{Diehl} R.,  et~al., 2021, \mn@doi [\pasa] {10.1017/pasa.2021.48}, \href
  {https://ui.adsabs.harvard.edu/abs/2021PASA...38...62D} {38, e062}

\bibitem[\protect\citeauthoryear{{Ebinger}, {Curtis}, {Fr{\"o}hlich}, {Hempel},
  {Perego}, {Liebend{\"o}rfer}  \& {Thielemann}}{{Ebinger}
  et~al.}{2019}]{2019ApJ...870....1E}
{Ebinger} K.,  {Curtis} S.,  {Fr{\"o}hlich} C.,  {Hempel} M.,  {Perego} A.,
  {Liebend{\"o}rfer} M.,   {Thielemann} F.-K.,  2019, \mn@doi [\apj]
  {10.3847/1538-4357/aae7c9}, \href
  {https://ui.adsabs.harvard.edu/abs/2019ApJ...870....1E} {870, 1}

\bibitem[\protect\citeauthoryear{{Ebinger}, {Curtis}, {Ghosh}, {Fr{\"o}hlich},
  {Hempel}, {Perego}, {Liebend{\"o}rfer}  \& {Thielemann}}{{Ebinger}
  et~al.}{2020}]{2020ApJ...888...91E}
{Ebinger} K.,  {Curtis} S.,  {Ghosh} S.,  {Fr{\"o}hlich} C.,  {Hempel} M.,
  {Perego} A.,  {Liebend{\"o}rfer} M.,   {Thielemann} F.-K.,  2020, \mn@doi
  [\apj] {10.3847/1538-4357/ab5dcb}, \href
  {https://ui.adsabs.harvard.edu/abs/2020ApJ...888...91E} {888, 91}

\bibitem[\protect\citeauthoryear{{Ertl}, {Janka}, {Woosley}, {Sukhbold}  \&
  {Ugliano}}{{Ertl} et~al.}{2016}]{2016ApJ...818..124E}
{Ertl} T.,  {Janka} H.~T.,  {Woosley} S.~E.,  {Sukhbold} T.,   {Ugliano} M.,
  2016, \mn@doi [\apj] {10.3847/0004-637X/818/2/124}, \href
  {https://ui.adsabs.harvard.edu/abs/2016ApJ...818..124E} {818, 124}

\bibitem[\protect\citeauthoryear{{Ertl}, {Woosley}, {Sukhbold}  \&
  {Janka}}{{Ertl} et~al.}{2020}]{2020ApJ...890...51E}
{Ertl} T.,  {Woosley} S.~E.,  {Sukhbold} T.,   {Janka} H.~T.,  2020, \mn@doi
  [\apj] {10.3847/1538-4357/ab6458}, \href
  {https://ui.adsabs.harvard.edu/abs/2020ApJ...890...51E} {890, 51}

\bibitem[\protect\citeauthoryear{{Farmer}, {Laplace}, {de Mink}  \&
  {Justham}}{{Farmer} et~al.}{2021}]{Farmer+2021}
{Farmer} R.,  {Laplace} E.,  {de Mink} S.~E.,   {Justham} S.,  2021, \mn@doi
  [\apj] {10.3847/1538-4357/ac2f44}, \href
  {https://ui.adsabs.harvard.edu/abs/2021ApJ...923..214F} {923, 214}

\bibitem[\protect\citeauthoryear{{Hashimoto}, {Nomoto}  \&
  {Shigeyama}}{{Hashimoto} et~al.}{1989}]{1989A&A...210L...5H}
{Hashimoto} M.,  {Nomoto} K.,   {Shigeyama} T.,  1989, \aap, \href
  {https://ui.adsabs.harvard.edu/abs/1989A&A...210L...5H} {210, L5}

\bibitem[\protect\citeauthoryear{{Hayden} et~al.,}{{Hayden}
  et~al.}{2015}]{2015ApJ...808..132H}
{Hayden} M.~R.,  et~al., 2015, \mn@doi [\apj] {10.1088/0004-637X/808/2/132},
  \href {https://ui.adsabs.harvard.edu/abs/2015ApJ...808..132H} {808, 132}

\bibitem[\protect\citeauthoryear{{Hix} et~al.,}{{Hix}
  et~al.}{2014}]{2014AIPA....4d1013H}
{Hix} W.~R.,  et~al., 2014, \mn@doi [AIP Advances] {10.1063/1.4870009}, \href
  {https://ui.adsabs.harvard.edu/abs/2014AIPA....4d1013H} {4, 041013}

\bibitem[\protect\citeauthoryear{{Hunter}}{{Hunter}}{2007}]{2007CSE.....9...90H}
{Hunter} J.~D.,  2007, \mn@doi [Computing in Science and Engineering]
  {10.1109/MCSE.2007.55}, \href
  {https://ui.adsabs.harvard.edu/abs/2007CSE.....9...90H} {9, 90}

\bibitem[\protect\citeauthoryear{{Iwamoto}, {Nomoto}, {H{\"o}flich}, {Yamaoka},
  {Kumagai}  \& {Shigeyama}}{{Iwamoto} et~al.}{1994}]{1994ApJ...437L.115I}
{Iwamoto} K.,  {Nomoto} K.,  {H{\"o}flich} P.,  {Yamaoka} H.,  {Kumagai} S.,
  {Shigeyama} T.,  1994, \mn@doi [\apjl] {10.1086/187696}, \href
  {https://ui.adsabs.harvard.edu/abs/1994ApJ...437L.115I} {437, L115}

\bibitem[\protect\citeauthoryear{{Janka}}{{Janka}}{2012}]{2012ARNPS..62..407J}
{Janka} H.-T.,  2012, \mn@doi [Annual Review of Nuclear and Particle Science]
  {10.1146/annurev-nucl-102711-094901}, \href
  {https://ui.adsabs.harvard.edu/abs/2012ARNPS..62..407J} {62, 407}

\bibitem[\protect\citeauthoryear{{Janka} \& {M{\"u}ller}}{{Janka} \&
  {M{\"u}ller}}{1996}]{1996A&A...306..167J}
{Janka} H.~T.,  {M{\"u}ller} E.,  1996, \aap, \href
  {https://ui.adsabs.harvard.edu/abs/1996A&A...306..167J} {306, 167}

\bibitem[\protect\citeauthoryear{{Janka}, {Langanke}, {Marek},
  {Mart{\'\i}nez-Pinedo}  \& {M{\"u}ller}}{{Janka}
  et~al.}{2007}]{2007PhR...442...38J}
{Janka} H.~T.,  {Langanke} K.,  {Marek} A.,  {Mart{\'\i}nez-Pinedo} G.,
  {M{\"u}ller} B.,  2007, \mn@doi [\physrep] {10.1016/j.physrep.2007.02.002},
  \href {https://ui.adsabs.harvard.edu/abs/2007PhR...442...38J} {442, 38}

\bibitem[\protect\citeauthoryear{{Janka}, {Melson}  \& {Summa}}{{Janka}
  et~al.}{2016}]{2016ARNPS..66..341J}
{Janka} H.-T.,  {Melson} T.,   {Summa} A.,  2016, \mn@doi [Annual Review of
  Nuclear and Particle Science] {10.1146/annurev-nucl-102115-044747}, \href
  {https://ui.adsabs.harvard.edu/abs/2016ARNPS..66..341J} {66, 341}

\bibitem[\protect\citeauthoryear{{Jerkstrand} et~al.,}{{Jerkstrand}
  et~al.}{2015}]{2015ApJ...807..110J}
{Jerkstrand} A.,  et~al., 2015, \mn@doi [\apj] {10.1088/0004-637X/807/1/110},
  \href {https://ui.adsabs.harvard.edu/abs/2015ApJ...807..110J} {807, 110}

\bibitem[\protect\citeauthoryear{{Khokhlov}, {H{\"o}flich}, {Oran}, {Wheeler},
  {Wang}  \& {Chtchelkanova}}{{Khokhlov} et~al.}{1999}]{1999ApJ...524L.107K}
{Khokhlov} A.~M.,  {H{\"o}flich} P.~A.,  {Oran} E.~S.,  {Wheeler} J.~C.,
  {Wang} L.,   {Chtchelkanova} A.~Y.,  1999, \mn@doi [\apjl] {10.1086/312305},
  \href {https://ui.adsabs.harvard.edu/abs/1999ApJ...524L.107K} {524, L107}

\bibitem[\protect\citeauthoryear{{Kifonidis}, {Plewa}, {Janka}  \&
  {M{\"u}ller}}{{Kifonidis} et~al.}{2000}]{2000ApJ...531L.123K}
{Kifonidis} K.,  {Plewa} T.,  {Janka} H.~T.,   {M{\"u}ller} E.,  2000, \mn@doi
  [\apjl] {10.1086/312541}, \href
  {https://ui.adsabs.harvard.edu/abs/2000ApJ...531L.123K} {531, L123}

\bibitem[\protect\citeauthoryear{{Kifonidis}, {Plewa}  \&
  {M{\"u}ller}}{{Kifonidis} et~al.}{2001}]{2001AIPC..561...21K}
{Kifonidis} K.,  {Plewa} T.,   {M{\"u}ller} E.,  2001, in {Arnould} M.,
  {Lewitowicz} M.,  {Oganessian} Y.~T.,  {Akimune} H.,  {Ohta} M.,
  {Utsunomiya} H.,  {Wada} T.,   {Yamagata} T.,  eds,  American Institute of
  Physics Conference Series Vol. 561, Symposium on Nuclear Physics IV. pp
  21--32 (\mn@eprint {arXiv} {astro-ph/0011206}), \mn@doi{10.1063/1.1372778}

\bibitem[\protect\citeauthoryear{{Kifonidis}, {Plewa}, {Janka}  \&
  {M{\"u}ller}}{{Kifonidis} et~al.}{2003}]{2003A&A...408..621K}
{Kifonidis} K.,  {Plewa} T.,  {Janka} H.~T.,   {M{\"u}ller} E.,  2003, \mn@doi
  [\aap] {10.1051/0004-6361:20030863}, \href
  {https://ui.adsabs.harvard.edu/abs/2003A&A...408..621K} {408, 621}

\bibitem[\protect\citeauthoryear{{Kifonidis}, {Plewa}, {Scheck}, {Janka}  \&
  {M{\"u}ller}}{{Kifonidis} et~al.}{2006}]{2006A&A...453..661K}
{Kifonidis} K.,  {Plewa} T.,  {Scheck} L.,  {Janka} H.~T.,   {M{\"u}ller} E.,
  2006, \mn@doi [\aap] {10.1051/0004-6361:20054512}, \href
  {https://ui.adsabs.harvard.edu/abs/2006A&A...453..661K} {453, 661}

\bibitem[\protect\citeauthoryear{{Kobayashi}, {Karakas}  \&
  {Lugaro}}{{Kobayashi} et~al.}{2020}]{2020ApJ...900..179K}
{Kobayashi} C.,  {Karakas} A.~I.,   {Lugaro} M.,  2020, \mn@doi [\apj]
  {10.3847/1538-4357/abae65}, \href
  {https://ui.adsabs.harvard.edu/abs/2020ApJ...900..179K} {900, 179}

\bibitem[\protect\citeauthoryear{{Limongi} \& {Chieffi}}{{Limongi} \&
  {Chieffi}}{2003}]{2003ApJ...592..404L}
{Limongi} M.,  {Chieffi} A.,  2003, \mn@doi [\apj] {10.1086/375703}, \href
  {https://ui.adsabs.harvard.edu/abs/2003ApJ...592..404L} {592, 404}

\bibitem[\protect\citeauthoryear{{Limongi} \& {Chieffi}}{{Limongi} \&
  {Chieffi}}{2006}]{2006ApJ...647..483L}
{Limongi} M.,  {Chieffi} A.,  2006, \mn@doi [\apj] {10.1086/505164}, \href
  {https://ui.adsabs.harvard.edu/abs/2006ApJ...647..483L} {647, 483}

\bibitem[\protect\citeauthoryear{{Limongi} \& {Chieffi}}{{Limongi} \&
  {Chieffi}}{2012}]{2012ApJS..199...38L}
{Limongi} M.,  {Chieffi} A.,  2012, \mn@doi [\apjs]
  {10.1088/0067-0049/199/2/38}, \href
  {https://ui.adsabs.harvard.edu/abs/2012ApJS..199...38L} {199, 38}

\bibitem[\protect\citeauthoryear{{Limongi} \& {Chieffi}}{{Limongi} \&
  {Chieffi}}{2018}]{2018ApJS..237...13L}
{Limongi} M.,  {Chieffi} A.,  2018, \mn@doi [\apjs] {10.3847/1538-4365/aacb24},
  \href {https://ui.adsabs.harvard.edu/abs/2018ApJS..237...13L} {237, 13}

\bibitem[\protect\citeauthoryear{{Lippuner} \& {Roberts}}{{Lippuner} \&
  {Roberts}}{2017}]{2017ApJS..233...18L}
{Lippuner} J.,  {Roberts} L.~F.,  2017, \mn@doi [\apjs]
  {10.3847/1538-4365/aa94cb}, \href
  {https://ui.adsabs.harvard.edu/abs/2017ApJS..233...18L} {233, 18}

\bibitem[\protect\citeauthoryear{{MacFadyen} \& {Woosley}}{{MacFadyen} \&
  {Woosley}}{1999}]{1999ApJ...524..262M}
{MacFadyen} A.~I.,  {Woosley} S.~E.,  1999, \mn@doi [\apj] {10.1086/307790},
  \href {https://ui.adsabs.harvard.edu/abs/1999ApJ...524..262M} {524, 262}

\bibitem[\protect\citeauthoryear{{Maeda} \& {Nomoto}}{{Maeda} \&
  {Nomoto}}{2003}]{2003ApJ...598.1163M}
{Maeda} K.,  {Nomoto} K.,  2003, \mn@doi [\apj] {10.1086/378948}, \href
  {https://ui.adsabs.harvard.edu/abs/2003ApJ...598.1163M} {598, 1163}

\bibitem[\protect\citeauthoryear{{Matteucci}}{{Matteucci}}{2003}]{2003Ap&SS.284..539M}
{Matteucci} F.,  2003, \mn@doi [\apss] {10.1023/A:1024089402368}, \href
  {https://ui.adsabs.harvard.edu/abs/2003Ap&SS.284..539M} {284, 539}

\bibitem[\protect\citeauthoryear{{Meskhi}, {Wolfe}, {Dai}, {Fr{\"o}hlich},
  {Miller}, {Wong}  \& {Vilalta}}{{Meskhi} et~al.}{2022}]{2022ApJ...932L...3M}
{Meskhi} M.~M.,  {Wolfe} N.~E.,  {Dai} Z.,  {Fr{\"o}hlich} C.,  {Miller} J.~M.,
   {Wong} R. K.~W.,   {Vilalta} R.,  2022, \mn@doi [\apjl]
  {10.3847/2041-8213/ac7054}, \href
  {https://ui.adsabs.harvard.edu/abs/2022ApJ...932L...3M} {932, L3}

\bibitem[\protect\citeauthoryear{{Mezzacappa}}{{Mezzacappa}}{2005}]{2005ARNPS..55..467M}
{Mezzacappa} A.,  2005, \mn@doi [Annual Review of Nuclear and Particle Science]
  {10.1146/annurev.nucl.55.090704.151608}, \href
  {https://ui.adsabs.harvard.edu/abs/2005ARNPS..55..467M} {55, 467}

\bibitem[\protect\citeauthoryear{{Moriya}, {Tominaga}, {Tanaka}, {Nomoto},
  {Sauer}, {Mazzali}, {Maeda}  \& {Suzuki}}{{Moriya}
  et~al.}{2010}]{2010ApJ...719.1445M}
{Moriya} T.,  {Tominaga} N.,  {Tanaka} M.,  {Nomoto} K.,  {Sauer} D.~N.,
  {Mazzali} P.~A.,  {Maeda} K.,   {Suzuki} T.,  2010, \mn@doi [\apj]
  {10.1088/0004-637X/719/2/1445}, \href
  {https://ui.adsabs.harvard.edu/abs/2010ApJ...719.1445M} {719, 1445}

\bibitem[\protect\citeauthoryear{{Morozova}, {Piro}, {Renzo}, {Ott}, {Clausen},
  {Couch}, {Ellis}  \& {Roberts}}{{Morozova}
  et~al.}{2015}]{2015ApJ...814...63M}
{Morozova} V.,  {Piro} A.~L.,  {Renzo} M.,  {Ott} C.~D.,  {Clausen} D.,
  {Couch} S.~M.,  {Ellis} J.,   {Roberts} L.~F.,  2015, \mn@doi [\apj]
  {10.1088/0004-637X/814/1/63}, \href
  {https://ui.adsabs.harvard.edu/abs/2015ApJ...814...63M} {814, 63}

\bibitem[\protect\citeauthoryear{{M{\"u}ller}}{{M{\"u}ller}}{1986}]{1986A&A...162..103M}
{M{\"u}ller} E.,  1986, \aap, \href
  {https://ui.adsabs.harvard.edu/abs/1986A&A...162..103M} {162, 103}

\bibitem[\protect\citeauthoryear{{M{\"u}ller}}{{M{\"u}ller}}{2016}]{2016PASA...33...48M}
{M{\"u}ller} B.,  2016, \mn@doi [\pasa] {10.1017/pasa.2016.40}, \href
  {https://ui.adsabs.harvard.edu/abs/2016PASA...33...48M} {33, e048}

\bibitem[\protect\citeauthoryear{{M{\"u}ller}}{{M{\"u}ller}}{2020}]{2020LRCA....6....3M}
{M{\"u}ller} B.,  2020, \mn@doi [Living Reviews in Computational Astrophysics]
  {10.1007/s41115-020-0008-5}, \href
  {https://ui.adsabs.harvard.edu/abs/2020LRCA....6....3M} {6, 3}

\bibitem[\protect\citeauthoryear{{M{\"u}ller}, {Heger}, {Liptai}  \&
  {Cameron}}{{M{\"u}ller} et~al.}{2016}]{2016MNRAS.460..742M}
{M{\"u}ller} B.,  {Heger} A.,  {Liptai} D.,   {Cameron} J.~B.,  2016, \mn@doi
  [\mnras] {10.1093/mnras/stw1083}, \href
  {https://ui.adsabs.harvard.edu/abs/2016MNRAS.460..742M} {460, 742}

\bibitem[\protect\citeauthoryear{{M{\"u}ller}, {Melson}, {Heger}  \&
  {Janka}}{{M{\"u}ller} et~al.}{2017a}]{2017MNRAS.472..491M}
{M{\"u}ller} B.,  {Melson} T.,  {Heger} A.,   {Janka} H.-T.,  2017a, \mn@doi
  [\mnras] {10.1093/mnras/stx1962}, \href
  {https://ui.adsabs.harvard.edu/abs/2017MNRAS.472..491M} {472, 491}

\bibitem[\protect\citeauthoryear{{M{\"u}ller}, {Prieto}, {Pejcha}  \&
  {Clocchiatti}}{{M{\"u}ller} et~al.}{2017b}]{2017ApJ...841..127M}
{M{\"u}ller} T.,  {Prieto} J.~L.,  {Pejcha} O.,   {Clocchiatti} A.,  2017b,
  \mn@doi [\apj] {10.3847/1538-4357/aa72f1}, \href
  {https://ui.adsabs.harvard.edu/abs/2017ApJ...841..127M} {841, 127}

\bibitem[\protect\citeauthoryear{{Nagataki}, {Hashimoto}, {Sato}  \&
  {Yamada}}{{Nagataki} et~al.}{1997}]{1997ApJ...486.1026N}
{Nagataki} S.,  {Hashimoto} M.-a.,  {Sato} K.,   {Yamada} S.,  1997, \mn@doi
  [\apj] {10.1086/304565}, \href
  {https://ui.adsabs.harvard.edu/abs/1997ApJ...486.1026N} {486, 1026}

\bibitem[\protect\citeauthoryear{{Nagataki}, {Mizuta}, {Yamada}, {Takabe}  \&
  {Sato}}{{Nagataki} et~al.}{2003}]{2003ApJ...596..401N}
{Nagataki} S.,  {Mizuta} A.,  {Yamada} S.,  {Takabe} H.,   {Sato} K.,  2003,
  \mn@doi [\apj] {10.1086/377530}, \href
  {https://ui.adsabs.harvard.edu/abs/2003ApJ...596..401N} {596, 401}

\bibitem[\protect\citeauthoryear{{Nagataki}, {Mizuta}  \& {Sato}}{{Nagataki}
  et~al.}{2006}]{2006ApJ...647.1255N}
{Nagataki} S.,  {Mizuta} A.,   {Sato} K.,  2006, \mn@doi [\apj]
  {10.1086/505618}, \href
  {https://ui.adsabs.harvard.edu/abs/2006ApJ...647.1255N} {647, 1255}

\bibitem[\protect\citeauthoryear{{Nakamura}, {Umeda}, {Iwamoto}, {Nomoto},
  {Hashimoto}, {Hix}  \& {Thielemann}}{{Nakamura}
  et~al.}{2001}]{2001ApJ...555..880N}
{Nakamura} T.,  {Umeda} H.,  {Iwamoto} K.,  {Nomoto} K.,  {Hashimoto} M.-a.,
  {Hix} W.~R.,   {Thielemann} F.-K.,  2001, \mn@doi [\apj] {10.1086/321495},
  \href {https://ui.adsabs.harvard.edu/abs/2001ApJ...555..880N} {555, 880}

\bibitem[\protect\citeauthoryear{{Nomoto}, {Tominaga}, {Umeda}, {Kobayashi}  \&
  {Maeda}}{{Nomoto} et~al.}{2006}]{2006NuPhA.777..424N}
{Nomoto} K.,  {Tominaga} N.,  {Umeda} H.,  {Kobayashi} C.,   {Maeda} K.,  2006,
  \mn@doi [\nphysa] {10.1016/j.nuclphysa.2006.05.008}, \href
  {https://ui.adsabs.harvard.edu/abs/2006NuPhA.777..424N} {777, 424}

\bibitem[\protect\citeauthoryear{{O'Connor} \& {Ott}}{{O'Connor} \&
  {Ott}}{2011}]{2011ApJ...730...70O}
{O'Connor} E.,  {Ott} C.~D.,  2011, \mn@doi [\apj]
  {10.1088/0004-637X/730/2/70}, \href
  {https://ui.adsabs.harvard.edu/abs/2011ApJ...730...70O} {730, 70}

\bibitem[\protect\citeauthoryear{{Ono}, {Nagataki}, {Ferrand}, {Takahashi},
  {Umeda}, {Yoshida}, {Orlando}  \& {Miceli}}{{Ono} et~al.}{2020}]{Ono+2020}
{Ono} M.,  {Nagataki} S.,  {Ferrand} G.,  {Takahashi} K.,  {Umeda} H.,
  {Yoshida} T.,  {Orlando} S.,   {Miceli} M.,  2020, \mn@doi [\apj]
  {10.3847/1538-4357/ab5dba}, \href
  {https://ui.adsabs.harvard.edu/abs/2020ApJ...888..111O} {888, 111}

\bibitem[\protect\citeauthoryear{{Orlando} et~al.,}{{Orlando}
  et~al.}{2020}]{Orlando+2020}
{Orlando} S.,  et~al., 2020, \mn@doi [\aap] {10.1051/0004-6361/201936718},
  \href {https://ui.adsabs.harvard.edu/abs/2020A&A...636A..22O} {636, A22}

\bibitem[\protect\citeauthoryear{{Paxton}, {Bildsten}, {Dotter}, {Herwig},
  {Lesaffre}  \& {Timmes}}{{Paxton} et~al.}{2011}]{2011ApJS..192....3P}
{Paxton} B.,  {Bildsten} L.,  {Dotter} A.,  {Herwig} F.,  {Lesaffre} P.,
  {Timmes} F.,  2011, \mn@doi [\apjs] {10.1088/0067-0049/192/1/3}, \href
  {https://ui.adsabs.harvard.edu/abs/2011ApJS..192....3P} {192, 3}

\bibitem[\protect\citeauthoryear{{Paxton} et~al.,}{{Paxton}
  et~al.}{2015}]{2015ApJS..220...15P}
{Paxton} B.,  et~al., 2015, \mn@doi [\apjs] {10.1088/0067-0049/220/1/15}, \href
  {https://ui.adsabs.harvard.edu/abs/2015ApJS..220...15P} {220, 15}

\bibitem[\protect\citeauthoryear{{Pejcha} \& {Thompson}}{{Pejcha} \&
  {Thompson}}{2015}]{2015ApJ...801...90P}
{Pejcha} O.,  {Thompson} T.~A.,  2015, \mn@doi [\apj]
  {10.1088/0004-637X/801/2/90}, \href
  {https://ui.adsabs.harvard.edu/abs/2015ApJ...801...90P} {801, 90}

\bibitem[\protect\citeauthoryear{{Perego}, {Hempel}, {Fr{\"o}hlich}, {Ebinger},
  {Eichler}, {Casanova}, {Liebend{\"o}rfer}  \& {Thielemann}}{{Perego}
  et~al.}{2015}]{2015ApJ...806..275P}
{Perego} A.,  {Hempel} M.,  {Fr{\"o}hlich} C.,  {Ebinger} K.,  {Eichler} M.,
  {Casanova} J.,  {Liebend{\"o}rfer} M.,   {Thielemann} F.~K.,  2015, \mn@doi
  [\apj] {10.1088/0004-637X/806/2/275}, \href
  {https://ui.adsabs.harvard.edu/abs/2015ApJ...806..275P} {806, 275}

\bibitem[\protect\citeauthoryear{{Ricks} \& {Dwarkadas}}{{Ricks} \&
  {Dwarkadas}}{2019}]{2019ApJ...880...59R}
{Ricks} W.,  {Dwarkadas} V.~V.,  2019, \mn@doi [\apj]
  {10.3847/1538-4357/ab287c}, \href
  {https://ui.adsabs.harvard.edu/abs/2019ApJ...880...59R} {880, 59}

\bibitem[\protect\citeauthoryear{{Sandoval}, {Hix}, {Messer}, {Lentz}  \&
  {Harris}}{{Sandoval} et~al.}{2021}]{2021ApJ...921..113S}
{Sandoval} M.~A.,  {Hix} W.~R.,  {Messer} O.~E.~B.,  {Lentz} E.~J.,   {Harris}
  J.~A.,  2021, \mn@doi [\apj] {10.3847/1538-4357/ac1d49}, \href
  {https://ui.adsabs.harvard.edu/abs/2021ApJ...921..113S} {921, 113}

\bibitem[\protect\citeauthoryear{{Sawada} \& {Maeda}}{{Sawada} \&
  {Maeda}}{2019}]{2019ApJ...886...47S}
{Sawada} R.,  {Maeda} K.,  2019, \mn@doi [\apj] {10.3847/1538-4357/ab4da3},
  \href {https://ui.adsabs.harvard.edu/abs/2019ApJ...886...47S} {886, 47}

\bibitem[\protect\citeauthoryear{{Scheck}, {Kifonidis}, {Janka}  \&
  {M{\"u}ller}}{{Scheck} et~al.}{2006}]{2006A&A...457..963S}
{Scheck} L.,  {Kifonidis} K.,  {Janka} H.-T.,   {M{\"u}ller} E.,  2006, \mn@doi
  [\aap] {10.1051/0004-6361:20064855}, \href
  {http://adsabs.harvard.edu/abs/2006A%26A...457..963S} {457, 963}

\bibitem[\protect\citeauthoryear{{Schneider}, {Podsiadlowski}  \&
  {M{\"u}ller}}{{Schneider} et~al.}{2021}]{2021A&A...645A...5S}
{Schneider} F.~R.~N.,  {Podsiadlowski} P.,   {M{\"u}ller} B.,  2021, \mn@doi
  [\aap] {10.1051/0004-6361/202039219}, \href
  {https://ui.adsabs.harvard.edu/abs/2021A&A...645A...5S} {645, A5}

\bibitem[\protect\citeauthoryear{{Shigeyama}, {Nomoto}  \&
  {Hashimoto}}{{Shigeyama} et~al.}{1988}]{1988A&A...196..141S}
{Shigeyama} T.,  {Nomoto} K.,   {Hashimoto} M.,  1988, \aap, \href
  {https://ui.adsabs.harvard.edu/abs/1988A&A...196..141S} {196, 141}

\bibitem[\protect\citeauthoryear{{Shimizu}, {Ebisuzaki}, {Sato}  \&
  {Yamada}}{{Shimizu} et~al.}{2001}]{2001ApJ...552..756S}
{Shimizu} T.~M.,  {Ebisuzaki} T.,  {Sato} K.,   {Yamada} S.,  2001, \mn@doi
  [\apj] {10.1086/320544}, \href
  {https://ui.adsabs.harvard.edu/abs/2001ApJ...552..756S} {552, 756}

\bibitem[\protect\citeauthoryear{{Siegel}, {Barnes}  \& {Metzger}}{{Siegel}
  et~al.}{2019}]{2019Natur.569..241S}
{Siegel} D.~M.,  {Barnes} J.,   {Metzger} B.~D.,  2019, \mn@doi [\nat]
  {10.1038/s41586-019-1136-0}, \href
  {https://ui.adsabs.harvard.edu/abs/2019Natur.569..241S} {569, 241}

\bibitem[\protect\citeauthoryear{{Sukhbold} \& {Woosley}}{{Sukhbold} \&
  {Woosley}}{2014}]{2014ApJ...783...10S}
{Sukhbold} T.,  {Woosley} S.~E.,  2014, \mn@doi [\apj]
  {10.1088/0004-637X/783/1/10}, \href
  {https://ui.adsabs.harvard.edu/abs/2014ApJ...783...10S} {783, 10}

\bibitem[\protect\citeauthoryear{{Sukhbold}, {Ertl}, {Woosley}, {Brown}  \&
  {Janka}}{{Sukhbold} et~al.}{2016}]{2016ApJ...821...38S}
{Sukhbold} T.,  {Ertl} T.,  {Woosley} S.~E.,  {Brown} J.~M.,   {Janka} H.~T.,
  2016, \mn@doi [\apj] {10.3847/0004-637X/821/1/38}, \href
  {https://ui.adsabs.harvard.edu/abs/2016ApJ...821...38S} {821, 38}

\bibitem[\protect\citeauthoryear{{Suwa}, {Tominaga}  \& {Maeda}}{{Suwa}
  et~al.}{2019}]{2019MNRAS.483.3607S}
{Suwa} Y.,  {Tominaga} N.,   {Maeda} K.,  2019, \mn@doi [\mnras]
  {10.1093/mnras/sty3309}, \href
  {https://ui.adsabs.harvard.edu/abs/2019MNRAS.483.3607S} {483, 3607}

\bibitem[\protect\citeauthoryear{{The LIGO Scientific Collaboration}
  et~al.,}{{The LIGO Scientific Collaboration}
  et~al.}{2021}]{2021arXiv211103606T}
{The LIGO Scientific Collaboration} et~al., 2021, arXiv e-prints, \href
  {https://ui.adsabs.harvard.edu/abs/2021arXiv211103606T} {p. arXiv:2111.03606}

\bibitem[\protect\citeauthoryear{{Thielemann}, {Hashimoto}  \&
  {Nomoto}}{{Thielemann} et~al.}{1990}]{1990ApJ...349..222T}
{Thielemann} F.-K.,  {Hashimoto} M.-A.,   {Nomoto} K.,  1990, \mn@doi [\apj]
  {10.1086/168308}, \href
  {https://ui.adsabs.harvard.edu/abs/1990ApJ...349..222T} {349, 222}

\bibitem[\protect\citeauthoryear{{Thielemann}, {Nomoto}  \&
  {Hashimoto}}{{Thielemann} et~al.}{1996}]{1996ApJ...460..408T}
{Thielemann} F.-K.,  {Nomoto} K.,   {Hashimoto} M.-A.,  1996, \mn@doi [\apj]
  {10.1086/176980}, \href
  {https://ui.adsabs.harvard.edu/abs/1996ApJ...460..408T} {460, 408}

\bibitem[\protect\citeauthoryear{{Timmes}, {Woosley}  \& {Weaver}}{{Timmes}
  et~al.}{1995}]{1995ApJS...98..617T}
{Timmes} F.~X.,  {Woosley} S.~E.,   {Weaver} T.~A.,  1995, \mn@doi [\apjs]
  {10.1086/192172}, \href
  {https://ui.adsabs.harvard.edu/abs/1995ApJS...98..617T} {98, 617}

\bibitem[\protect\citeauthoryear{{Ugliano}, {Janka}, {Marek}  \&
  {Arcones}}{{Ugliano} et~al.}{2012}]{2012ApJ...757...69U}
{Ugliano} M.,  {Janka} H.-T.,  {Marek} A.,   {Arcones} A.,  2012, \mn@doi
  [\apj] {10.1088/0004-637X/757/1/69}, \href
  {https://ui.adsabs.harvard.edu/abs/2012ApJ...757...69U} {757, 69}

\bibitem[\protect\citeauthoryear{{Umeda} \& {Nomoto}}{{Umeda} \&
  {Nomoto}}{2008}]{2008ApJ...673.1014U}
{Umeda} H.,  {Nomoto} K.,  2008, \mn@doi [\apj] {10.1086/524767}, \href
  {https://ui.adsabs.harvard.edu/abs/2008ApJ...673.1014U} {673, 1014}

\bibitem[\protect\citeauthoryear{{Valerin} et~al.,}{{Valerin}
  et~al.}{2022}]{Valerin+2022}
{Valerin} G.,  et~al., 2022, \mn@doi [\mnras] {10.1093/mnras/stac1182}, \href
  {https://ui.adsabs.harvard.edu/abs/2022MNRAS.513.4983V} {513, 4983}

\bibitem[\protect\citeauthoryear{{Weaver}, {Zimmerman}  \& {Woosley}}{{Weaver}
  et~al.}{1978}]{1978ApJ...225.1021W}
{Weaver} T.~A.,  {Zimmerman} G.~B.,   {Woosley} S.~E.,  1978, \mn@doi [\apj]
  {10.1086/156569}, \href
  {https://ui.adsabs.harvard.edu/abs/1978ApJ...225.1021W} {225, 1021}

\bibitem[\protect\citeauthoryear{{Wirth}, {Jerabkova}, {Yan}, {Kroupa}, {Haas}
  \& {{\v{S}}ubr}}{{Wirth} et~al.}{2021}]{2021MNRAS.506.4131W}
{Wirth} H.,  {Jerabkova} T.,  {Yan} Z.,  {Kroupa} P.,  {Haas} J.,
  {{\v{S}}ubr} L.,  2021, \mn@doi [\mnras] {10.1093/mnras/stab2011}, \href
  {https://ui.adsabs.harvard.edu/abs/2021MNRAS.506.4131W} {506, 4131}

\bibitem[\protect\citeauthoryear{{Wongwathanarat}, {Janka}  \&
  {M{\"u}ller}}{{Wongwathanarat} et~al.}{2010}]{2010ApJ...725L.106W}
{Wongwathanarat} A.,  {Janka} H.-T.,   {M{\"u}ller} E.,  2010, \mn@doi [\apjl]
  {10.1088/2041-8205/725/1/L106}, \href
  {https://ui.adsabs.harvard.edu/abs/2010ApJ...725L.106W} {725, L106}

\bibitem[\protect\citeauthoryear{{Wongwathanarat}, {Janka}  \&
  {M{\"u}ller}}{{Wongwathanarat} et~al.}{2013}]{2013A&A...552A.126W}
{Wongwathanarat} A.,  {Janka} H.~T.,   {M{\"u}ller} E.,  2013, \mn@doi [\aap]
  {10.1051/0004-6361/201220636}, \href
  {https://ui.adsabs.harvard.edu/abs/2013A&A...552A.126W} {552, A126}

\bibitem[\protect\citeauthoryear{{Wongwathanarat}, {M{\"u}ller}  \&
  {Janka}}{{Wongwathanarat} et~al.}{2015}]{2015A&A...577A..48W}
{Wongwathanarat} A.,  {M{\"u}ller} E.,   {Janka} H.~T.,  2015, \mn@doi [\aap]
  {10.1051/0004-6361/201425025}, \href
  {https://ui.adsabs.harvard.edu/abs/2015A&A...577A..48W} {577, A48}

\bibitem[\protect\citeauthoryear{{Wongwathanarat}, {Janka}, {M{\"u}ller},
  {Pllumbi}  \& {Wanajo}}{{Wongwathanarat} et~al.}{2017}]{2017ApJ...842...13W}
{Wongwathanarat} A.,  {Janka} H.-T.,  {M{\"u}ller} E.,  {Pllumbi} E.,
  {Wanajo} S.,  2017, \mn@doi [\apj] {10.3847/1538-4357/aa72de}, \href
  {https://ui.adsabs.harvard.edu/abs/2017ApJ...842...13W} {842, 13}

\bibitem[\protect\citeauthoryear{{Woosley}}{{Woosley}}{1988}]{1988ApJ...330..218W}
{Woosley} S.~E.,  1988, \mn@doi [\apj] {10.1086/166468}, \href
  {https://ui.adsabs.harvard.edu/abs/1988ApJ...330..218W} {330, 218}

\bibitem[\protect\citeauthoryear{{Woosley} \& {Heger}}{{Woosley} \&
  {Heger}}{2007}]{2007PhR...442..269W}
{Woosley} S.~E.,  {Heger} A.,  2007, \mn@doi [\physrep]
  {10.1016/j.physrep.2007.02.009}, \href
  {https://ui.adsabs.harvard.edu/abs/2007PhR...442..269W} {442, 269}

\bibitem[\protect\citeauthoryear{{Woosley} \& {Hoffman}}{{Woosley} \&
  {Hoffman}}{1992}]{1992ApJ...395..202W}
{Woosley} S.~E.,  {Hoffman} R.~D.,  1992, \mn@doi [\apj] {10.1086/171644},
  \href {https://ui.adsabs.harvard.edu/abs/1992ApJ...395..202W} {395, 202}

\bibitem[\protect\citeauthoryear{{Woosley} \& {Janka}}{{Woosley} \&
  {Janka}}{2005}]{2005NatPh...1..147W}
{Woosley} S.,  {Janka} T.,  2005, \mn@doi [Nature Physics] {10.1038/nphys172},
  \href {https://ui.adsabs.harvard.edu/abs/2005NatPh...1..147W} {1, 147}

\bibitem[\protect\citeauthoryear{{Woosley} \& {Weaver}}{{Woosley} \&
  {Weaver}}{1995}]{1995ApJS..101..181W}
{Woosley} S.~E.,  {Weaver} T.~A.,  1995, \mn@doi [\apjs] {10.1086/192237},
  \href {https://ui.adsabs.harvard.edu/abs/1995ApJS..101..181W} {101, 181}

\bibitem[\protect\citeauthoryear{{Woosley}, {Heger}  \& {Weaver}}{{Woosley}
  et~al.}{2002}]{2002RvMP...74.1015W}
{Woosley} S.~E.,  {Heger} A.,   {Weaver} T.~A.,  2002, \mn@doi [Reviews of
  Modern Physics] {10.1103/RevModPhys.74.1015}, \href
  {https://ui.adsabs.harvard.edu/abs/2002RvMP...74.1015W} {74, 1015}

\bibitem[\protect\citeauthoryear{{Woosley}, {Sukhbold}  \& {Janka}}{{Woosley}
  et~al.}{2020}]{2020ApJ...896...56W}
{Woosley} S.~E.,  {Sukhbold} T.,   {Janka} H.~T.,  2020, \mn@doi [\apj]
  {10.3847/1538-4357/ab8cc1}, \href
  {https://ui.adsabs.harvard.edu/abs/2020ApJ...896...56W} {896, 56}

\bibitem[\protect\citeauthoryear{{Yamamoto}, {Fujimoto}, {Nagakura}  \&
  {Yamada}}{{Yamamoto} et~al.}{2013}]{2013ApJ...771...27Y}
{Yamamoto} Y.,  {Fujimoto} S.-i.,  {Nagakura} H.,   {Yamada} S.,  2013, \mn@doi
  [\apj] {10.1088/0004-637X/771/1/27}, \href
  {https://ui.adsabs.harvard.edu/abs/2013ApJ...771...27Y} {771, 27}

\bibitem[\protect\citeauthoryear{{Yang} et~al.,}{{Yang}
  et~al.}{2021}]{2021A&A...655A..90Y}
{Yang} S.,  et~al., 2021, \mn@doi [\aap] {10.1051/0004-6361/202141244}, \href
  {https://ui.adsabs.harvard.edu/abs/2021A&A...655A..90Y} {655, A90}

\bibitem[\protect\citeauthoryear{{Young} \& {Fryer}}{{Young} \&
  {Fryer}}{2007}]{2007ApJ...664.1033Y}
{Young} P.~A.,  {Fryer} C.~L.,  2007, \mn@doi [\apj] {10.1086/518081}, \href
  {https://ui.adsabs.harvard.edu/abs/2007ApJ...664.1033Y} {664, 1033}

\bibitem[\protect\citeauthoryear{{Zhang}, {Woosley}  \& {Heger}}{{Zhang}
  et~al.}{2008}]{2008ApJ...679..639Z}
{Zhang} W.,  {Woosley} S.~E.,   {Heger} A.,  2008, \mn@doi [\apj]
  {10.1086/526404}, \href
  {https://ui.adsabs.harvard.edu/abs/2008ApJ...679..639Z} {679, 639}

\bibitem[\protect\citeauthoryear{{van der Walt}, {Colbert}  \&
  {Varoquaux}}{{van der Walt} et~al.}{2011}]{2011CSE....13b..22V}
{van der Walt} S.,  {Colbert} S.~C.,   {Varoquaux} G.,  2011, \mn@doi
  [Computing in Science and Engineering] {10.1109/MCSE.2011.37}, \href
  {https://ui.adsabs.harvard.edu/abs/2011CSE....13b..22V} {13, 22}

\makeatother
\end{thebibliography}








\bsp	
\label{lastpage}
\end{document}